\documentclass[10pt,twocolumn]{article}

\usepackage[a4paper,%
            margin=1.5cm,     % 1.5cm margins all around
            includehead,      % include header in margin calculation
            includefoot]{geometry}
\usepackage[utf8]{inputenc}
\usepackage[T1]{fontenc}
\usepackage[numbers,sort&compress]{natbib}
\usepackage{graphicx}
\usepackage{amsmath,amssymb}
\usepackage[colorlinks]{hyperref}
\usepackage[ruled,vlined]{algorithm2e}
\usepackage{listings}
\usepackage{subfig}
\usepackage{mathtools}
\usepackage{float}
\usepackage{cancel}
\usepackage[percent]{overpic}
\usepackage{authblk}           % <-- for multiple authors/affiliations

\usepackage{tablefootnote}

\makeatletter
\renewcommand\@makefnmark{}
\makeatother

\newcommand{\inlinecite}[1]{\citenum{#1}}

\lstdefinelanguage{pseudo}{
    morekeywords={Function,If,Else,EndIf,While,EndWhile,Return},
    sensitive=false
}
\title{Orbit-averaging and deposition accuracy for runaway electron beams in hybrid kinetic-MHD simulations of the runaway plateau} 

\author[*1]{O. E. L\'opez}
\author[2]{D. Vargun}
\author[2]{C. D. Hauck}
\author[1]{M. T. Beidler}

\affil[1]{Fusion Energy Division, Oak Ridge National Laboratory, Oak Ridge, TN 37831, USA}
\affil[2]{Computer Science and Mathematics Division, Oak Ridge National Laboratory, Oak Ridge, TN 37831, USA}

\date{}

\begin{document}

% -------------------------------------------------
% Single‐column title + abstract
% -------------------------------------------------
\twocolumn[
  \maketitle
  \vspace{-3em}
  %— manually typeset the abstract title to match \section (article class uses \Large\bfseries) —
  {\centering\Large\bfseries Abstract\par}
  \vspace{1em}
  We develop a new procedure that combines the Kinetic Orbit Runaway electrons Code (KORC) and the NIMROD extended-MHD code to simulate runaway electrons (REs) in the post-disruption plateau. 
  KORC integrates guiding-center orbits using a barycentric-based binary search strategy to generate initial guesses for the Newton-Raphson logical-to-physical coordinate inversion, 
  guaranteeing reliable particle-to-mesh mapping in NIMROD, whose fields remain static for the present study. 
  Samples are drawn in accord with experimental parallel current profiles of RE beams during the plateau phase. 
  Deposition in NIMROD is verified through comparison with a Python-based finite element code that ensures periodicity in the poloidal direction and continuity at the magnetic axis. 
  Accurate representation of near-axis fields requires finer mesh resolution to prevent under- and overshoots in current density from orbit inaccuracies. 
  Yet, at a fixed particle count, increasing mesh resolution amplifies statistical noise in the deposited fields. 
  An orbit-averaging method accumulates partial current deposits over multiple kinetic steps and reduces the statistical noise with little added computational cost. 
  By coupling kinetic routines from KORC directly into the NIMROD codebase, these developments lay essential groundwork for future self-consistent KORC-NIMROD coupling.

  \vspace{2em}
]

\section{Introduction}

\footnotetext{This manuscript has been authored by UT-Battelle, LLC, under contract DE-AC05-00OR22725 with the US Department of Energy (DOE). 
The US government retains and the publisher, by accepting the work for publication, acknowledges that the US government retains a non-exclusive, paid-up, irrevocable, 
world-wide license to publish or reproduce the submitted manuscript version of this work, or allow others to do so, for US government purposes. DOE will provide public access to these results of 
federally sponsored research in accordance with the DOE Public Access Plan (http://energy.gov/downloads/doe-public-access-plan). \\ *Corresponding author: lopezortizoe@ornl.gov}

In present-day tokamaks, runaway electrons (REs) are created when a critical electric field accelerates a fraction of the electron population to multi-MeV energies, overcoming collisional drag \cite{dreicer1959i,dreicer1959ii,connor1975}.
This process can occur during plasma disruptions, where rapid cooling and the resulting increase in plasma resistivity induce strong toroidal electric fields that drive RE beam formation. This initial RE beam can then be amplified via an avalanche process
\cite{sokolov1979,jayakumar1993,rosenbluth1997,chiu1998}. 
After the thermal and current quenches, a fraction of the pre-disruption current, instead of decaying to zero, is sustained by REs forming a quiescent phase known as the RE plateau. 
In DIII-D, active control of the RE beam can sustain a plateau lasting hundreds of milliseconds \cite{eidietis2012}. Modeling RE dynamics during the plateau phase is essential for predicting beam evolution and devising mitigation strategies to safeguard plasma-facing components and blankets.

RE dynamics in magnetohydrodynamic (MHD) plasmas require numerical models that capture relativistic kinetic behavior
and self-consistently couple to bulk plasma. Fluid-based RE models \cite{cai2015,matsuyama2017,bandaru2019,zhao2021,liu21,sainterme2024} 
evolve a RE bulk density while using a reduced momentum-space description. 
By contrast, $\delta f$ methods devised for energetic ions \cite{todo98,todo04,kim2004hybrid,kim2008,todo21,liu22} have evolved into full-$f$ approaches, such as MEGA \cite{bierwage13} and JOREK \cite{bogaarts22}. Full-$f$ approaches are now being repurposed for REs in JOREK \cite{bergstrom2025} and M3D-C1-K \cite{liu23}. 

Ultimately, we aim to build a self-consistent kinetic-MHD capability for RE mitigation in evolving MHD fields, coupling relativistic guiding-center RE dynamics from the Kinetic Orbit Runaway electrons Code (KORC)\cite{beidler20} with NIMROD \cite{sovinec04}, an extended-MHD finite element code that evolves the bulk plasma's fluid equations. 
While fully self-consistent KORC-NIMROD coupling remains a long-term objective, the RE plateau, with its steady-state fields, provides an ideal environment for testing and refining specific aspects of the coupling. Focusing on a representative post-disruption equilibrium allows us to develop and verify essential elements of the coupling before attempting dynamic field evolution. This approach also provides initial conditions for future kinetic-MHD simulations that will incorporate evolving fields.

In this work, we construct a relativistic guiding-center RE distribution using particles sampled from post-thermal-quench experimental reconstructions of DIII-D shot 184602. Before evolving the distribution, we verify that our deposition routines accurately match the experimentally inferred current profiles. We then observe that coarse finite element meshes can yield unphysical under- or overshoots near the magnetic axis when tracing guiding-center orbits. This effect necessitates finer mesh resolution to maintain fidelity. Such refinement suppresses the artificial peaks and troughs in current density but increases statistical noise if the total number of particles remains fixed. To address this trade-off, we reduce the required particle count without compromising accuracy by accumulating partial deposits of the current over multiple guiding-center steps (an orbit-averaging method). We also implement a barycentric search algorithm to accelerate and improve the reliability of the particle-to-mesh interpolation. Although this study keeps the MHD fields fixed in time, these developments enable dynamic kinetic-MHD coupling between KORC and NIMROD by developing methods to limit numerical noise that could otherwise restrict the convergence of numerical solvers.

The remainder of the paper is organized as follows: Section\,\ref{sec:sampleMHD} presents a methodology to sample RE beam distributions based on an MHD equilibrium. Section\,\ref{sec:barycentric} introduces the barycentric search algorithm and its implementation in NIMROD. Section\,\ref{sec:deposition} discusses the particle deposition methodology, including mathematical formulations and verification tests. Section\,\ref{sec:nimrod_dep} discusses the guiding-center orbit-averaging method applied to DIII-D RE beams, together with error quantification and efficiency analysis. Finally, in Section\,\ref{sec:conclusions} we summarize our efforts to model REs and outline directions for future work.

%%%%%%%%%%%%%%%%%%%%%%%%%%%%%%%%%%%%%%%%%%%%%%%%%%%%%%%%%%%%%%%%%%%%%%%%%%%%%%%
%%%%%%%%%%%%%%%%%%%%%%%%%%%%%%%%%%%%%%%%%%%%%%%%%%%%%%%%%%%%%%%%%%%%%%%%%%%%%%%
%%%%%%%%%%%%%%%%%%%%%%%%%%%%%%%%%%%%%%%%%%%%%%%%%%%%%%%%%%%%%%%%%%%%%%%%%%%%%%%
%%%%%%%%%%%%%%%%%%%%%%%%%%%%%%%%%%%%%%%%%%%%%%%%%%%%%%%%%%%%%%%%%%%%%%%%%%%%%%%

\section{Runaway sampling based on equilibrium profiles}
\label{sec:sampleMHD}

We develop a method to sample the MHD equilibrium so that the initial distribution of relativistic guiding-centers reproduces the experimentally reconstructed parallel-current profile in the RE plateau. This ensures that the net RE current matches the experiment in a nearly force-free regime and places the beam in a known Grad-Shafranov \cite{grad1958,shafranov1966} solution. Our approach thus provides consistent initial conditions before evolving the runaway beam to a quasi-static configuration in Sec.\,\ref{sec:nimrod_dep}. 

References \inlinecite{bierwage12,bierwage13} introduced an orbit-based initialization that discretizes phase space along constants of motion and places markers on unperturbed guiding-center orbits, distributing them uniformly in time. This procedure yields a time-independent marker distribution that reduces statistical fluctuations in hybrid kinetic-MHD simulations of energetic particles. Our particle initialization approach parallels Ref. \inlinecite{bierwage12,bierwage13} in the sense that we combine Metropolis-Hastings-based sampling (Sec.\,\ref{sec:MH}) plus orbit-averaging (Sec.\,\ref{sec:orbit-averaging}) to achieve a similarly low-noise representation of the distribution. The core idea of initializing markers according to physical trajectories to capture dynamics applies to both methods, although the details of how we sample the orbits and manage weights differ.

Because the KORC-NIMROD coupling is not yet fully self-consistent, as fields with kinetic RE effects are not presently being evolved in time, additional work is required to confirm our initialization's performance in fully coupled kinetic-MHD simulations. However, the orbit-averaging method of Sec.\,\ref{sec:orbit-averaging} significantly reduces noise levels in the modeling of the RE plateau in this work.

\subsection{Relativistic guiding-center equations}\label{subsec:RelativisticGuidingCenter}

Throughout this work, we adopt cylindrical coordinates $(R, \phi, Z)$ referenced to the symmetry axis of a tokamak, with $R$ measured outward from the axis, $\phi$ the toroidal angle, and $Z = 0$ corresponding to the midplane.
We employ the relativistic guiding-center (RGC) equations to evolve the guiding-centers of REs. These equations determine the guiding-center position $\boldsymbol{X}=\boldsymbol{X}(t)=(R(t), \phi(t), Z(t))$, and the momentum parallel to the magnetic field  $p_{\parallel}=p_{\parallel}(t)$, according to:
\begin{align}\label{eq:drgc}
    \frac{d\boldsymbol{X}}{dt} = \frac{1}{\hat{\boldsymbol{b}} \cdot \boldsymbol{B}^*} \bigg( & q_e\,\boldsymbol{E} \times \hat{\boldsymbol{b}} + \frac{m\,\mu\,\hat{\boldsymbol{b}} \times \nabla B + p_{\parallel}\,\boldsymbol{B}^*}{m\,\gamma} \bigg),
\end{align}
\begin{equation}\label{eq:dpllgc}
    \frac{d p_{\parallel}}{dt} = \frac{\boldsymbol{B}^*}{\hat{\boldsymbol{b}} \cdot \boldsymbol{B}^*} \cdot \left( q_e\,\boldsymbol{E}  - \frac{\mu\,\nabla B}{\gamma} \right),
\end{equation}
while the magnetic moment $\mu$, defined in Eq.\,\eqref{eq:magnetic_moment_in_text}, is an invariant of the motion:
\begin{equation}\label{eq:dmu}
    \frac{d \mu}{dt} = 0.
\end{equation}
The initial conditions $\boldsymbol{X}(0)$,  $p_{\parallel}(0) $, and $\mu(0)$ are determined by the sampling procedure described in Sec.\,\ref{sec:MH}. Here, $\boldsymbol{E}=\boldsymbol{E}(\boldsymbol{X})$ and $\boldsymbol{B}=\boldsymbol{B}(\boldsymbol{X})$ are prescribed electric and magnetic fields, respectively. However, note that all simulations presented in this work assume no loop-voltage acceleration ($\boldsymbol{E}=0$). The unitary vector along the magnetic field is $\hat{\boldsymbol{b}}(\boldsymbol{X}) = \boldsymbol{B}(\boldsymbol{X}) / B(\boldsymbol{X})$ where 
$B(\boldsymbol{X}) = ||\boldsymbol{B}(\boldsymbol{X})||$ denotes the magnetic field magnitude. The Lorentz factor is
\begin{equation}\label{eq:gamma}
    \gamma \;=\; \sqrt{\,1 \;+\; \Bigl(\tfrac{p_{\parallel}}{m\,c}\Bigr)^2 \;+\; \tfrac{2\,\mu\,B}{m\,c^2}\,},
\end{equation}
where $c$ is the speed of light, and $m$ is the electron mass.  The relativistic momentum is
\begin{equation}
   \boldsymbol{p} \;=\; \gamma\,m\,\boldsymbol{V},
\end{equation}
where $\boldsymbol{V}$ is the electron velocity in the laboratory frame. The electron charge is $q_e <0$. The magnetic moment is
\begin{equation}\label{eq:magnetic_moment_in_text}
    \mu \;=\; \frac{p_{\perp}^2}{2\,m\,B},
\end{equation}
where $p_{\perp}$ is the momentum perpendicular to $\boldsymbol{B}$.  The pitch angle $\eta \in [0, \pi]$ is defined as the angle between the momentum vector $\boldsymbol{p}$ and the magnetic field direction $\hat{\boldsymbol{b}}$. It determines the decomposition:
\begin{equation}
  p_{\parallel} \;=\;  |\boldsymbol{p}| \,\cos(\eta), 
  \quad
  p_{\perp} \;=\;  |\boldsymbol{p}| \,\sin(\eta).
\end{equation}
We also define
\begin{equation}\label{eq:Bstar}
    \boldsymbol{B}^*
    \;=\;
    q_e\,\boldsymbol{B}
    \;+\;
    p_{\parallel}\;\nabla \times \hat{\boldsymbol{b}}.
\end{equation}

We build on the original work of Ref.~\inlinecite{izzo11}, which first implemented a high aspect-ratio approximation of the RGC equations in NIMROD and solved them using the numerical integrator LSODE \cite{Hindmarsh1980}. Following Ref. \inlinecite{beidler20}, we re-implemented the RGC equations without aspect-ratio assumptions and using a Cash-Karp fifth-order Runge-Kutta integrator \cite{cash1990variable}. Consistent with Ref.~\inlinecite{izzo11}, we employ domain decomposition where each MPI rank stores a local copy of the global MHD fields and advances a subset of REs throughout the entire simulation. This strategy, similarly adopted in $\delta f$ hybrid kinetic-MHD particle-in-cell (PIC) simulations in NIMROD~\cite{taylor2022} and GPU-accelerated M3D-C1-K simulations~\cite{liu22}, reduces MPI-based and GPU-based data-transfer overhead. In contrast, approaches such as the original $\delta f$ kinetic-MHD PIC code in NIMROD of Ref.~\inlinecite{kim2004hybrid}, which distribute subsets of MHD finite element data across ranks, require more involved search algorithms (e.g., bucket sorting). Our work also generalizes the approaches of Ref.~\inlinecite{diego18}, in which standalone KORC advanced NIMROD-generated fields stored in a mesh file, and Ref.~\inlinecite{cornille21}, where KORC accessed NIMROD fields via the Fusion-IO interface \cite{fusionio}.

\subsection{Runaway plateau sampling for DIII-D shot 184602}\label{sec:MH}

\begin{figure}[ht]
    \centering
    \subfloat{
        \begin{overpic}[width=0.4\columnwidth]{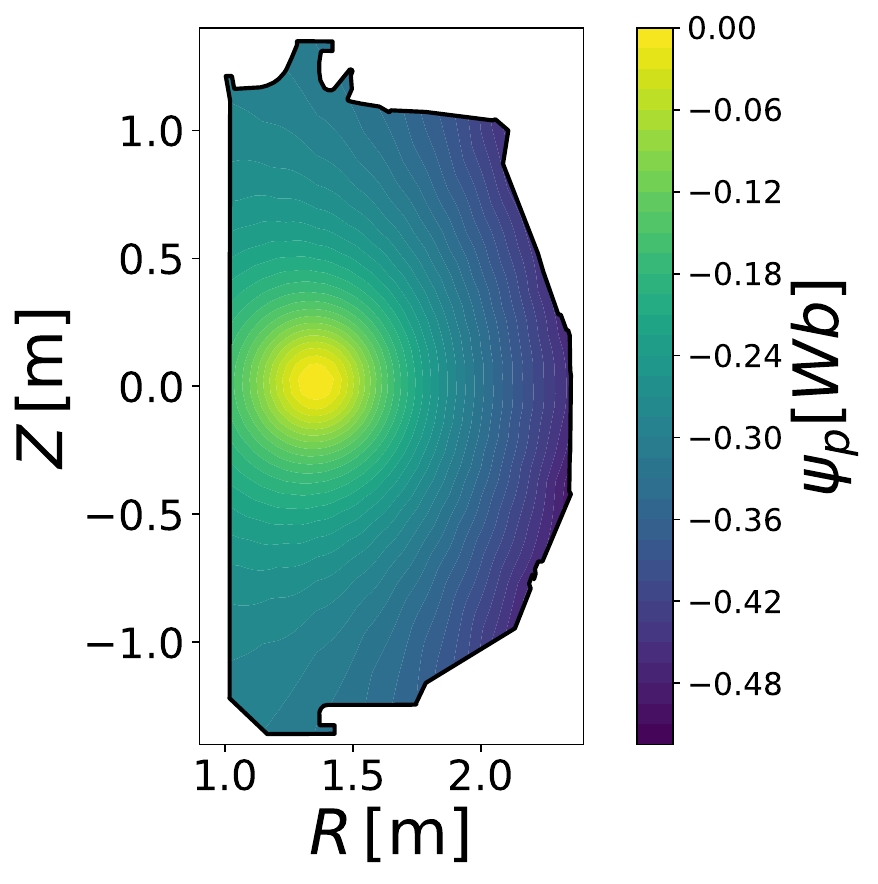}
            \put(2,86){\small{(a)}}
        \end{overpic}
        \label{fig:j_psi_184602}
    }
    \hspace{0.005\textwidth}
    \subfloat{
        \begin{overpic}[width=0.4\columnwidth]{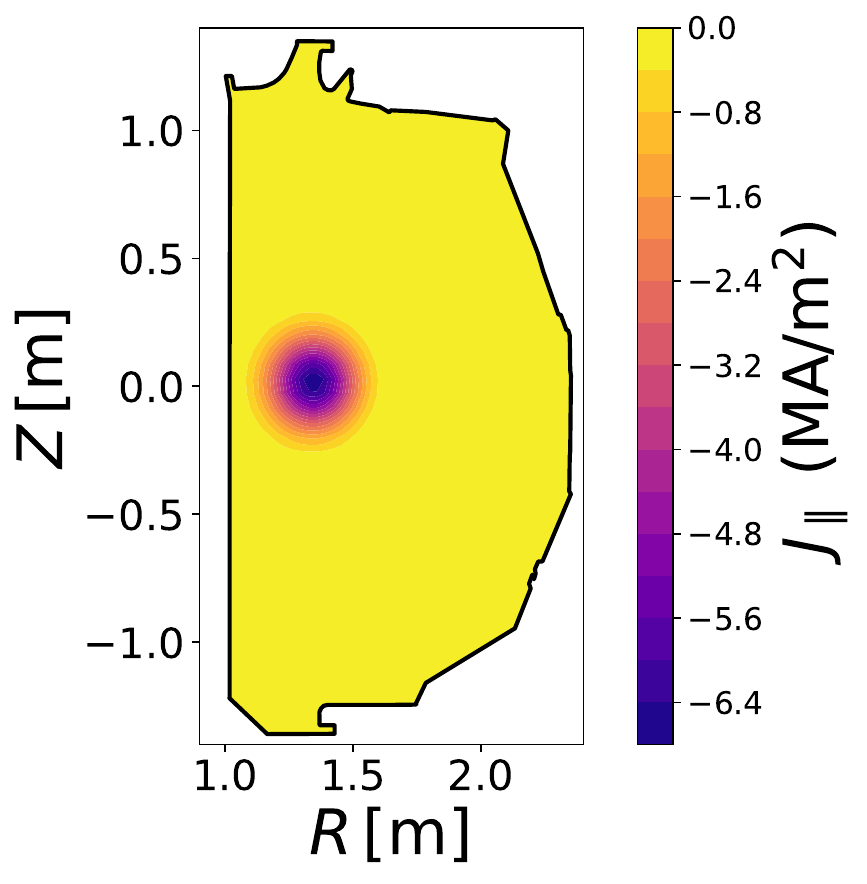}
            \put(2,86){\small{(b)}}
        \end{overpic}
        \label{fig:j_par_184602}
    }\\[1ex]
    \subfloat{
        \begin{overpic}[width=0.4\columnwidth]{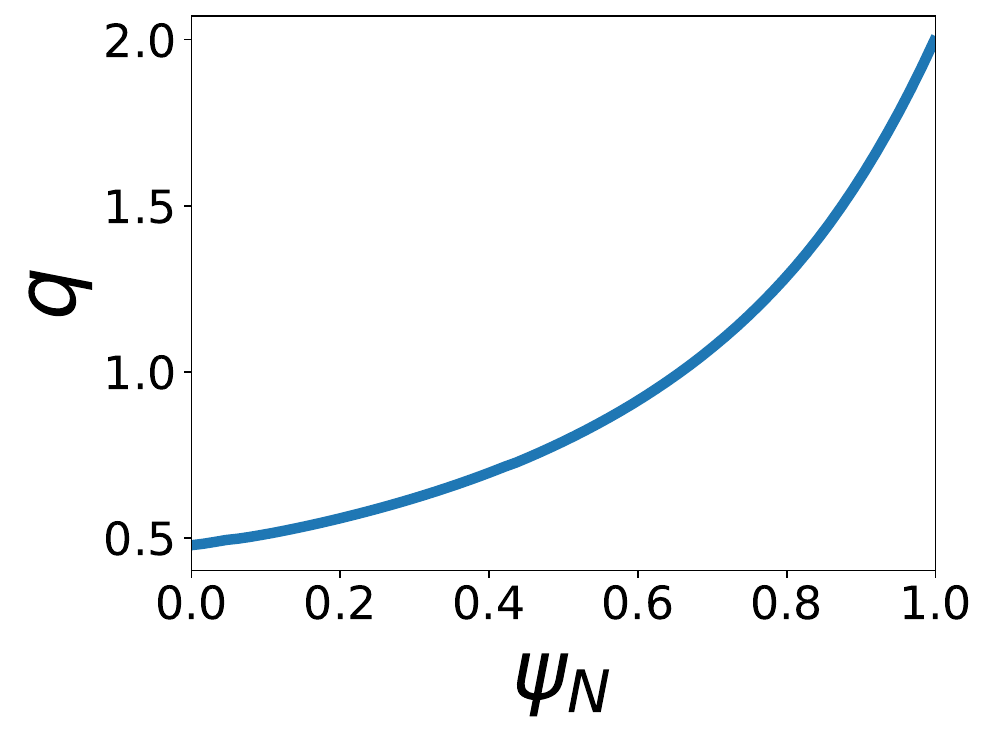}
            \put(2,80){\small{(c)}}
        \end{overpic}
        \label{fig:q_184602}
    }
    \hspace{0.005\textwidth}
    \subfloat{%
        \begin{overpic}[width=0.4\columnwidth]{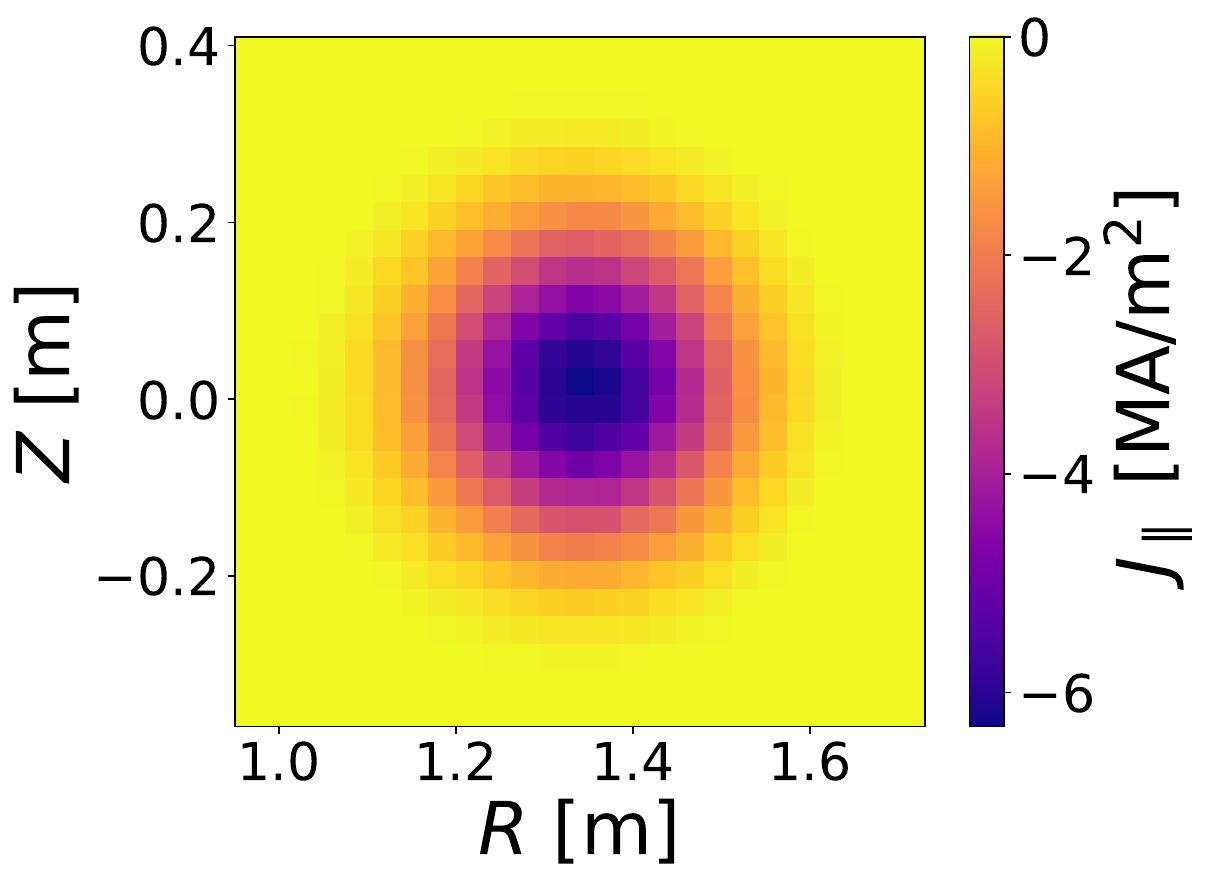}
            \put(2,80){\small{(d)}}
        \end{overpic}
        \label{fig:j_par_sampled_184602}
    }\\[1ex]
    \subfloat{%
        \begin{overpic}[width=0.4\columnwidth]{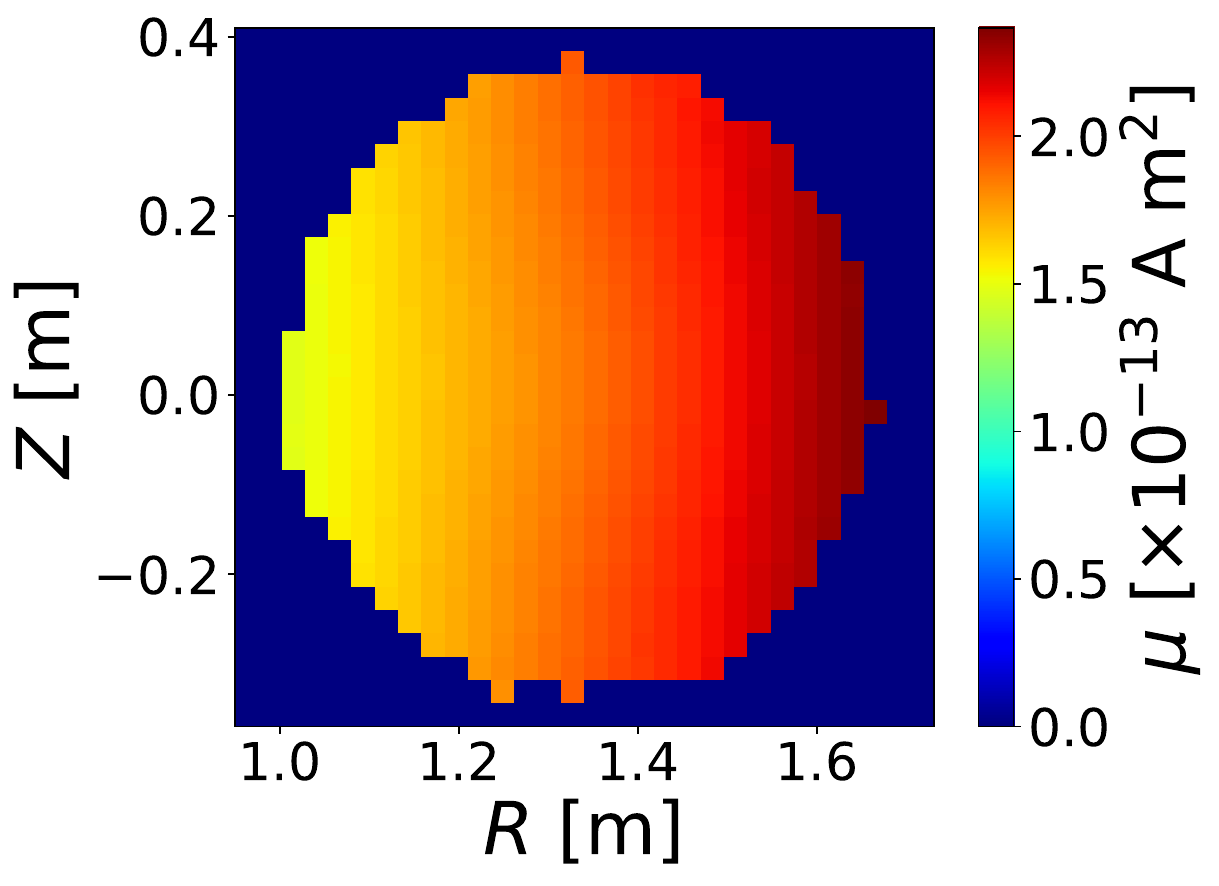}
            \put(2,78){\small{(e)}}
        \end{overpic}
        \label{fig:j_mu_184602}
    }
    \hspace{0.005\textwidth}
    \subfloat{%
        \begin{overpic}[width=0.4\columnwidth]{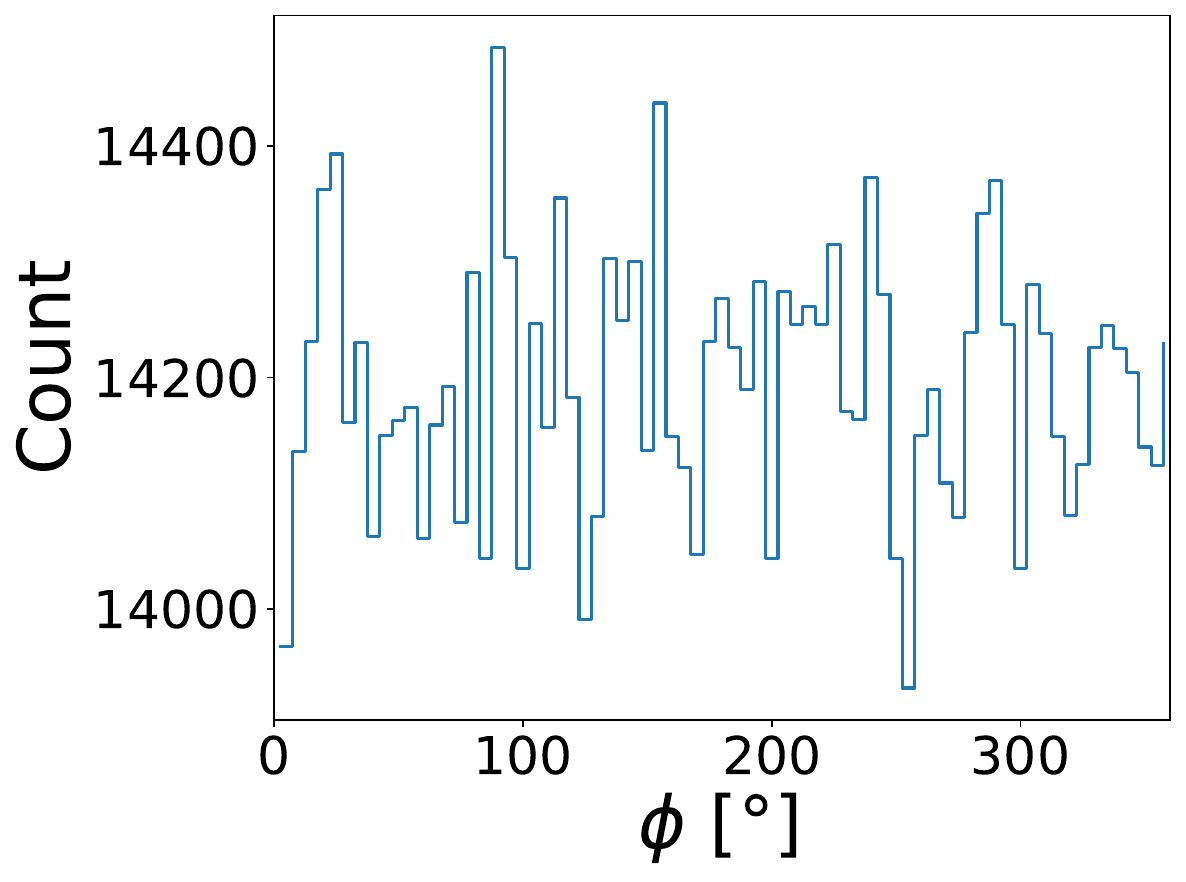}
            \put(2,78){\small{(f)}}
        \end{overpic}
        \label{fig:j_phi_184602}
    }
    \caption{DIII-D shot~184602 post-thermal-quench equilibrium and RE beam diagnostics. 
    (a) Poloidal flux ($\psi_p$) contours from the EFIT Grad-Shafranov solution. 
    (b) Parallel current density. 
    (c) Safety-factor as function of a normalized poloidal flux variable ($\psi_N$), with on-axis $q<1$. 
    (d) Sampled parallel current density. 
    (e) Sampled magnetic moment. 
    (f) Particle count vs.\ toroidal coordinate.}
    \label{fig:184602}
\end{figure}

Our study focuses on modeling the RE beam generated during DIII-D shot~184602, 
which features a limited plasma (Fig.\,\ref{fig:j_psi_184602}) with a high-field-side RE beam and involves low-$Z$ ($\mathrm{D}_2$) material injection (not included in our simulations). High-$Z$ impurities undergo recombination and are expelled from the plasma~\cite{hollmann2020}, causing a pronounced plasma density reduction that allows us to track RE dynamics without explicitly incorporating collisional effects, thereby simplifying our analysis.
This scenario has counter-current flow and an on-axis safety factor of less than one (Fig.\,\ref{fig:q_184602}), 
making it susceptible to sawtooth instabilities \cite{vonGoeler1974}. 
Despite being sawtooth unstable, we hold NIMROD fields fixed to model a static plateau. The dynamic coupling of REs to evolving MHD fields is beyond the scope of this study but remains a key target for ongoing development.

During the post-thermal-quench (cold) phase, the background plasma pressure is far lower than before the quench, and the plasma may be considered force-free in a first approximation. The equilibrium current density is therefore fully specified by the parallel current density $J_{\parallel}(R,Z)$. We obtain $J_{\parallel}(R,Z)$ from an MHD equilibrium fitting (EFIT) \cite{lao1985} reconstruction (Fig.\,\ref{fig:j_par_184602}). In this plateau regime, the plasma current is largely carried by REs, so the EFIT-derived MHD parallel current is assumed to be completely due to the current carried by REs.

For a single electron (1e) with parallel momentum $p_{\parallel}$ and magnetic moment $\mu$, the parallel current in the relativistic guiding-center approximation is given in Appendix\,\ref{sec:CurrentApp} as
\begin{equation}
I_{\parallel}^{(1e)}(R,Z,p_{\parallel},\mu)\;=\;\frac{1}{2\pi R}\biggl(
    \frac{q_e\,p_{\parallel}}{m\,\gamma}
    - \mu\,(\hat{\mathbf{b}}\cdot\nabla\times\hat{\mathbf{b}})
\biggr),
\label{eq:single_j}
\end{equation}
consisting of streaming along the magnetic field and magnetization contributions \cite{todo98,todo04,bierwage13}. 
The absence of an $E \times B$ drift term in Eq.\,\eqref{eq:single_j} follows from quasi-neutrality \cite{todo98}.

We construct an ensemble of relativistic guiding-center markers with individual currents $I_{\parallel}^{(1e)}$ to reproduce the experimental current density $J_{\parallel}(R,Z)$ from Fig.\,\ref{fig:j_par_184602}. We assume the kinetic energy and pitch angle are fixed for all REs (i.e., mono-energetic and mono-pitch) with values set to $KE = 10\,\mathrm{MeV}$ and $\eta = 10^\circ $, with $\phi$ chosen uniformly from $[0,2\pi)$ ensuring an axisymmetric initial distribution. The kinetic energy is given by $KE = (\gamma - 1)mc^2$. All markers share an initial parallel momentum of $p_{\parallel} = 5.525 \times 10^{-21} \,\text{kg}\cdot\text{m/s}$, while their magnetic moment $\mu$ varies spatially, depending on the local field strength $B(R,Z)$. Hence, under these conditions, $I_{\parallel}^{(1e)}$ becomes a function solely of $(R,Z)$. Since $I_{\parallel}^{(1e)}(R,Z)$ in Eq.\,\eqref{eq:single_j} now depends on $(R,Z)$ only, 
we can directly sample two-dimensional space via a Metropolis-Hastings (MH) algorithm adapted from Refs.~\inlinecite{metropolis1953,hastings1970}. The Markov chain is constructed by proposing new particle positions $(R'_\ell, Z'_\ell)$ and accepting or rejecting them based on the standard MH acceptance ratio $A\big((R_\ell,Z_\ell)\rightarrow(R'_\ell,Z'_\ell)\big)$, given by
\begin{equation}\label{eq:MH_acceptance_ratio}
    A\big((R_\ell,Z_\ell)\rightarrow(R'_\ell,Z'_\ell)\big)=
    \min\left\{1,\frac{\frac{J_{\parallel}(R'_\ell,Z'_\ell)}{I_{\parallel}^{(1e)}(R'_\ell,Z'_\ell)}}{\frac{J_{\parallel}(R_\ell,Z_\ell)}{I_{\parallel}^{(1e)}(R_\ell,Z_\ell)}}\right\}.
\end{equation}
Figure\,\ref{fig:j_par_sampled_184602} shows a $10^6$-particle sample, in which a single global scaling factor is applied to all particles so that their total current corresponds to the experimental value. The poloidal distribution of the markers' magnetic moment $\mu$ is shown in Fig.\,\ref{fig:j_mu_184602}, and the toroidal coordinate distribution of the markers is illustrated in Fig.\,\ref{fig:j_phi_184602}.

Recent work \cite{bandaru2023} derived self-consistent kinetic-MHD equilibria of RE beams and argued that the equilibria cannot be interpreted as force-free. Ref. \inlinecite{bandaru2023} retains pressure gradient effects and the curvature drift motion of REs, showing that the radial shifts between flux surfaces and RE orbit surfaces are more significant at high RE energies ($>40 \mathrm{MeV}$). However, most simulations in our study focus on $10 \mathrm{MeV}$ RE beams, where such shifts remain modest (Sec.\,\ref{sec:nimrod_dep}). The effects of RE curvature drift motion are not included in this manuscript and will be addressed in future work. Additionally, Ref.\,\inlinecite{marini2024} introduced a method using visible synchrotron and Ar-II line emission to refine the RE plateau current profile, revealing significant discrepancies from EFIT-derived currents (Fig. 11 of Ref.\,\inlinecite{marini2024}). In particular, they find that the current is less centrally-peaked, increasing the safety factor to marginally above 1. In our work, we rely on EFIT-derived currents, as they more easily interface with the existing NIMROD codebase for equilibrium computations. Our primary focus is testing algorithm components for future self-consistent KORC-NIMROD coupling.

%%%%%%%%%%%%%%%%%%%%%%%%%%%%%%%%%%%%%%%%%%%%%%%%%%%%%%%%%%%%%%%%%%%%%%%%%%%%%%%
%%%%%%%%%%%%%%%%%%%%%%%%%%%%%%%%%%%%%%%%%%%%%%%%%%%%%%%%%%%%%%%%%%%%%%%%%%%%%%%
%%%%%%%%%%%%%%%%%%%%%%%%%%%%%%%%%%%%%%%%%%%%%%%%%%%%%%%%%%%%%%%%%%%%%%%%%%%%%%%
%%%%%%%%%%%%%%%%%%%%%%%%%%%%%%%%%%%%%%%%%%%%%%%%%%%%%%%%%%%%%%%%%%%%%%%%%%%%%%%

\section{ Efficient particle-to-element mapping for REs}\label{sec:barycentric}

In the hybrid kinetic-MHD simulations of this work, mapping particle coordinates from physical $(R,Z)$ space onto the finite element mesh used by NIMROD is fundamental for particle advancement (Sec.\,\ref{sec:MH}) and deposition onto the mesh procedures (Sec.\,\ref{sec:deposition}). Both processes will also be essential for modeling RE dynamics in future time-dependent self-consistent KORC-NIMROD simulations. NIMROD constructs a nonuniform physical mesh in $(R,Z)$ coordinates by mapping a rectangular logical mesh divided into $n\times m$ elements in logical $(\xi,\upsilon)$ coordinates, where $n$ and $m$ are the numbers of divisions in the radial and poloidal directions, respectively. Mapping a particle from physical coordinates $\mathbf{P}=(R^*,Z^*)$ to logical coordinates $(\xi^*,\upsilon^*)$ involves a Newton-Raphson solver, whose convergence critically depends on the accuracy of its initial guess. 

To ensure  reliable convergence of the Newton-Raphson solver, we introduce an efficient and robust initialization strategy based on barycentric coordinates and binary search. This approach provides a good initial guess, often referred to as a warm start, which is crucial for the solver's performance. Without such a strategy, the Newton-Raphson method may fail to converge, especially for particles whose logical coordinates are far from the initial guess. First by identifying the correct poloidal sector through a binary search \cite{cormen2009introduction} over polar angles and then locating the radial index via barycentric coordinates within triangular subdivisions, we determine the finite element $e_{i^*,j^*}$ containing the particle with computational complexity $\mathcal{O}(\log n+\log m)$. This significantly improves both the accuracy and robustness of the coordinate inversion process with reduced computational cost, especially simulations involving large populations of REs.

\begin{figure}[ht]
    \centering
            \includegraphics[scale=0.25]{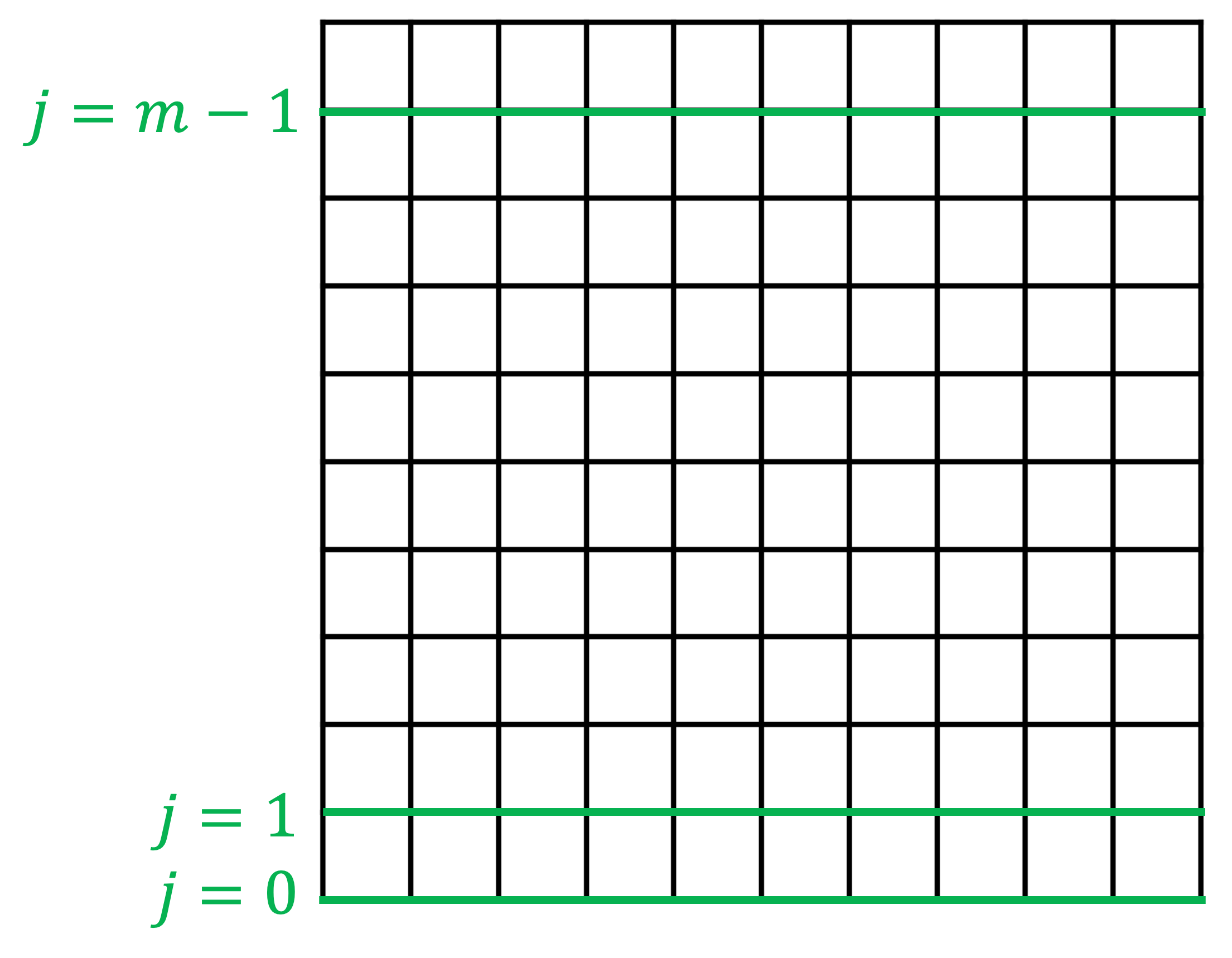}
    \includegraphics[scale=0.3]{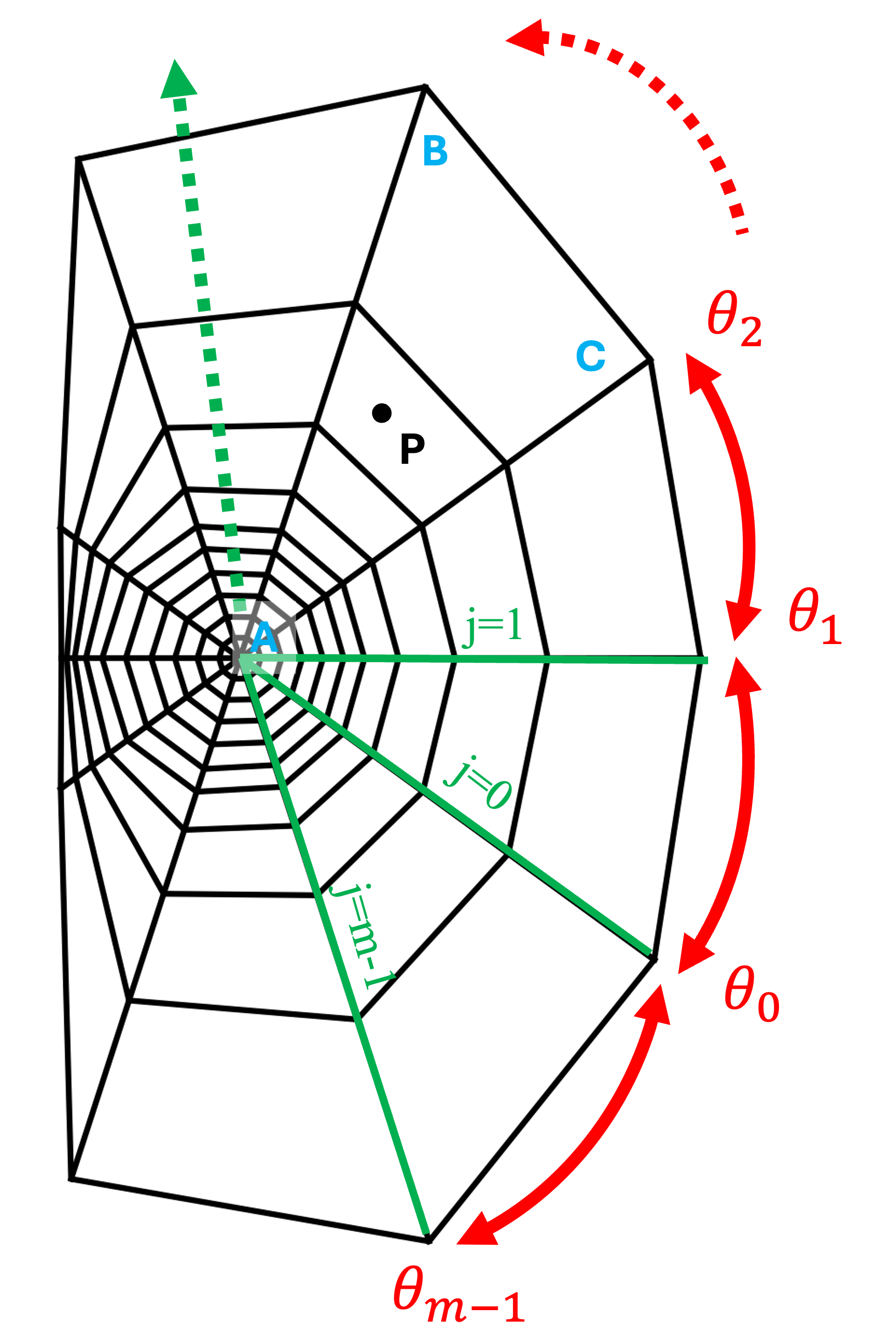}
    \caption{NIMROD mesh in the poloidal cross-section (right) and its corresponding logical mesh (left). Green (red) dotted lines indicate the radial (poloidal) direction. Three poloidal sectors are shown, and the vertices of the primary triangle $\triangle \mathbf{ABC}$ for a particle at $\mathbf{P}$ are marked.}
    \label{fig:poloidal_radial}
\end{figure}

The logical domain consists of unit-square elements defined as $e_{i,j}=[\xi_i,\xi_{i+1}]\times[\upsilon_j,\upsilon_{j+1}]$, with strictly increasing sequences $\xi_i<\xi_{i+1}$ and $\upsilon_j<\upsilon_{j+1}$, where indices $i=0,\dots,n-1$ and $j=0,\dots,m-1$. In NIMROD, as the mesh resolution is increased by raising $n$ and $m$, the logical domain scales accordingly, maintaining unit-square logical elements. The line $\xi=\xi_0=0$ maps to the magnetic axis in physical space. Moving along the $\xi$ direction in logical space corresponds to moving in the radial direction (outward from the magnetic axis)
 in physical space and so $i$ is referred to as the \emph{radial index}.
Moving along the $\upsilon$ direction in logical space corresponds to moving in the poloidal direction in physical space 
and so $j$ is referred to as the \emph{poloidal index}. Dotted lines in Fig.\,\ref{fig:poloidal_radial} indicate these directions.

The mapping from logical $(\xi,\upsilon)$ to physical $(R,Z)$ coordinates is expanded over finite element nodes:
\begin{equation}\label{eq:mapping}
  \mathcal{R}(\xi,\upsilon)
    = \sum_k R_k\,\alpha_k(\xi,\upsilon), \qquad
  \mathcal{Z}(\xi,\upsilon)
    = \sum_k Z_k\,\alpha_k(\xi,\upsilon),
\end{equation}
where each $\alpha_k(\xi,\upsilon): [0,\xi_{n}] \times [0,\upsilon_{m}]\to \mathbb{R}$ is a compactly supported two-dimensional Lagrange polynomial basis function (consult, e.g., Sec. 3.2 of Ref.~\inlinecite{hughes2012finite}). The polynomial order of these basis functions is denoted by $\textit{poly-deg}$. This work specifically uses bilinear elements ($\text{poly-deg} = 1$). For a mapping function with bilinear elements in NIMROD, each nodal coefficient pair $(R_k, Z_k)$ directly corresponds to the physical coordinates at a node of the mesh: $(R_k, Z_k) = \bigl(\mathcal{R}(\xi_k,\upsilon_k),\, \mathcal{Z}(\xi_k,\upsilon_k)\bigr)$, where $(\xi_k, \upsilon_k)$ enumerates the vertices of all squares in the logical mesh domain.

This study employs bilinear basis functions within a nonuniform physical mesh composed of triangular elements adjacent to the magnetic axis and quadrilateral elements elsewhere. Presently, bilinear elements are used for both mapping and field representation, but future studies could adopt a non-isoparametric approach \cite{strang1987}, employing higher-order polynomial approximations for fields while retaining bilinear mappings (a possibility already supported by NIMROD). The following discussion introduces a barycentric search algorithm specifically tailored to the aforementioned meshes combining triangular and quadrilateral elements, with further details provided in Appendix\,\ref{sec:BarycentricApp}. 

\subsection{Poloidal index determination}
\label{sec:poloidal_index}

We partition the domain in physical space into $m$ poloidal sectors indexed by 
$j=0,\cdots,m-1$. For each integer $j$, we consider the boundary vertex 
\begin{equation}
\bigl(\mathcal{R}(\xi_{n},\upsilon_{j}),\;\mathcal{Z}(\xi_{n},\upsilon_{j})\bigr),
\end{equation}
and compute
\begin{equation}
  \theta_{j} 
  \;=\; 
  \arctan\!\Bigl(\tfrac{\mathcal{Z}(\xi_{n},\upsilon_{j})-Z_{mag}}{\mathcal{R}(\xi_{n},\upsilon_{j})-R_{mag}}\Bigr),
\end{equation}
where $(R_{mag},Z_{mag})$ are the physical coordinates corresponding to the magnetic axis. Similarly, 
for a particle at $\mathbf{P}$ we compute the angle $\theta^* = \arctan( ( Z^*-Z_{mag})/(R^*-R_{mag}))$.
A binary search (Algorithm\,\ref{alg:bisectsort} of Appendix\,\ref{sec:BarycentricApp}) identifies the sector whose bounds bracket $\theta^*$ (
$\theta_{j^*}\le\theta^{*}<\theta_{j^*+1}$). This procedure yields the poloidal index, 
$j^*$, and selects the \emph{primary triangle} $\triangle \mathbf{A B C}$, where $\mathbf{A}=(R_{mag},Z_{mag})$ is the magnetic axis and the other vertices are
\begin{subequations}
    \begin{equation}
        \mathbf{B} = \bigl(\mathcal{R}(\xi_{n},\, \upsilon_{j^*}),\; \mathcal{Z}(\xi_{n},\, \upsilon_{ j^*})\bigr),
    \end{equation}
    \begin{equation}
        \mathbf{C} = \bigl(\mathcal{R}(\xi_{n},\,\upsilon_{j^*+1}),\; \mathcal{Z}(\xi_{n},\,\upsilon_{ j^*+1})\bigr).
    \end{equation}
\end{subequations} Figure\,\ref{fig:poloidal_radial} illustrates different poloidal sectors and the primary triangle for a particle at $\mathbf{P}$.

\subsection{Barycentric coordinates and triangle containment}

Barycentric coordinates \cite{ciarlet2002finite} allow us to determine whether a particle at $\mathbf{P}$ lies inside a given triangle $\triangle \mathbf{XYZ}$ with vertices at
$\mathbf{X}=(R_{\mathbf{X}}, Z_{\mathbf{X}})$, $\mathbf{Y}=(R_{\mathbf{Y}}, Z_{\mathbf{Y}})$, and $\mathbf{Z}=(R_{\mathbf{Z}}, Z_{\mathbf{Z}})$. 
The signed area of $\triangle\mathbf{XYZ}$ is defined as
\begin{equation}
    \mathcal{A}_{{\mathbf{XYZ}}} 
    = \frac{1}{2}
    \begin{vmatrix}
      R_{\mathbf{Y}} - R_{\mathbf{X}} & R_{\mathbf{Z}} - R_{\mathbf{X}} \\
      Z_{\mathbf{Y}} - Z_{\mathbf{X}} & Z_{\mathbf{Z}} - Z_{\mathbf{X}}
    \end{vmatrix},
\end{equation}
which is half the determinant of the vectors 
of $\mathbf{Y}-\mathbf{X}$ and $\mathbf{Z}-\mathbf{X}$; 
its sign follows a counterclockwise orientation of the vertices.

\begin{figure}[ht]
    \centering
    \includegraphics[scale=0.35]{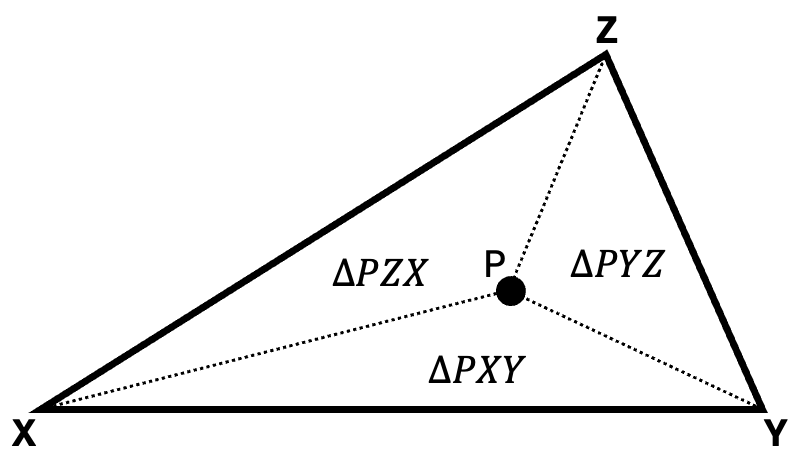}
    \caption{Barycentric decomposition of a triangle $\triangle \mathbf{XYZ}$ in the $(R,Z)$ plane. 
    The point $\mathbf{P}$ partitions the triangle into $\triangle \mathbf{PXY}$, $\triangle \mathbf{PXZ}$, and $\triangle \mathbf{PYZ}$. 
    Barycentric coordinates for $\mathbf{P}$ confirm whether this point lies inside, 
    on the boundary, or outside $\triangle \mathbf{XYZ}$.}
    \label{fig:barycentricdivision}
\end{figure}

The \emph{barycentric coordinates} of a particle at $\mathbf{P}$ with respect to 
$\triangle \mathbf{XYZ}$ are defined as
\begin{equation}
    x = \frac{\mathcal{A}_{\mathbf{PYZ}}}{\mathcal{A}_{\mathbf{XYZ}}}, \quad 
    y = \frac{\mathcal{A}_{\mathbf{PXZ}}}{\mathcal{A}_{\mathbf{XYZ}}}, \quad 
    z = 1 - x - y,
\end{equation}
where $\mathcal{A}_{\mathbf{XYZ}}$ is the signed area of $\triangle \mathbf{\mathbf{XYZ}}$, and
$\mathcal{A}_{\mathbf{PYZ}},\,\mathcal{A}_{\mathbf{PXZ}}$ are computed by 
replacing $\mathbf{X},\mathbf{Y}$ with $\mathbf{P}$ in turn.
Figure\,\ref{fig:barycentricdivision} depicts the special case where $\mathbf{P}$
lies strictly within $\triangle \mathbf{XYZ}$, so that $\triangle \mathbf{PYZ}$, $\triangle \mathbf{PXZ}$, and $\triangle \mathbf{PXY}$ partition $\triangle \mathbf{XYZ}$.
If all barycentric coordinates are positive, then a particle at $\mathbf{P}$ lies inside $\triangle \mathbf{XYZ}$; if at least one is zero, it lies on the triangle's boundary;
otherwise, it lies outside; see Algorithm\,\ref{alg:barycentric} of Appendix\,\ref{sec:BarycentricApp} for a complete description.

\subsection{Radial index determination}

Within each poloidal sector, $n$ nested triangles fan outward from 
the magnetic axis to the boundaries of the physical mesh. To determine the radial index $i^*$ of a particle located at $ \mathbf{P}$, we construct the primary triangle
defined by the innermost and outermost radial indices in the $j^*$th poloidal sector and perform a bracketed binary search using barycentric coordinates.

%If $ \mathcal{A}_{\mathbf{PBC}} = 0 $, $ \mathbf{P} $ lies on the boundary, and we assign $i^*= n$. Otherwise, we perform a bracketed binary search using barycentric coordinates.

\begin{figure}[ht]
    \centering
    % Row 1 (single figure)
    \subfloat{%
        \begin{minipage}{0.48\columnwidth}
            \centering
            \begin{overpic}[width=\linewidth]{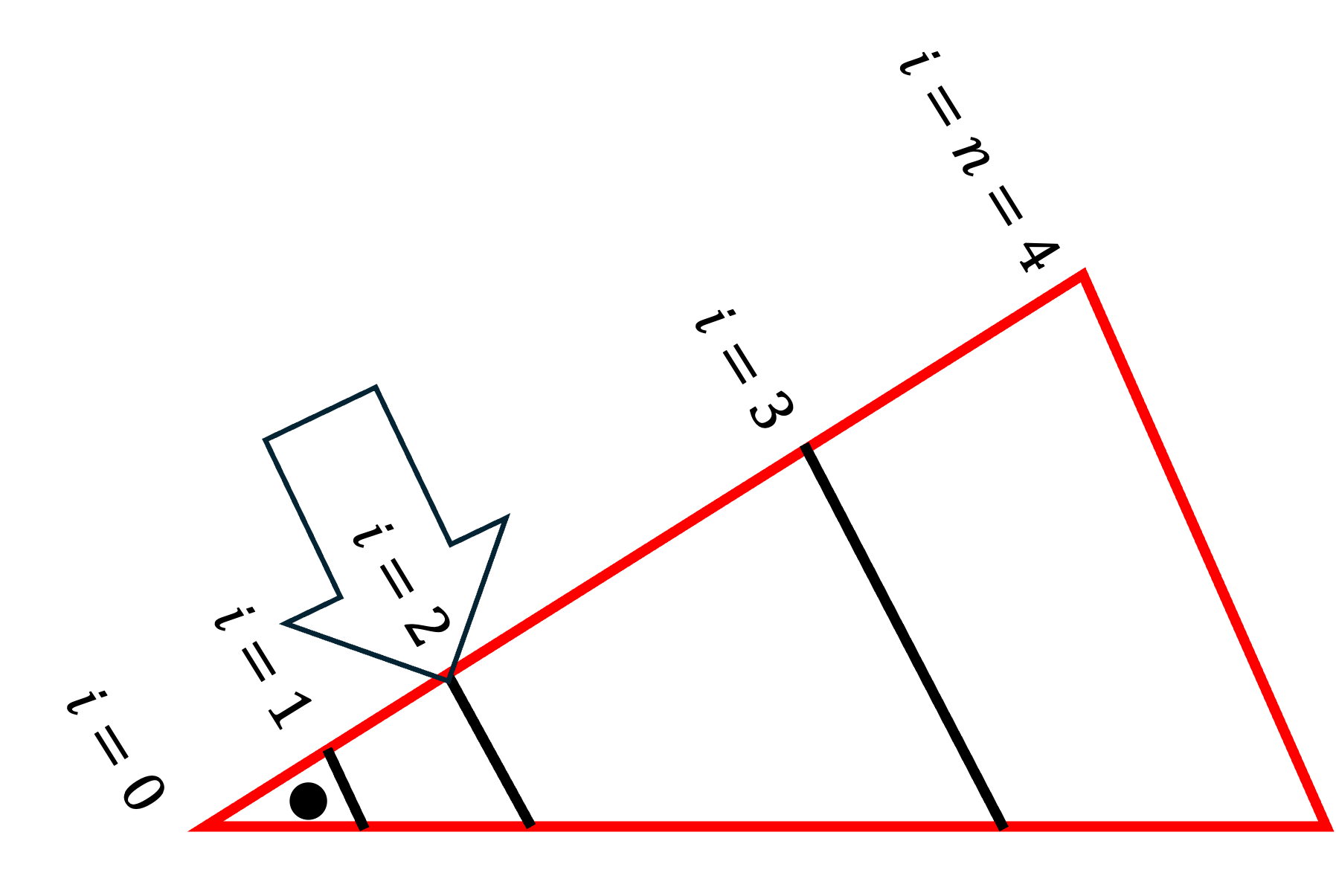}
                %\put(30,55){\small{(a)}}
            \end{overpic}
        \end{minipage}
    }\\[1ex]
    % Row 2 (single figure)
    \subfloat{%
        \begin{minipage}{0.48\columnwidth}
            \centering
            \begin{overpic}[width=\linewidth]{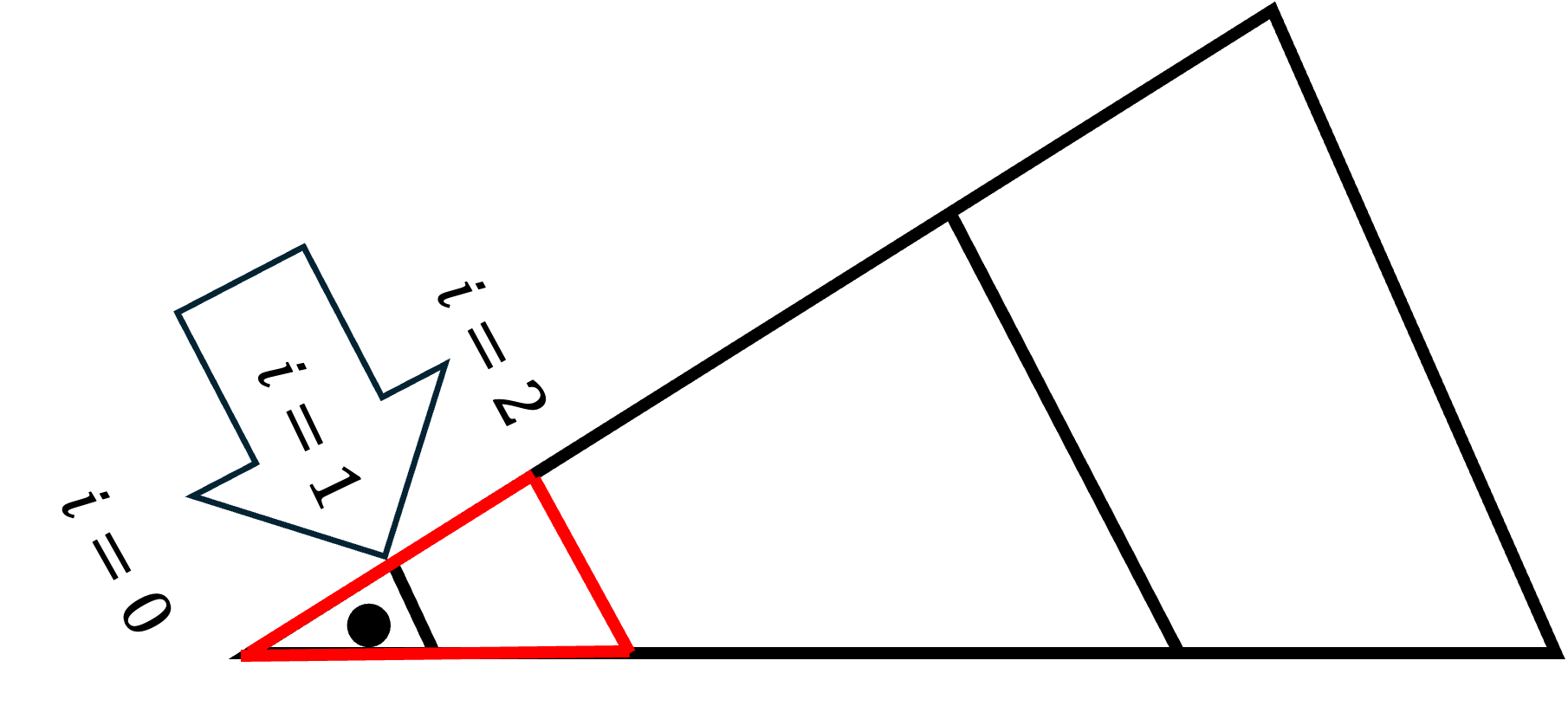}
                %\put(30,55){\small{(b)}}
            \end{overpic}
        \end{minipage}
    }\\[1ex]
    % Row 3 (single figure)
    \subfloat{%
        \begin{minipage}{0.48\columnwidth}
            \centering
            \begin{overpic}[width=\linewidth]{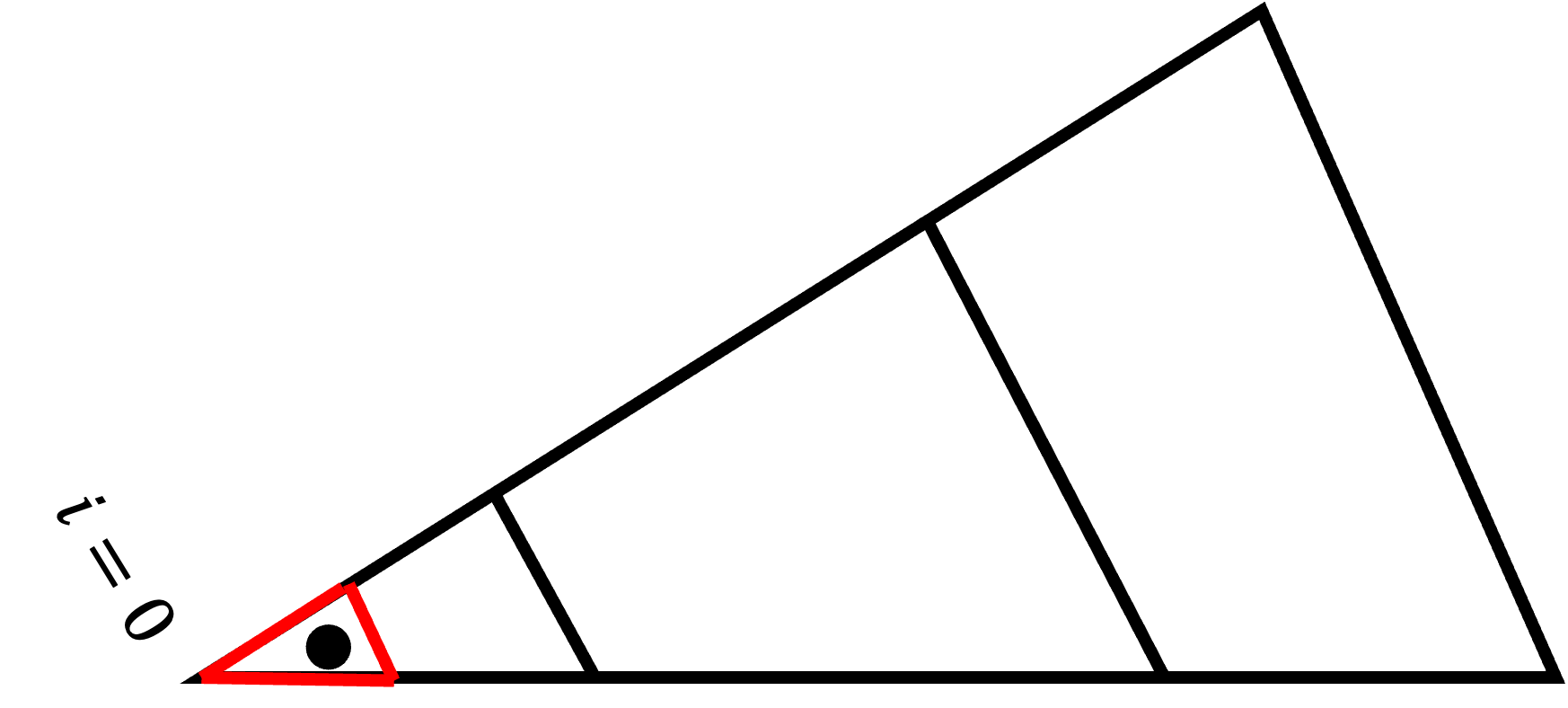}
                %\put(30,55){\small{(c)}}
            \end{overpic}
        \end{minipage}
    }
    \caption{Binary search in the radial direction.
This setup divides the search space into $n$ segments, and the binary search algorithm iteratively narrows down the range based on whether the particle lies within the intermediate triangle associated with the current midpoint. The process continues until the correct finite element $e_{i^*, j^*}$ is identified.}
    \label{fig:barycentricsort}
\end{figure}

We initialize bracket indices $i^\text{low}=0$ and $i^\text{high}=n$ and refine them via binary search. At each iteration, we take the middle index of the current bracket:
\begin{equation}
   i^\text{mid} = \left\lfloor \frac{i^\text{low} + i^\text{high}}{2}  \right\rfloor,
\end{equation}
where $\lfloor \cdot \rfloor$ denotes the floor function. We then form an intermediate triangle that joins the magnetic axis to the middle index's boundary and check containment with barycentric coordinates. 
If $\mathbf{P}$ lies inside an intermediate triangle, we reduce the bracket's upper limit to the middle index, effectively ``moving inward.'' 
Otherwise, we ``move outward'' by increasing the bracket's lower limit to the middle index. Once only two adjacent radial subdivisions remain
($i^\text{high} = i^\text{low} + 1$), the particle's radial index is set to the lower bracket index: $ i^*=i^\text{low} $.
Figure\,\ref{fig:barycentricsort} illustrates the binary search in the radial direction. This procedure is described in Algorithm\,\ref{alg:barycentricsort} of Appendix\,\ref{sec:BarycentricApp}. 

To address floating-point precision errors in NIMROD's mesh, which can cause $\pm1$ index misidentifications near finite element boundaries, we have implemented flags to ensure that, if the radial index algorithm fails to converge, it retries the radial index determination with $\pm1$ neighboring poloidal indices. This method has proven effective in all RE simulations conducted so far.

%%%%%%%%%%%%%%%%%%%%%%%%%%%%%%%%%%%%%%%%%%%%%%%%%%%%%%%%%%%%%%%%%%%%%%%%%%%%%%%
%%%%%%%%%%%%%%%%%%%%%%%%%%%%%%%%%%%%%%%%%%%%%%%%%%%%%%%%%%%%%%%%%%%%%%%%%%%%%%%
%%%%%%%%%%%%%%%%%%%%%%%%%%%%%%%%%%%%%%%%%%%%%%%%%%%%%%%%%%%%%%%%%%%%%%%%%%%%%%%
%%%%%%%%%%%%%%%%%%%%%%%%%%%%%%%%%%%%%%%%%%%%%%%%%%%%%%%%%%%%%%%%%%%%%%%%%%%%%%%

\section{Particle deposition}\label{sec:deposition}

This section describes how guiding-center weights are deposited onto NIMROD's finite element mesh to form macroscopic fields, such as the parallel current density. First, the axisymmetric formulation is developed, showing how each particle's contribution is incorporated through a weak form that produces a system of linear equations.  We then verify the deposition approach using a Gaussian density test.  Both NIMROD's finite element solution and a Python-based solver match the analytic profile and show decreasing relative error as the number of particles increases, although their difference stalls beyond roughly $10^6$ markers.

\subsection{Mathematical formulation for the axisymmetric case}

This paper examines the assembly of axisymmetric scalar fields (e.g., parallel current or density) from individual particles. 
The process of assembling fields from particles in NIMROD was introduced in Ref.~\inlinecite{kim2004hybrid}, and it is also included here for completeness.
In physical space, axisymmetric fields are represented as $\mathcal{K}(R,Z)$, and in logical space as $ \mathcal{F}(\xi,\upsilon)$. The field representation $\mathcal{K}(R,Z)$ is given by:
\begin{equation}\label{eq:k_delta_axi}
    \mathcal{K}(\xi,\upsilon) = \sum_{l=1}^{N_{p}} w_l \frac{ \delta(\mathcal{R}(\xi,\upsilon) - R_l) \delta(\mathcal{Z}(\xi,\upsilon) - Z_l)}{2\pi \mathcal{R}(\xi,\upsilon)},
\end{equation} where $N_p$ is the number of particles, $w_l$ is the weight of the $l$-th particle, and $(R_l,Z_l)$ are the particle's physical coordinates.
The mapping from logical to physical coordinates is given by Eq.\,\eqref{eq:mapping}.
We approximate $\mathcal{F}(\xi,\upsilon)$ using a basis expansion:
\begin{equation}\label{eq:k_basis_axi}
   \mathcal{F}(\xi,\upsilon) = \sum_{s=1}^{N_{\text{dof}}} k_s \alpha_s(\xi,\upsilon),
\end{equation}
with $N_{\text{dof}}$ degrees of freedom and coefficients $k_s$ to be determined.
The coefficients $k_s$ are determined by enforcing equality of $\mathcal{K}$ and $\mathcal{F}$ in the following weak sense:
\begin{eqnarray}\label{eq:weak}
    &&\int \mathcal{F}(\xi,\upsilon)\alpha_\lambda(\xi,\upsilon) \mathcal{R}(\xi,\upsilon) \mathcal{J}(\xi,\upsilon) \, d\xi \, d\upsilon \nonumber\\
    &&= \int \mathcal{K}(\xi,\upsilon)\alpha_\lambda(\xi,\upsilon) \mathcal{R}(\xi,\upsilon) \mathcal{J}(\xi,\upsilon) \, d\xi \, d\upsilon,
\end{eqnarray} for each basis function $\alpha_\lambda(\xi,\upsilon)$, $\lambda = 1, \dots, N_{\text{dof}}$. 
This weak formulation includes an $ \mathcal{R}(\xi,\upsilon) $ weighting factor, unlike Ref.~\inlinecite{kim2004hybrid}, which omits it; both formulations are valid.
Here, the absolute value of the Jacobian determinant is given by:
\begin{equation}
\mathcal{J}(\xi,\upsilon) := \left| \frac{\partial(\mathcal{R},\mathcal{Z})}{\partial(\xi,\upsilon)} \right|.
\end{equation}
Substituting the expressions for $\mathcal{K}(\xi,\upsilon)$ and $\mathcal{F}(\xi,\upsilon)$ into the weak formulation, and utilizing the property of the Dirac-delta functions:
\begin{equation}
    \delta(\mathcal{R}(\xi,\upsilon) - R_l) \delta(\mathcal{Z}(\xi,\upsilon) - Z_l)  = \frac{\delta(\xi - \xi_l)\delta(\upsilon - \upsilon_l)}{{\cal J}(\xi_l,\upsilon_l)},
\end{equation} where $(\xi_l,\upsilon_l)$ are logical coordinates for $(R_l,Z_l)$, we obtain
\begin{equation}\label{eq:mass_matrix_system}
   \sum_{s=1}^{N_{\text{dof}}} k_s m_{s \lambda}  = \sum_{l=1}^{N_{p}} \frac{w_l}{2\pi} \alpha_\lambda(\xi_l,\upsilon_l) ,
\end{equation}
with the mass matrix given by
\begin{equation}\label{eq:mass_matrix}
     m_{s \lambda} = \int \alpha_s(\xi,\upsilon) \alpha_\lambda(\xi,\upsilon)  \mathcal{R}(\xi,\upsilon){\cal J}(\xi,\upsilon)\, d\xi \, d\upsilon.
\end{equation}

%%%%%%%%%%%%%%%%%%%%%%%%%%%%%%%%%%%%%%%%%%%%%%%%%%%%%%%%%%%%%%%%%%%%%%%%%%%%%%%
%%%%%%%%%%%%%%%%%%%%%%%%%%%%%%%%%%%%%%%%%%%%%%%%%%%%%%%%%%%%%%%%%%%%%%%%%%%%%%%

\subsection{NIMROD vs DepoPy deposition verification}\label{sec:DepoPy}

Having established the weak formulation, we turn to its numerical realization in NIMROD. We build upon the $\delta f$ PIC implementation of Ref.\,\inlinecite{taylor2022}. We solve Eq.\,\eqref{eq:mass_matrix_system}, subject to poloidal periodicity, by introducing an explicit mass-matrix inversion that is carried out with an iterative method. Continuity at the magnetic axis is secured by regularizing Eq.\,\eqref{eq:mass_matrix_system} right-hand side before the linear solve. We verify NIMROD's calculations by constructing a Python-based finite element code, DepoPy. DepoPy achieves continuity at the axis through an approach introduced here as the Magnetic-Axis-Basis (MAB). The MAB combines multiple basis functions into a reduced set of degrees of freedom, so that the finite elements adjacent to the magnetic axis are treated as a single entity in logical space. While NIMROD employs iterative methods, DepoPy uses a direct solve. Presently, DepoPy supports a bilinear MAB-based deposition scheme; extending it to higher-order elements is planned for future work.

To verify the deposition methods, we compare NIMROD's with DepoPy using a synthetic Gaussian distribution. Tests are performed on an axisymmetric torus with a $n=18$, $m=18$ mesh, where $n$ is the number of radial elements and $m$ is the number of poloidal elements, over the same computational domain as the DIII-D case of Sec.\,\ref{sec:MH}, using increasing particle counts: $1.5 \times 10^{4}$, $1.5 \times 10^{5}$, $1.5 \times 10^{6}$, and $1.5 \times 10^{7}$ particles. 

The Gaussian probability density function (PDF) is defined as
\begin{equation}
    n^{\mathrm{Analytic}}(R, Z) 
    = n_0 \exp\!\Bigl(-\Bigl(\tfrac{(R - R_{\mathrm{mag}})^2}{2\sigma_R^2} 
    + \tfrac{(Z - Z_{\mathrm{mag}})^2}{2\sigma_Z^2} \Bigr)\Bigr),
\end{equation}
where $\sigma_R = 0.12\,\mathrm{m}$, $\sigma_Z = 0.15\,\mathrm{m}$, and $(R_{\mathrm{mag}}, Z_{\mathrm{mag}})=(1.35\,\mathrm{m}, 0.016\,\mathrm{m})$ is the magnetic axis location.
The normalization constant $n_0$ ensures that each Gaussian integrates to $1$ over the torus volume. Samples of size $N_p$ are generated with rejection sampling, and the weights required for the NIMROD and DepoPy finite element solutions in Eq.\,\eqref{eq:k_delta_axi} are set to $w_l = N_{p}^{-1}$. The tests use bilinear finite elements in NIMROD to match DepoPy's bilinear MAB-based deposition.

 We define a relative error between two density distributions $n^{(1)}$ and $n^{(2)}$ at each mesh vertex $(R_j, Z_j)$ as
\begin{equation}\label{eq:pointwise_error}
    \text{Error}_j 
    = 
    \frac{\bigl|\,n^{(1)}(R_j, Z_j) - n^{(2)}(R_j, Z_j)\bigr|}
         {\tfrac{1}{N}\sum_{k=1}^{N} \bigl| n^{(1)}(R_k, Z_k) \bigr|},
\end{equation}
where $N$ is the total number of mesh vertices. 
Since both PDFs approach zero far from the magnetic axis, dividing by the point-wise value $n^{(1)}(R_j, Z_j)$ could lead to undefined or large errors; instead, we normalize by the average absolute value of the reference distribution across all mesh points. When comparing NIMROD to the analytic Gaussian we set $n^{(1)} = n^{\mathrm{Analytic}}$ and $n^{(2)} = n^{\mathrm{DepoPy}}$. When comparing NIMROD to DepoPy we set $n^{(1)} = n^{\mathrm{NIMROD}}$ and $n^{(2)} = n^{\mathrm{DepoPy}}$.  

From these point-wise errors, we define two global errors that measure the overall performance across the mesh. The first is the average of the relative error across all vertices:
\begin{subequations}\label{eq:error_metrics}
    \begin{equation} \label{eq:mean_error}
        \text{Error}_{\text{average}}  = \frac{1}{N} \sum_{j=1}^{N} \text{Error}_j.
    \end{equation}
    The second global error measure is largest relative error found at the mesh vertices:
    \begin{equation} \label{eq:max_error}
        \text{Error}_{\max} = \max \bigl\{ \text{Error}_j \colon j = 1,\ldots,N \bigr\}.
    \end{equation}
\end{subequations}

\begin{figure}[ht]
    \centering
    % Row 1
    \subfloat{%
        \begin{overpic}[width=0.75\columnwidth]{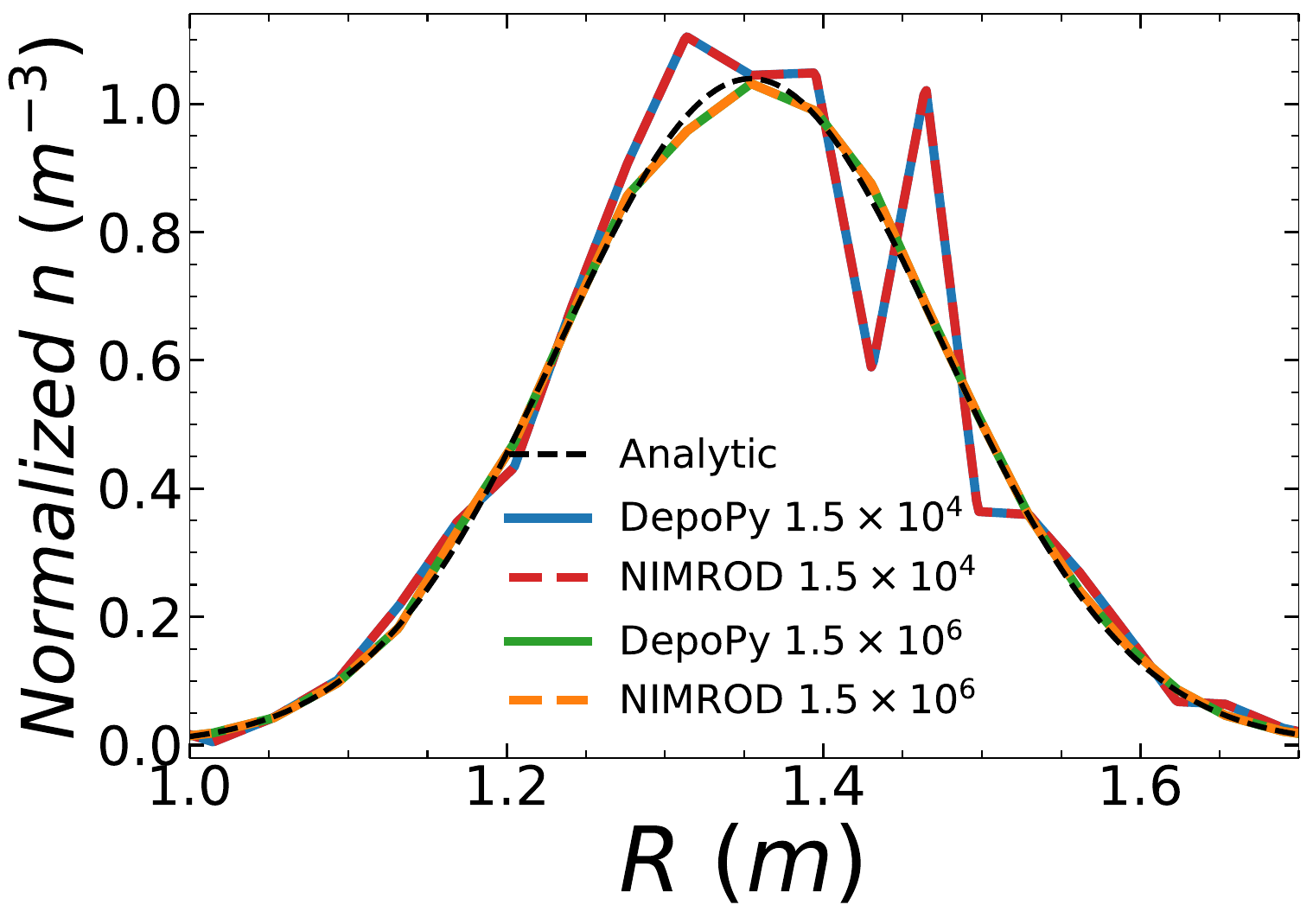}
            \put(92,64){\small{(a)}}
        \end{overpic}
        \label{fig:Gaussian_vs_NIMROD_vs_DepoPy_a}
    }
    \hspace{0.005\textwidth}
    \subfloat{%
        \begin{overpic}[width=0.75\columnwidth]{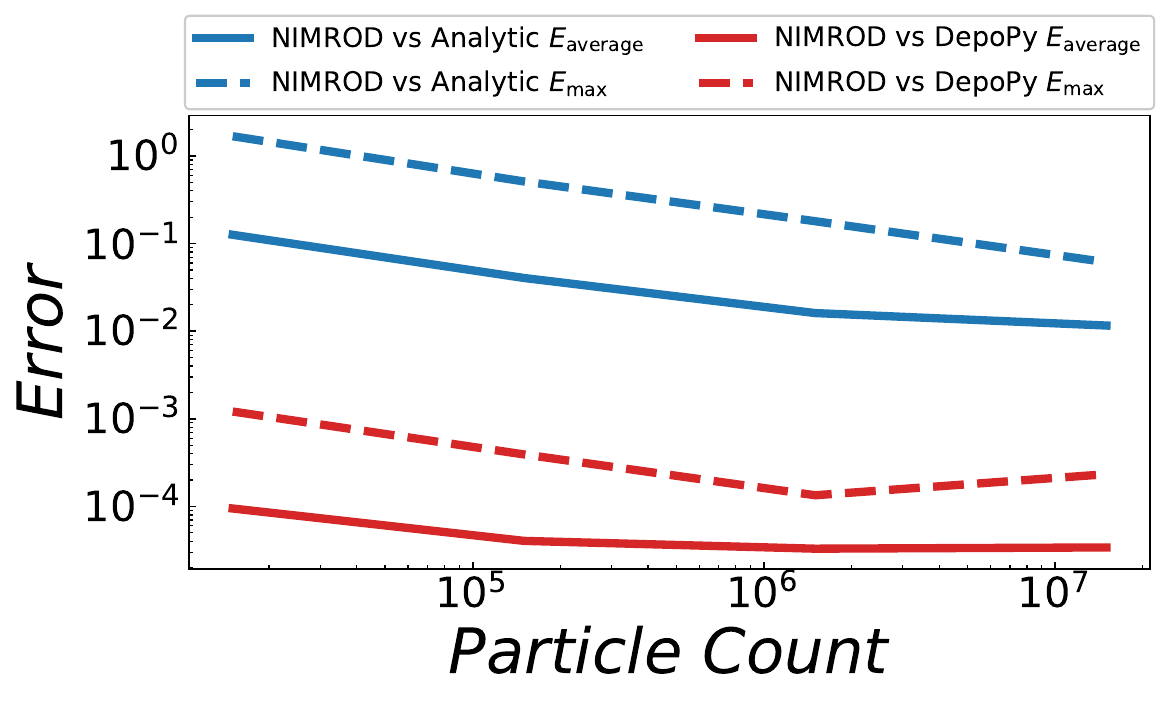}
            \put(92,46){\small{(b)}}
        \end{overpic}
        \label{fig:Gaussian_vs_NIMROD_vs_DepoPy_b}
    }
    \caption{Comparison of density profiles: NIMROD vs. analytic Gaussian and vs. DepoPy. (a) Radial cut through the magnetic axis (b) Errors for NIMROD vs. analytic Gaussian and vs. DepoPy at selected particle counts. NIMROD's error relative to the analytic Gaussian decreases with higher particle counts, while its discrepancy with DepoPy remains small but levels off beyond $10^6$ particles.
    }
    \label{fig:Gaussian_vs_NIMROD_vs_DepoPy}
\end{figure}

Figures\,\ref{fig:Gaussian_vs_NIMROD_vs_DepoPy}(a,b) show that finite element solutions closely reproduce the analytic Gaussian, and 
provide errors for both NIMROD vs.~analytic Gaussian and NIMROD vs.~DepoPy. 
NIMROD's average error relative to the Gaussian decreases from $1.3\times 10^{-1}$ at $1.5\times 10^{4}$ particles 
to $1.2\times 10^{-2}$ at $1.5\times 10^{7}$, while its maximum error drops from $1.7\times 10^{0}$ to $6.1\times 10^{-2}$. 
However, the discrepancy between NIMROD and DepoPy remains small but does not improve beyond $1.5\times 10^{6}$ particles. 
In fact, the average error between NIMROD and DepoPy marginally increases from $3.3\times 10^{-5}$ to $3.4\times 10^{-5}$, 
while the maximum error grows from $1.3\times 10^{-4}$ to $2.3\times 10^{-4}$. 

%%%%%%%%%%%%%%%%%%%%%%%%%%%%%%%%%%%%%%%%%%%%%%%%%%%%%%%%%%%%%%%%%%%%%%%%%%%%%%%

\section{Guiding-center orbit-averaging method for DIII-D RE beams}\label{sec:nimrod_dep}

In this section we apply the techniques from earlier sections to the post-disruption RE beam of DIII-D shot 184602. We start by illustrating how coarse spatial resolution near the magnetic axis can give rise to spurious peaks or troughs in the deposited current while refined meshes avoid these issues. An orbit-averaging strategy is then presented, depositing partial current contributions over multiple kinetic steps in order to reduce numerical noise without a prohibitive increase in particle number and without significantly increasing the computational cost. In closing, we examine how higher-energy (tens of MeV) REs exhibit progressively larger radial displacements on microsecond timescales.

\subsection{Error analysis for parallel current deposition at the initial time}\label{sec:error_analysis_parallel_current}

In this subsection, we quantify relative errors in the parallel current by comparing the experimentally reconstructed profile for DIII-D shot 184602 (Fig.\,\ref{fig:j_par_184602}) with parallel currents obtained through deposition onto the NIMROD mesh (Sec.\,\ref{sec:deposition}). These deposited currents are generated from particle ensembles produced by the Metropolis-Hastings method of Sec.\,\ref{sec:MH}, which directly samples a mono-energetic, mono-pitch beam from the reconstructed parallel current profile. The parallel current deposition accuracy is examined using what we refer to as the ``coarse mesh'' ($n = 32$, $m = 32$) and the ``refined mesh'' ($n = 64$, $m= 32$), where the radial direction is resolved double that of the coarse mesh. Particle counts range from $1.0 \times 10^5$ to $1.0 \times 10^7$. Each sampled RE has kinetic energy of $10\,\mathrm{MeV}$ and a pitch angle of $10^{\circ}$.
The weights $w_l$ in Eq.\,\eqref{eq:k_delta_axi} for the deposition of RE parallel current onto the NIMROD mesh are set to:
\begin{equation}
        w_l = g_\textsubscript{RE} \left( \frac{q_e p_{\parallel}}{m \gamma} - \mu (\hat{\mathbf{b}} \cdot \nabla \times \hat{\mathbf{b}}) \right)_l,
        \label{eq:weight_parallel_current}
\end{equation}
where the expression in parentheses (units $A \cdot m$) represents, up to an inverse $2\pi R$ factor, the parallel current contribution from a single electron. The prefactor $g_\textsubscript{RE}$ is chosen to equate the total parallel current from all computational particles with the experimentally reconstructed MHD parallel current. The weight in Eq.\,\eqref{eq:weight_parallel_current} matches the single-particle parallel current in Eq.\,\eqref{eq:single_j}, with the factor $2\pi R$ already accounted for in Eq.\,\eqref{eq:k_delta_axi}.

\begin{figure}[ht]
    \centering
    \subfloat{%
        \begin{minipage}{0.75\columnwidth}
            \centering
            \textbf{~~~~~~~~~Coarse Mesh}\\[0.5em]
            \begin{overpic}[width=\linewidth]{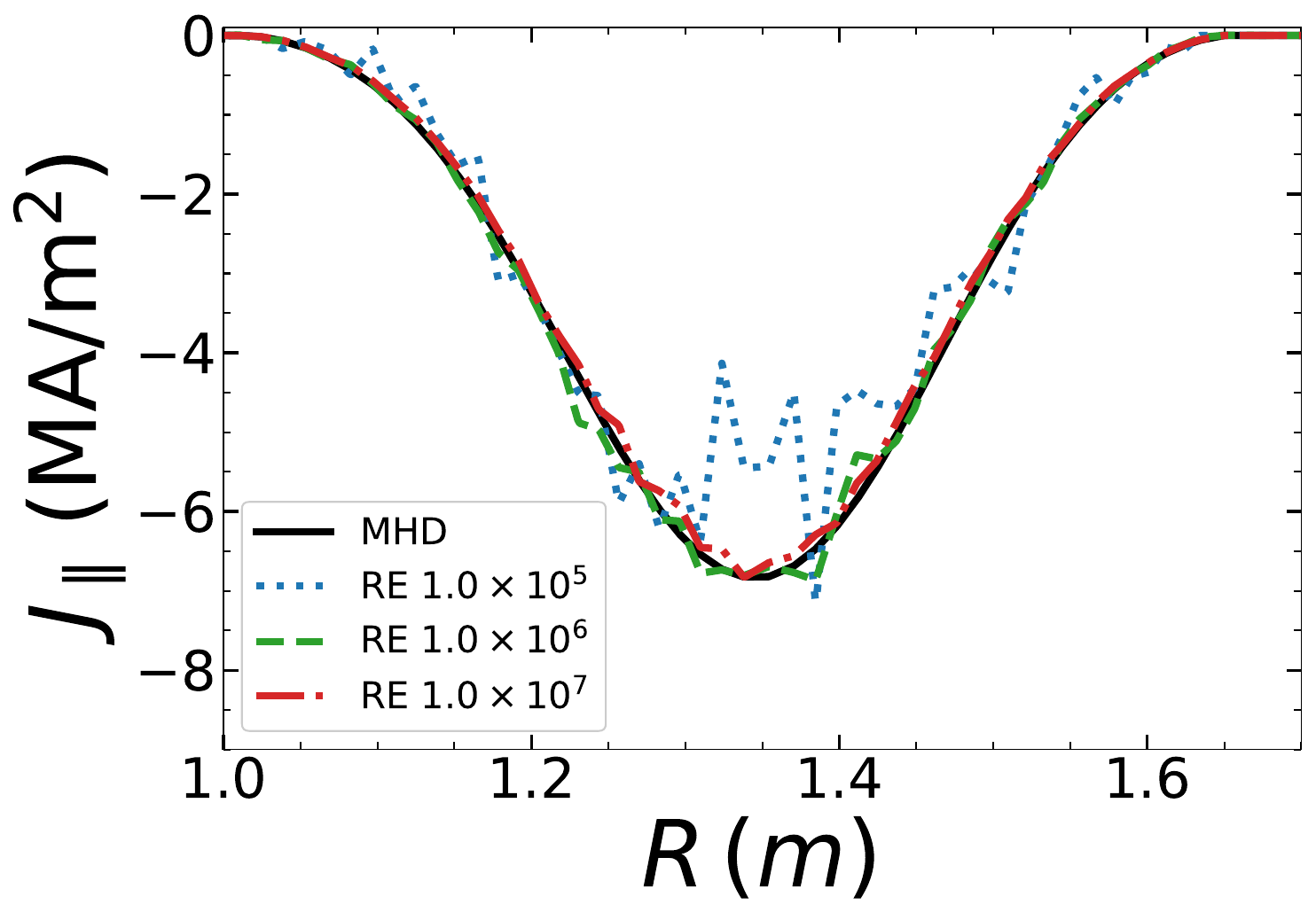}
                \put(18,58){\small{(a)}}
            \end{overpic}
        \end{minipage}
    }
    \hspace{0.005\textwidth}
    \subfloat{%
        \begin{minipage}{0.75\columnwidth}
            \centering
            \textbf{~~~~~~~~Refined Mesh}\\[0.5em]
            \begin{overpic}[width=\linewidth]{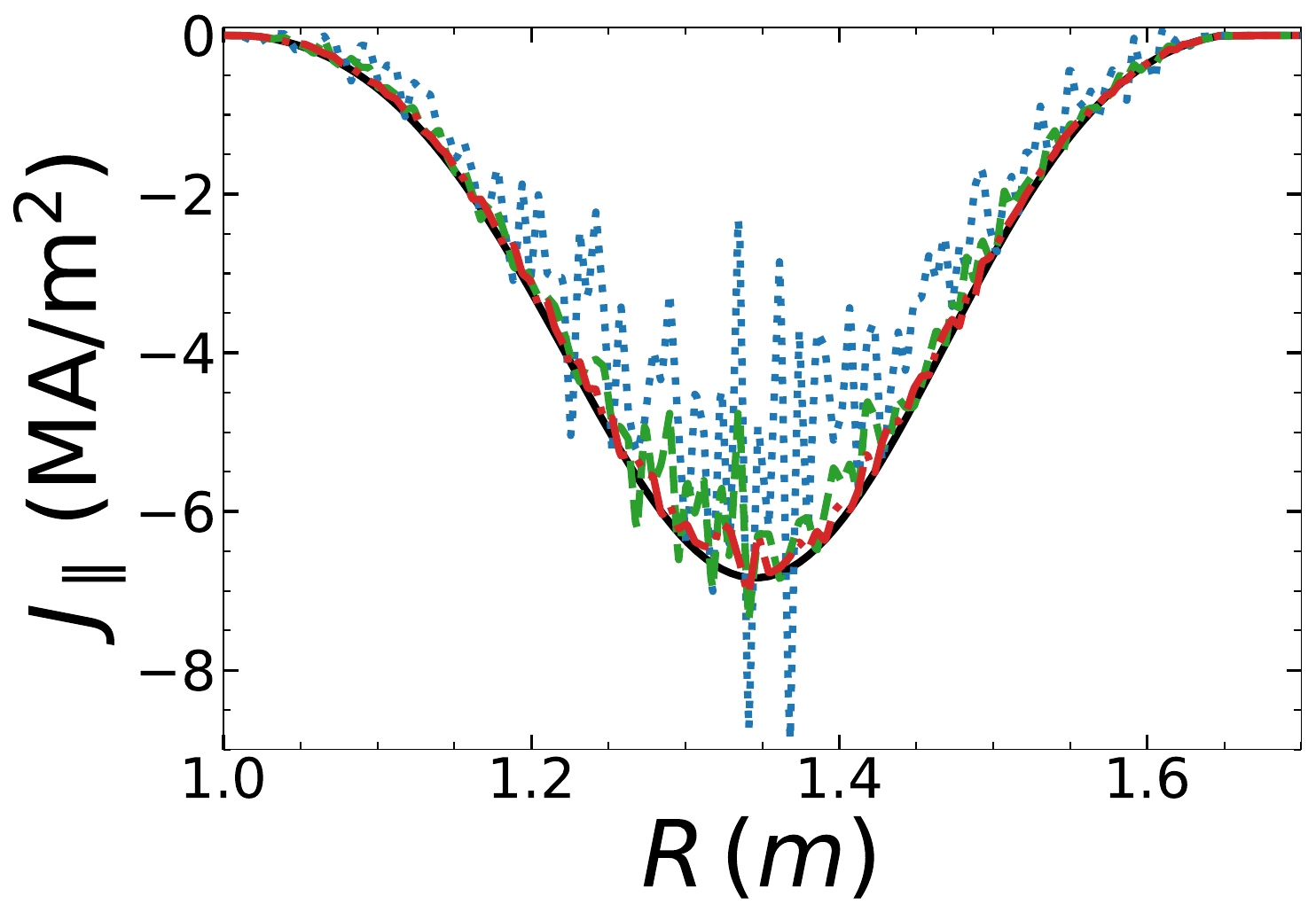}
                \put(18,58){\small{(b)}}
            \end{overpic}
        \end{minipage}
    }
    \caption{Cuts through the magnetic axis of the deposited RE parallel current density (colored curves) versus the MHD reference (black dashed line) for two meshes: 
    (a) Coarse $(n=32,m=32)$. (b) Refined $(n=64,m=32)$. As the number of particles increases, the RE profile converges to the EFIT-derived $J_{\parallel}$. 
    The refined mesh demands more particles for comparable accuracy.}
    \label{fig:compare_mesh}
\end{figure}

\begin{table}[ht]
    \centering
    \begin{tabular}{|c|c|c|c|}
    \hline
    $\mathbf{N_p}$  & $\mathbf{Error_{max}}$ & $\mathbf{Error_{average}}$ & \textbf{$k$-estimate} \\
    \hline
    \multicolumn{4}{|c|}{\textbf{Coarse mesh ($32\times32$)}} \\ \hline
    $1.0\times10^{5}$ & $1.5\times10^{0}$ & $1.4\times10^{-1}$ & --     \\
    $1.0\times10^{6}$ & $5.8\times10^{-1}$ & $4.6\times10^{-2}$ & $-0.48$ \\
    $1.0\times10^{7}$ & $2.5\times10^{-1}$ & $3.0\times10^{-2}$ & $-0.19$ \\
    \hline
    \multicolumn{4}{|c|}{\textbf{Refined mesh ($64\times32$)}} \\ \hline
    $1.0\times10^{5}$ & $3.4\times10^{0}$ & $2.7\times10^{-1}$ & --     \\
    $1.0\times10^{6}$ & $8.3\times10^{-1}$ & $8.7\times10^{-2}$ & $-0.49$ \\
    $1.0\times10^{7}$ & $3.1\times10^{-1}$ & $3.4\times10^{-2}$ & $-0.41$ \\
    \hline
    \end{tabular}
    \caption{Global parallel current deposition errors for the coarse and refined meshes at increasing particle counts. Both the average and maximum errors, calculated via Eqs.\,\eqref{eq:mean_error}--\eqref{eq:max_error}, decrease as the number of particles ($N_p$) increases. Assuming the power law $Error_{\text{average}}=C\,N_p^{k}$, two consecutive particle counts $N_p^{(1)}<N_p^{(2)}$ give $k=\frac{\log_{10}\!\bigl(Error_{\text{average}}(N_p^{(2)})/Error_{\text{average}}(N_p^{(1)})\bigr)}{\log_{10}(N_p^{(2)}/N_p^{(1)})}$. Because each refinement multiplies $N_p$ by $10$, the denominator equals unity, i.e.\ $\log_{10}(N_p^{(2)}/N_p^{(1)})=1$.}
    \label{tab:mesh_comparison}
    \end{table}

Figures\,\ref{fig:compare_mesh}(a,b) illustrate the convergence of the deposited RE parallel current to the MHD reference as the particle count increases. 
Correspondingly, Table\,\ref{tab:mesh_comparison} summarizes the global errors in parallel current for these same cases, 
calculated using Eqs.\,\eqref{eq:mean_error}--\eqref{eq:max_error}, which are determined by setting $n^{(1)} = J_{\parallel}^{\mathrm{MHD}}$
as the reference MHD parallel current and $n^{(2)} = J_{\parallel}^{\mathrm{RE}}$ as the particle-based RE parallel current. The deposition onto the refined mesh exhibits larger errors than that onto the coarse mesh at the same particle count due to the presence of fewer particles per element. Increasing the particle count reduces both the average and maximum errors for both mesh types.

The exponents $k$ in Table\,\ref{tab:mesh_comparison} quantify the empirical scaling of the mean relative deposition error with particle count ($N_p$), modeled by $Error_{\text{average}} = C\,N_{p}^{k}$, where $C$ is a constant. Both meshes converge close to the ideal Monte-Carlo rate ($k=-0.5$). For the coarse mesh, the exponent remains near ideal between $10^{5}$ and $10^{6}$ particles ($k=-0.48$) but weakens to $k=-0.19$ at $10^{7}$ particles, signalling a shift from statistical-noise-dominated to grid-error-dominated behavior. The refined mesh, by contrast, sustains near-ideal scaling throughout ($k=-0.49$ followed by $k=-0.41$).

\subsection{Single particle orbit dynamics near the magnetic axis}\label{sec:orbit_dynamics}

\begin{figure}[H]
    \centering
    \subfloat{%
        \begin{minipage}{0.4\columnwidth}
            \centering
            \textbf{~~~~~~~~~Coarse mesh, \text{poly-deg}=1} \\[0.5em]
            \begin{overpic}[width=\linewidth]{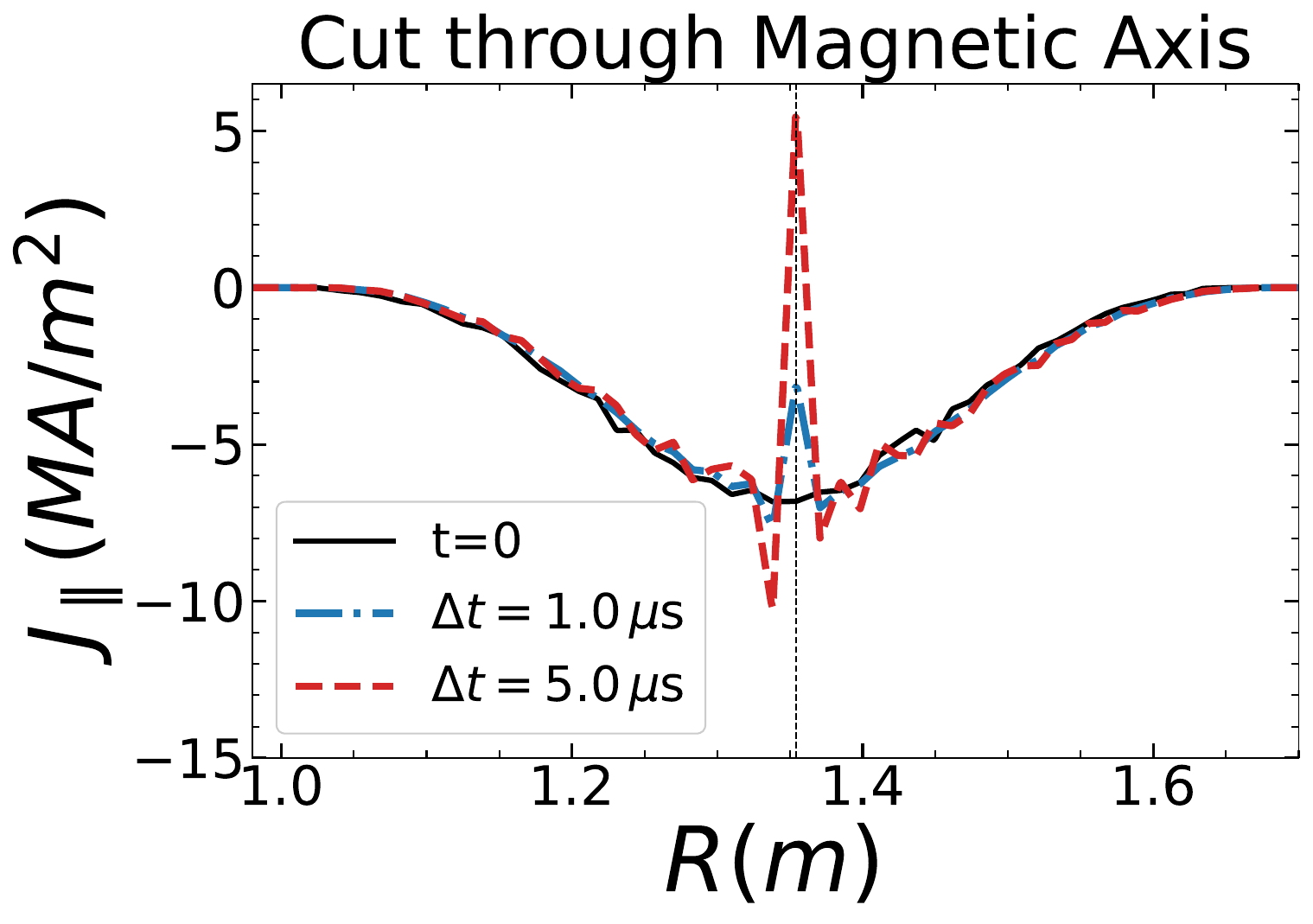}
                \put(22,57){\small{(a)}}
            \end{overpic}
            \label{fig:evol_dep_mesh_comp_a}
        \end{minipage}
    }
    \hspace{0.005\textwidth}
    \subfloat{%
        \begin{minipage}{0.4\columnwidth}
            \centering
            \textbf{~~~~~~~~Refined mesh, \text{poly-deg}=1} \\[0.5em]
            \begin{overpic}[width=\linewidth]{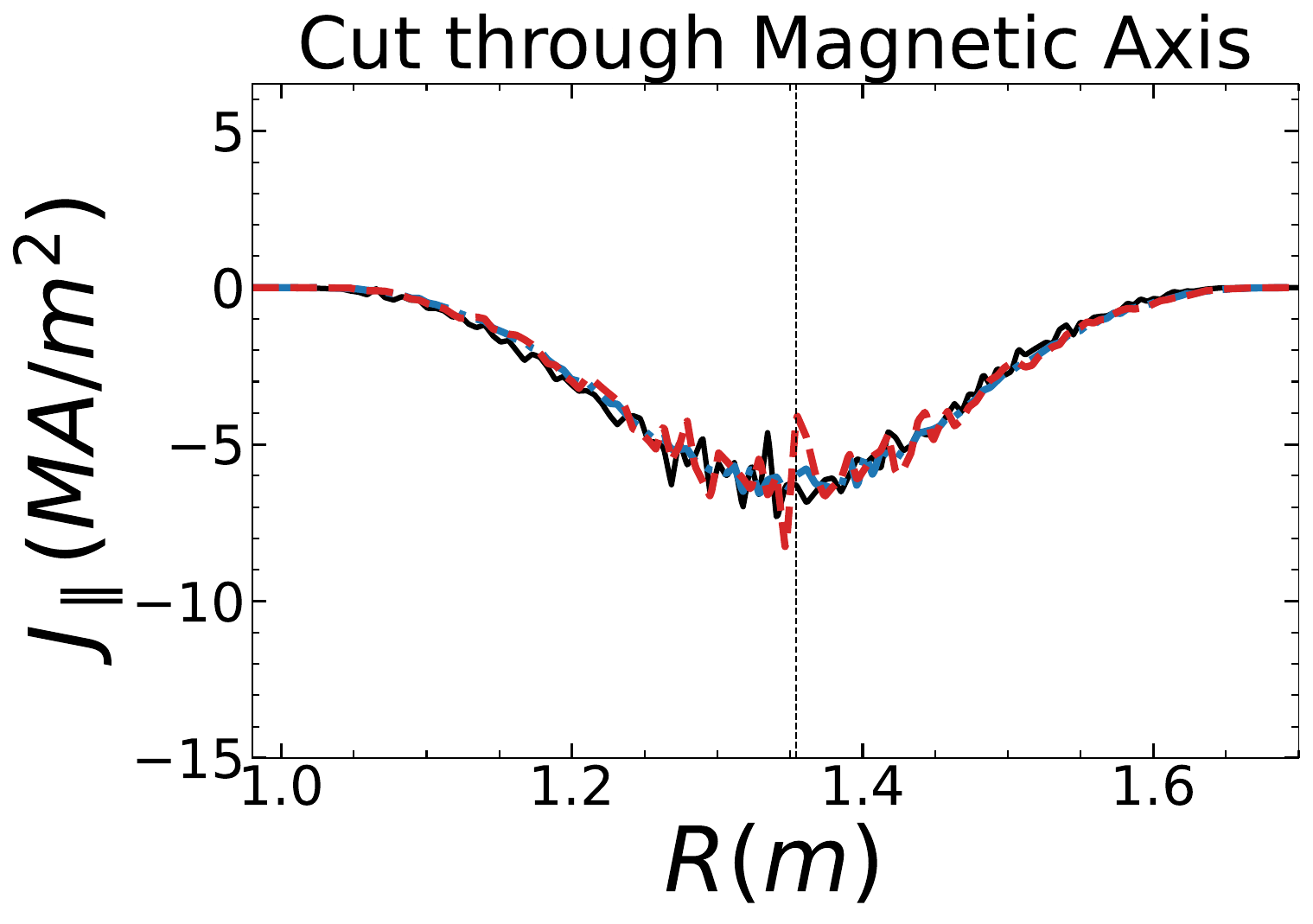}
                \put(22,57){\small{(b)}}
            \end{overpic}
            \label{fig:evol_dep_mesh_comp_b}
        \end{minipage}
    }\\[1ex]
    \subfloat{%
        \begin{overpic}[width=0.4\columnwidth]{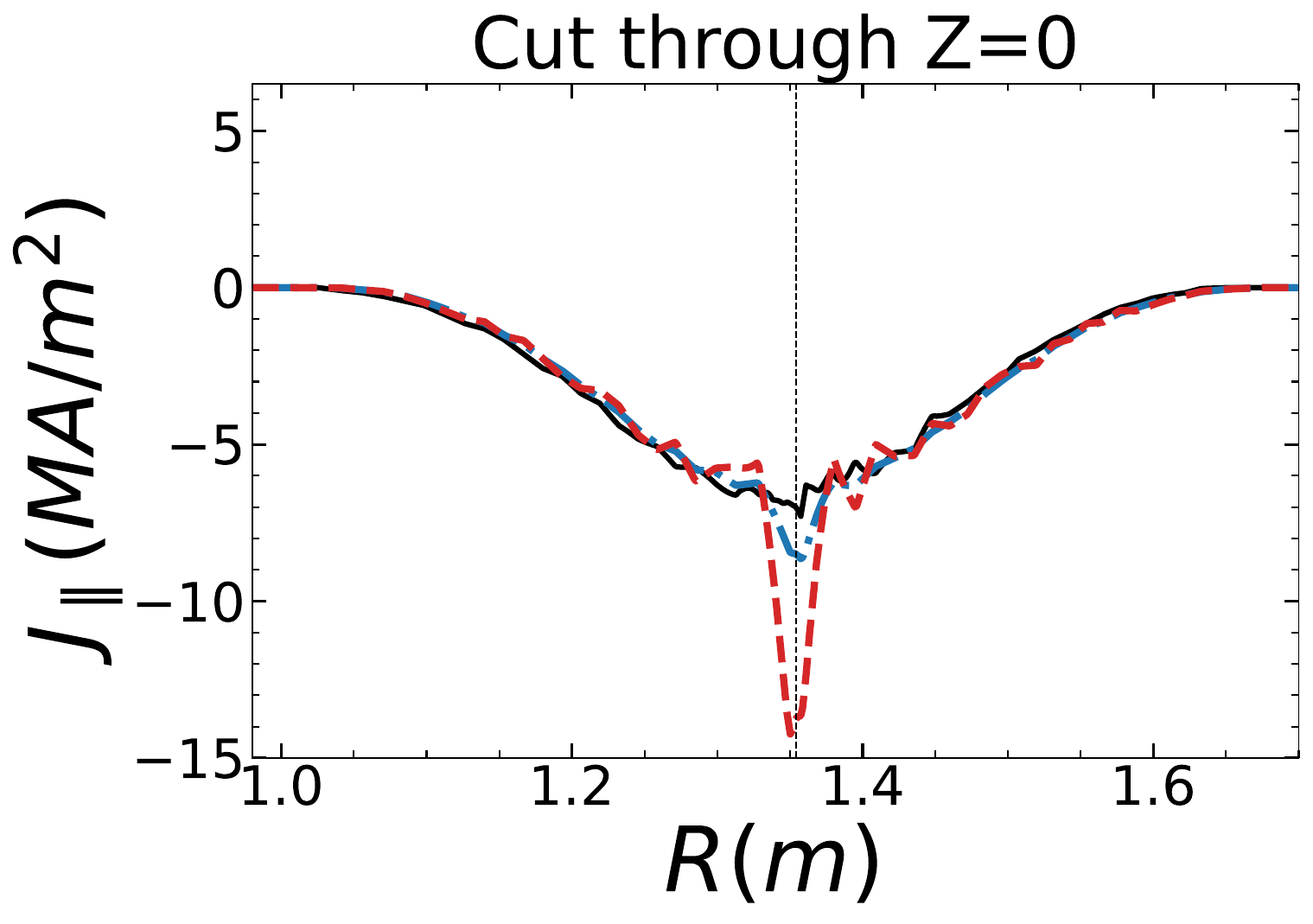}
            \put(22,57){\small{(c)}}
        \end{overpic}
        \label{fig:evol_dep_mesh_comp_c}
    }
    \hspace{0.005\textwidth}
    \subfloat{%
        \begin{overpic}[width=0.4\columnwidth]{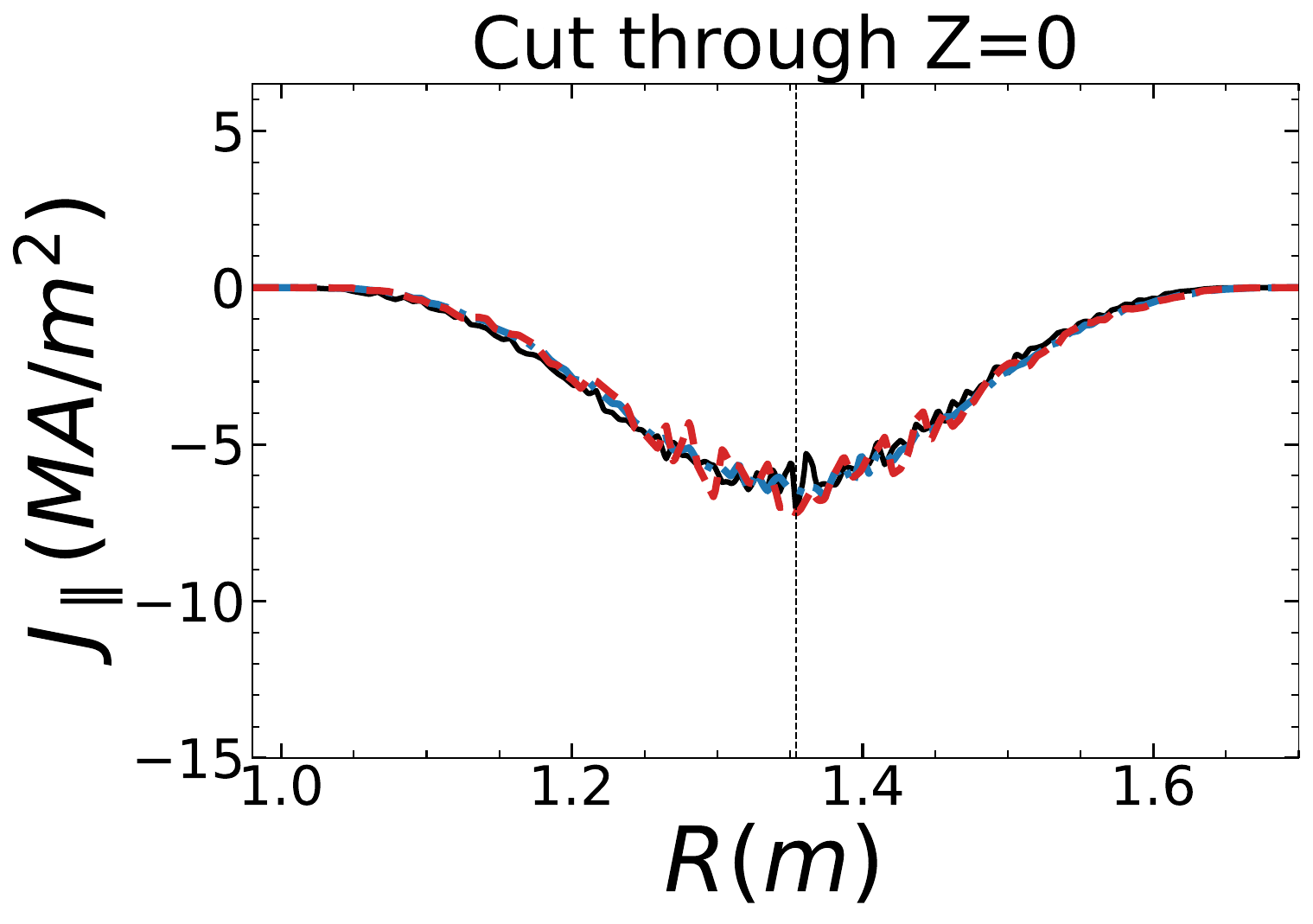}
            \put(22,57){\small{(d)}}
        \end{overpic}
        \label{fig:evol_dep_mesh_comp_d}
    }
    \caption{Time evolution of deposited $J_{\parallel}$ at $t=0$, $1\,\mu s$, and $5\,\mu s$. 
    (a,b) Cuts through the magnetic axis. (c,d) Cuts through $Z=0$. 
    Coarser-field representations (left panels) develop large under- and overshoot near the magnetic axis, 
    whereas refined meshes (right panels) prevent central artifacts. Vertical dashed lines indicate the magnetic axis.
    The bilinear (\text{poly-deg}=1) deposition employs orbit-averaging with $c_\textsubscript{step}=1$, $\Delta t_{\mathrm{dump}}=1.0 \times 10^{-8}\,\mathrm{s}$ 
    and $\Delta t_{\mathrm{KORC}}=1.0 \times 10^{-11}\,\mathrm{s}$ (Sec.\,\ref{sec:orbit-averaging}.)}
    \label{fig:evol_dep_mesh_comp}
\end{figure}

\begin{figure}[H]
    \centering
    \subfloat{%
        \begin{minipage}{0.4\columnwidth}
            \begin{overpic}[width=\linewidth]{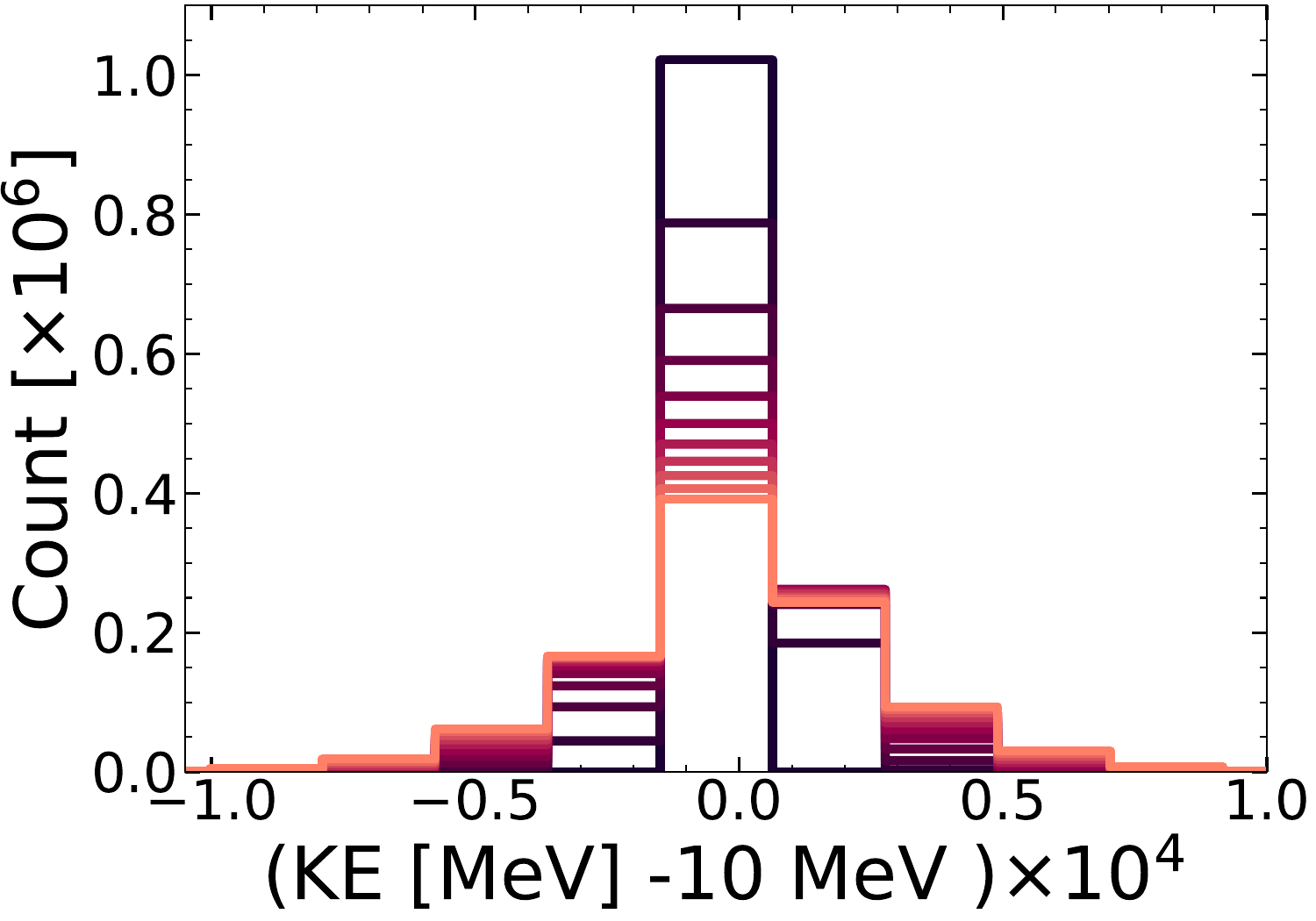}
                \put(16,62){\small{(a)}}
            \end{overpic}
            \label{fig:1d_hist_a}
        \end{minipage}
    }
    \hspace{0.005\textwidth}
    \subfloat{%
        \begin{minipage}{0.4\columnwidth}
            \begin{overpic}[width=\linewidth]{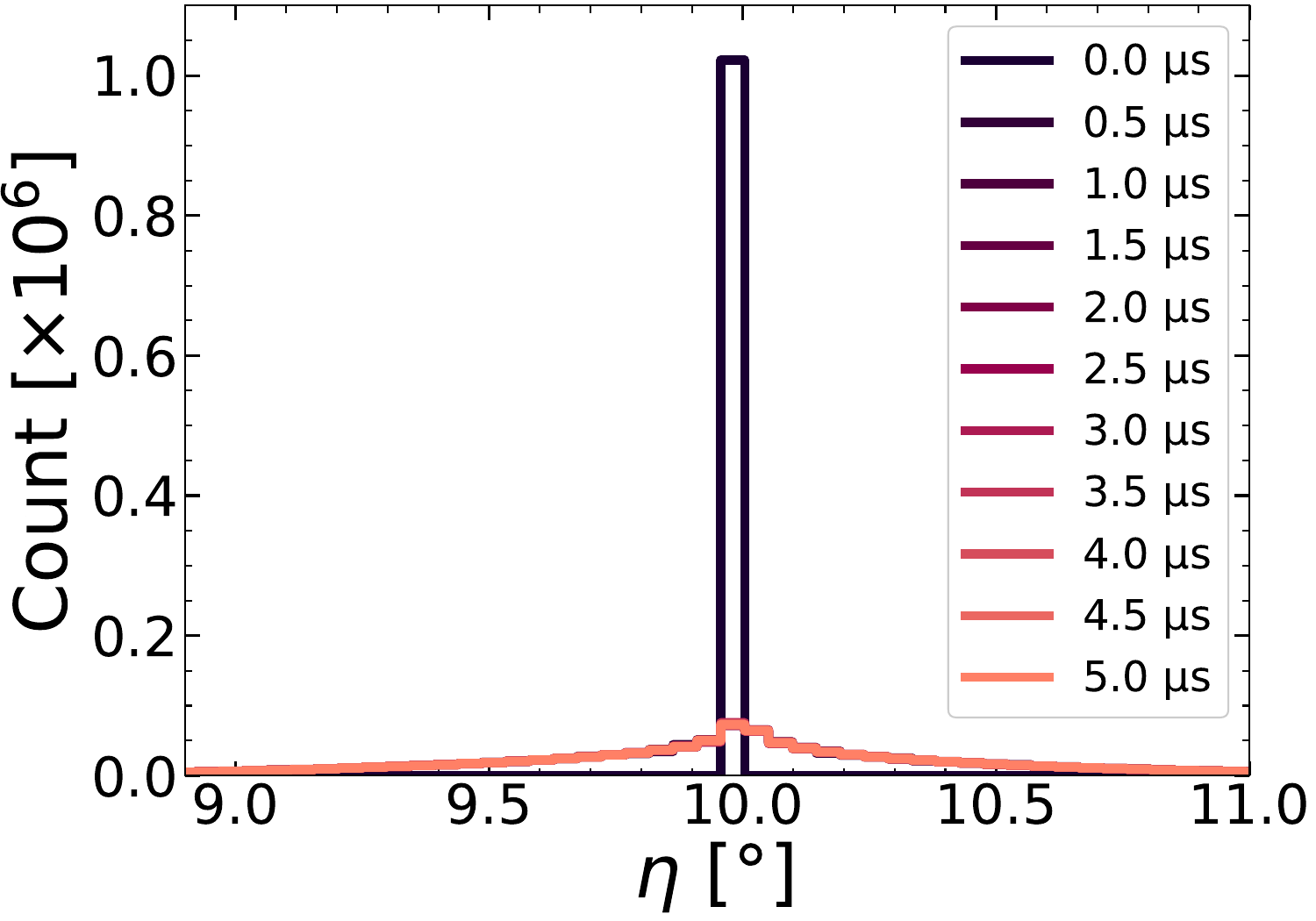}
                \put(16,62){\small{(b)}}
            \end{overpic}
            \label{fig:1d_hist_b}
        \end{minipage}
    }\\[1ex]
    \subfloat{%
        \begin{overpic}[width=0.4\columnwidth]{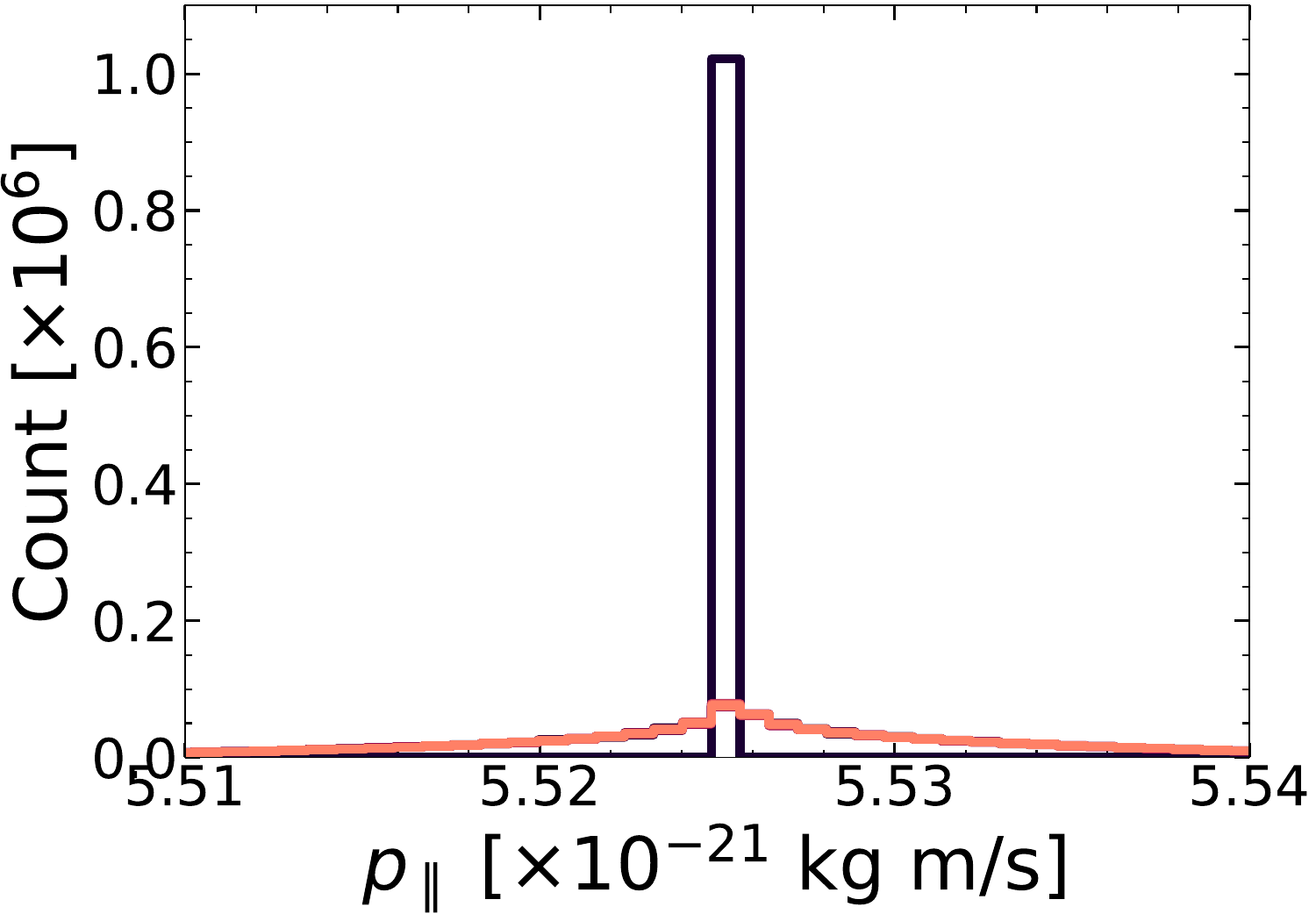}
            \put(16,62){\small{(c)}}
        \end{overpic}
        \label{fig:1d_hist_c}
    }
    \hspace{0.005\textwidth}
    \subfloat{%
        \begin{overpic}[width=0.4\columnwidth]{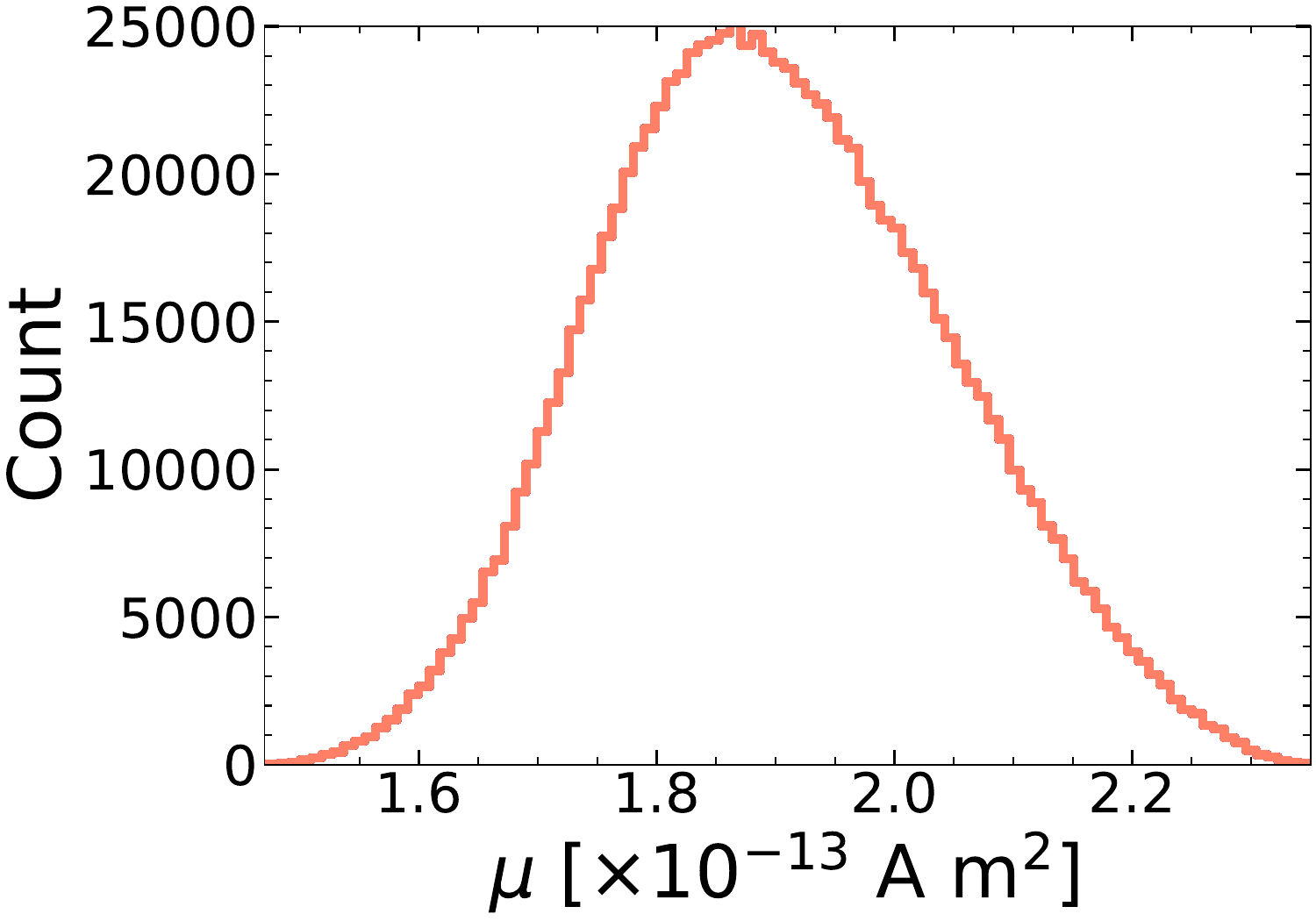}
            \put(22,60){\small{(d)}}
        \end{overpic}
        \label{fig:1d_hist_d}
    }
    \caption{1D histograms from the coarse-mesh simulation. 
    (a) Energy: starting from a mono-energetic (10~MeV) sample, the energy peak progressively decreases over time. 
    (b) Pitch angle: an initial mono-pitch of $10^\circ$ broadens to span roughly $9^\circ-11^\circ$ within 0.5~$\mu$s. 
    (c) Parallel momentum: the initially mono-distributed $p_\parallel$ (stemming from the mono-energy and mono-pitch conditions) evolves within 0.5~$\mu$s to a range of $5.51$--$5.54 \times 10^{-21}$~kg\,m/s. 
    (d) Magnetic moment: the distribution remains constant over time.}
    \label{fig:1d_hist}
\end{figure}

\begin{figure}[H]
    \centering
    \subfloat{%
        \begin{minipage}{0.4\columnwidth}
            \centering
            \textbf{~~~~~~~Coarse mesh, \text{poly-deg}=1} \\[0.5em]
            \begin{overpic}[width=\linewidth]{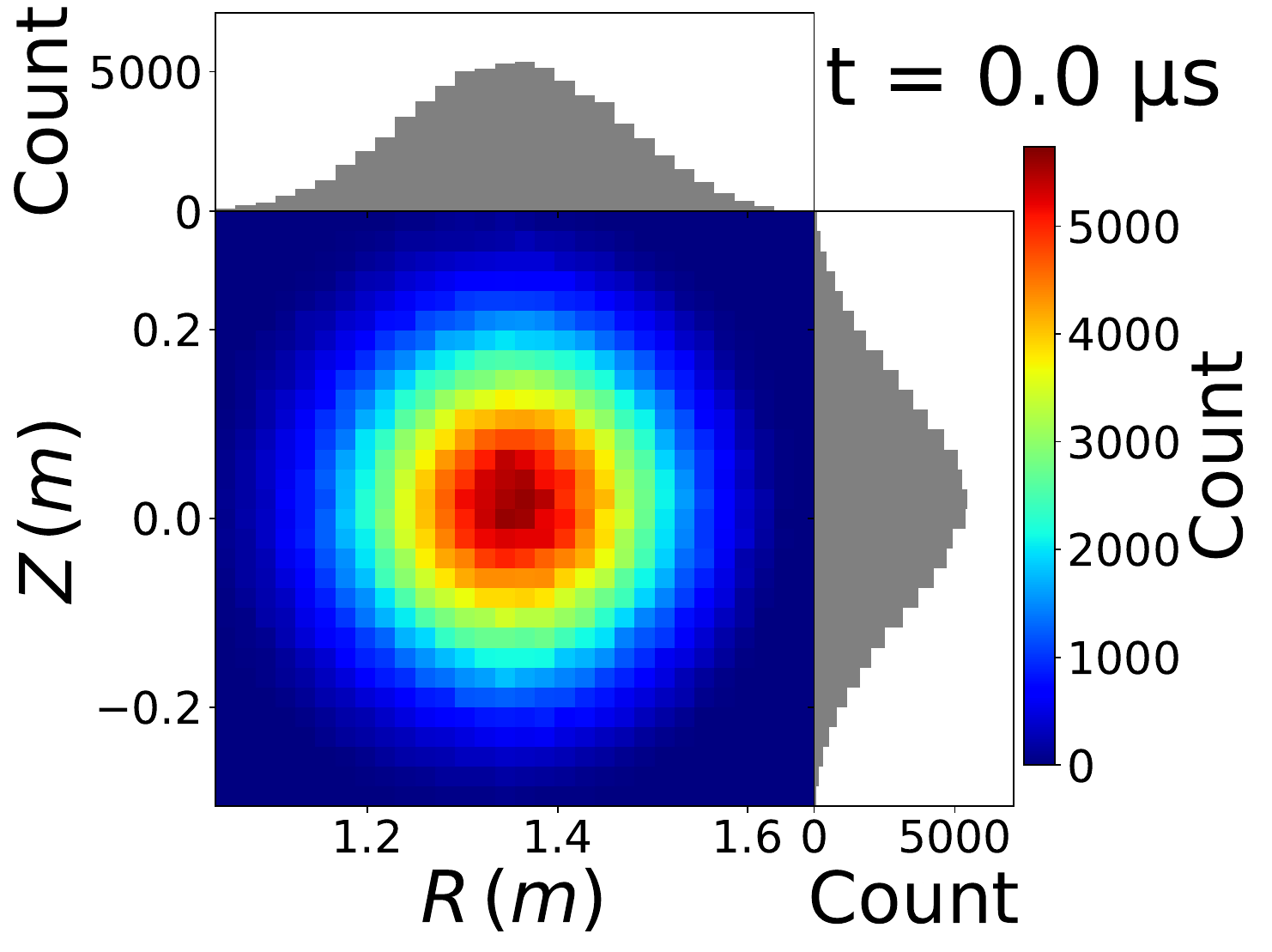}
                \put(20,65){\small{(a)}}
            \end{overpic}
            \label{fig:2d_sample_a}
        \end{minipage}
    }
    \hspace{0.005\textwidth}
    \subfloat{%
        \begin{minipage}{0.4\columnwidth}
            \centering
            \textbf{~~~~~~Refined mesh, \text{poly-deg}=1} \\[0.5em]
            \begin{overpic}[width=\linewidth]{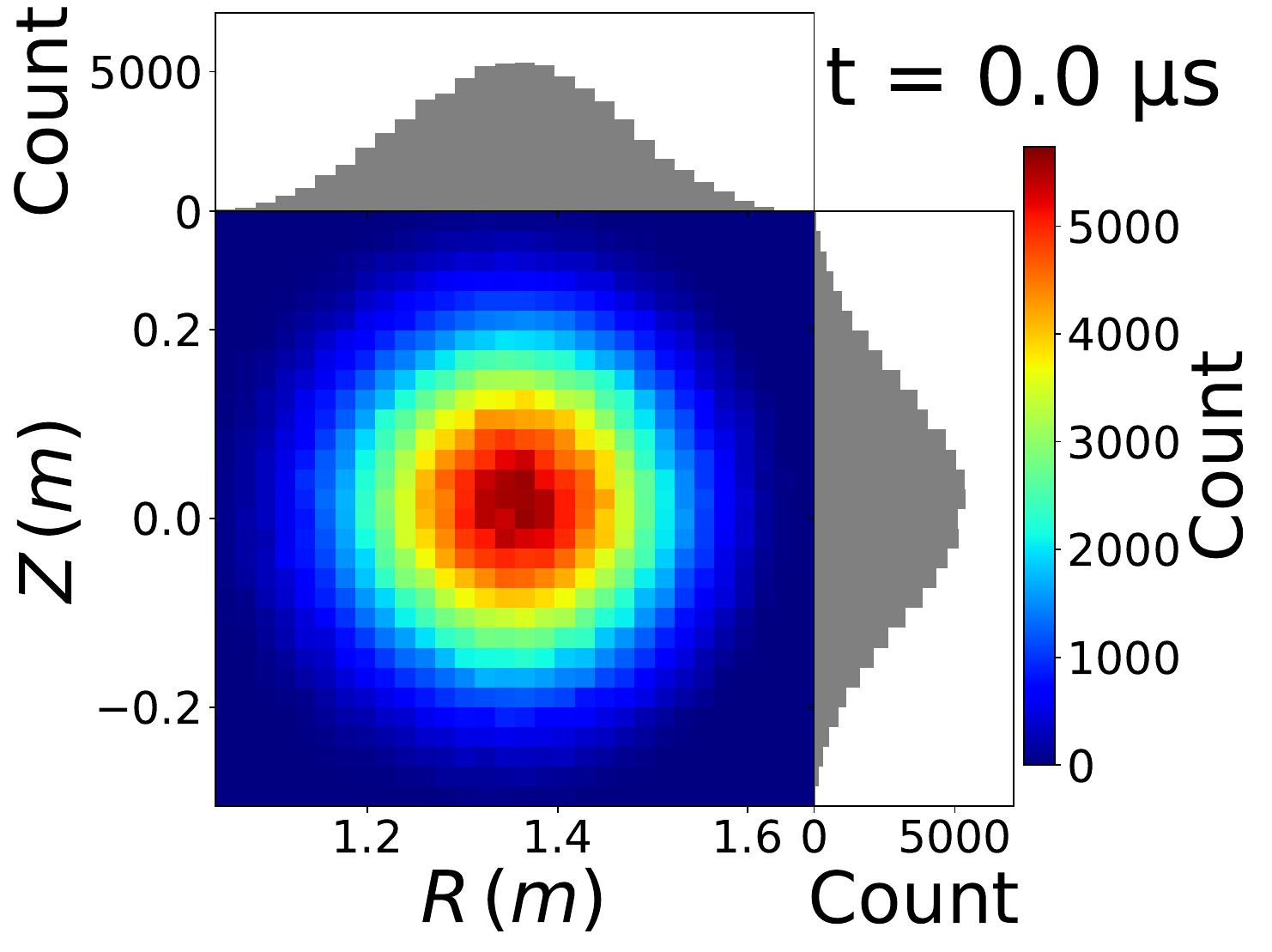}
                \put(20,65){\small{(b)}}
            \end{overpic}
            \label{fig:2d_sample_b}
        \end{minipage}
    }\\[1ex]
    \subfloat{%
    \begin{minipage}{0.4\columnwidth}
        \begin{overpic}[width=\linewidth]{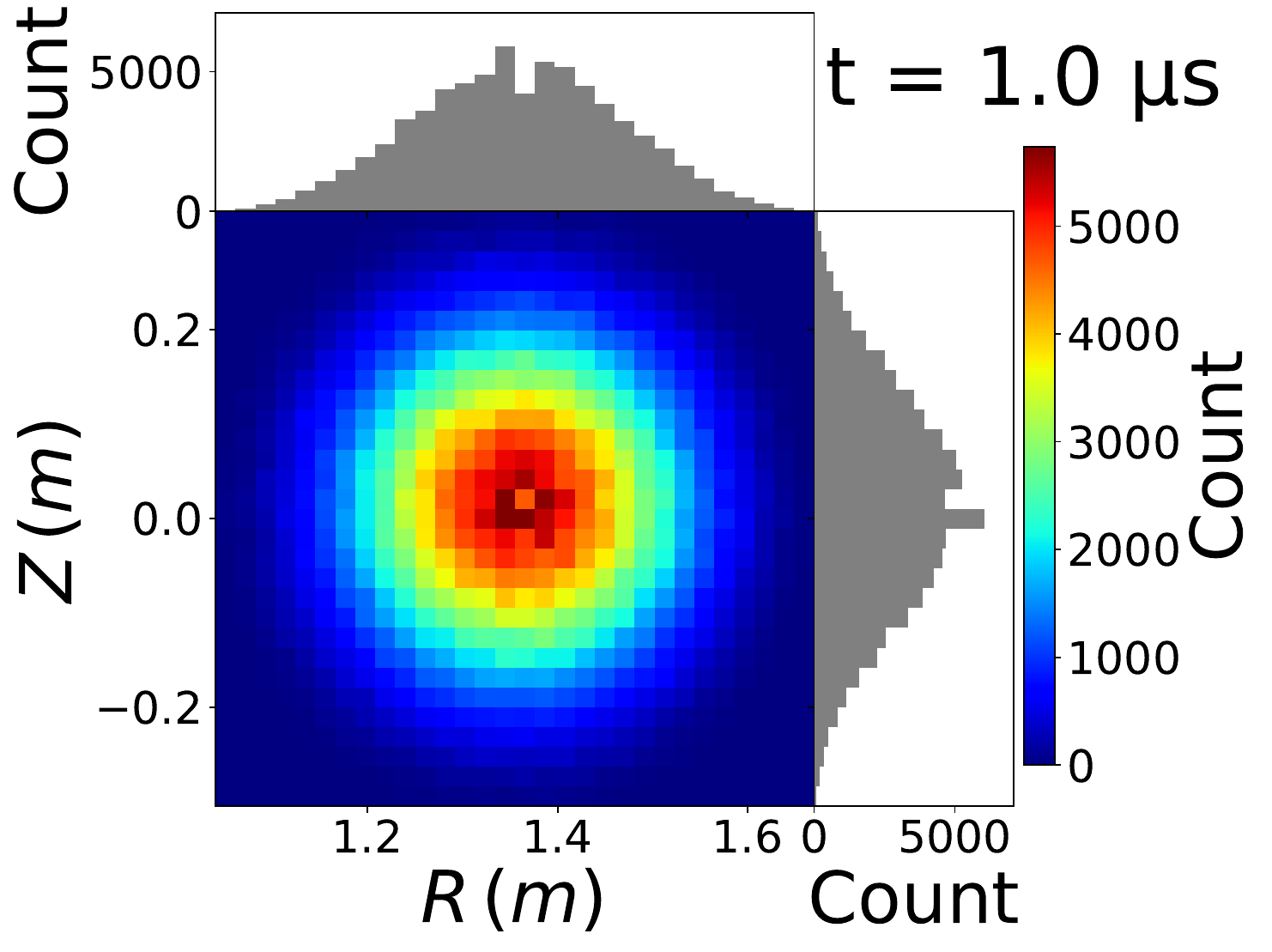}
            \put(20,65){\small{(c)}}
        \end{overpic}
        \label{fig:2d_sample_c}
    \end{minipage}
    }
    \hspace{0.005\textwidth}
    \subfloat{%
    \begin{minipage}{0.4\columnwidth}
        \begin{overpic}[width=\linewidth]{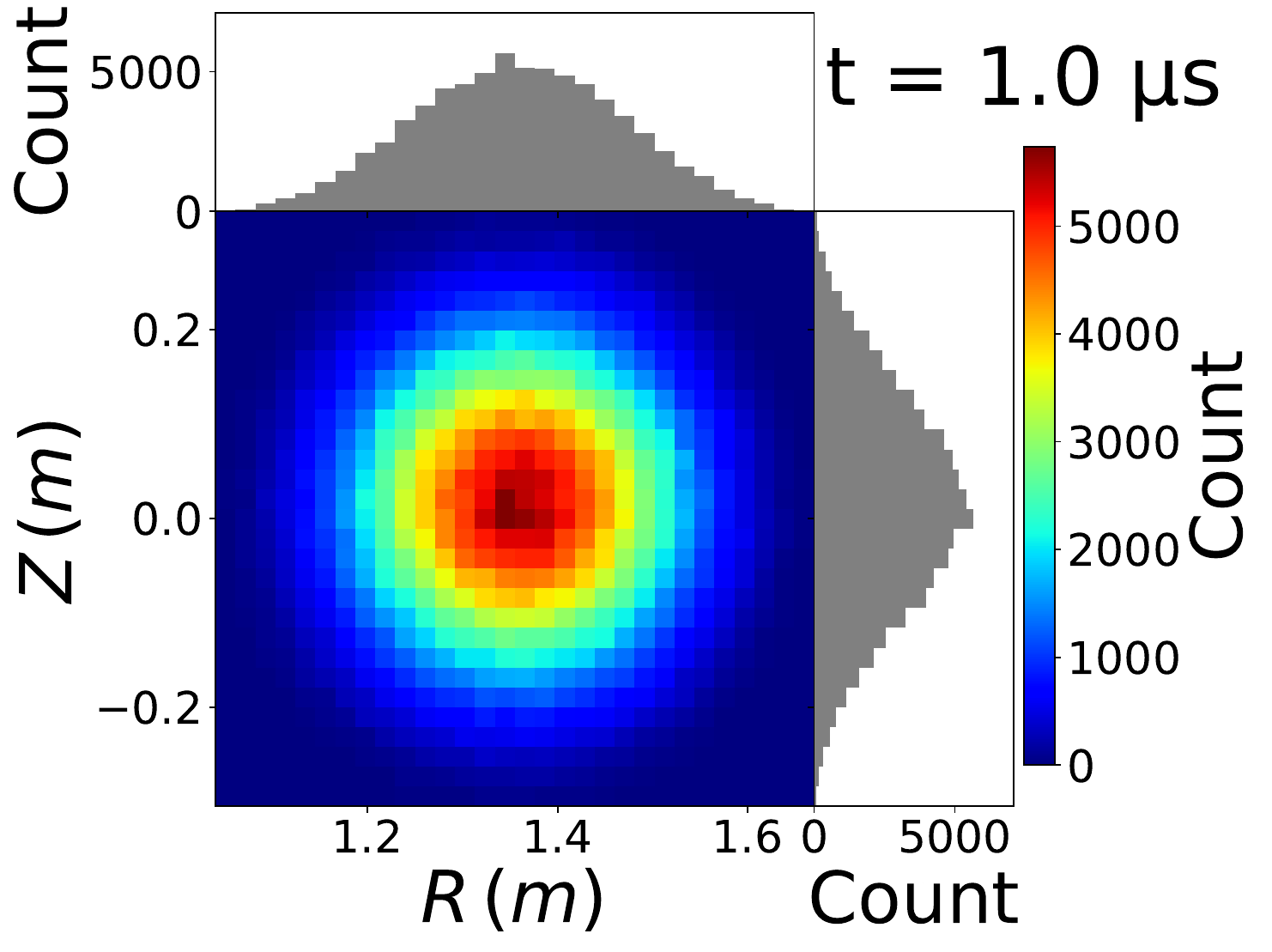}
            \put(20,65){\small{(d)}}
        \end{overpic}
        \label{fig:2d_sample_d}
    \end{minipage}
    }\\[1ex]
    \subfloat{%
        \begin{overpic}[width=0.4\columnwidth]{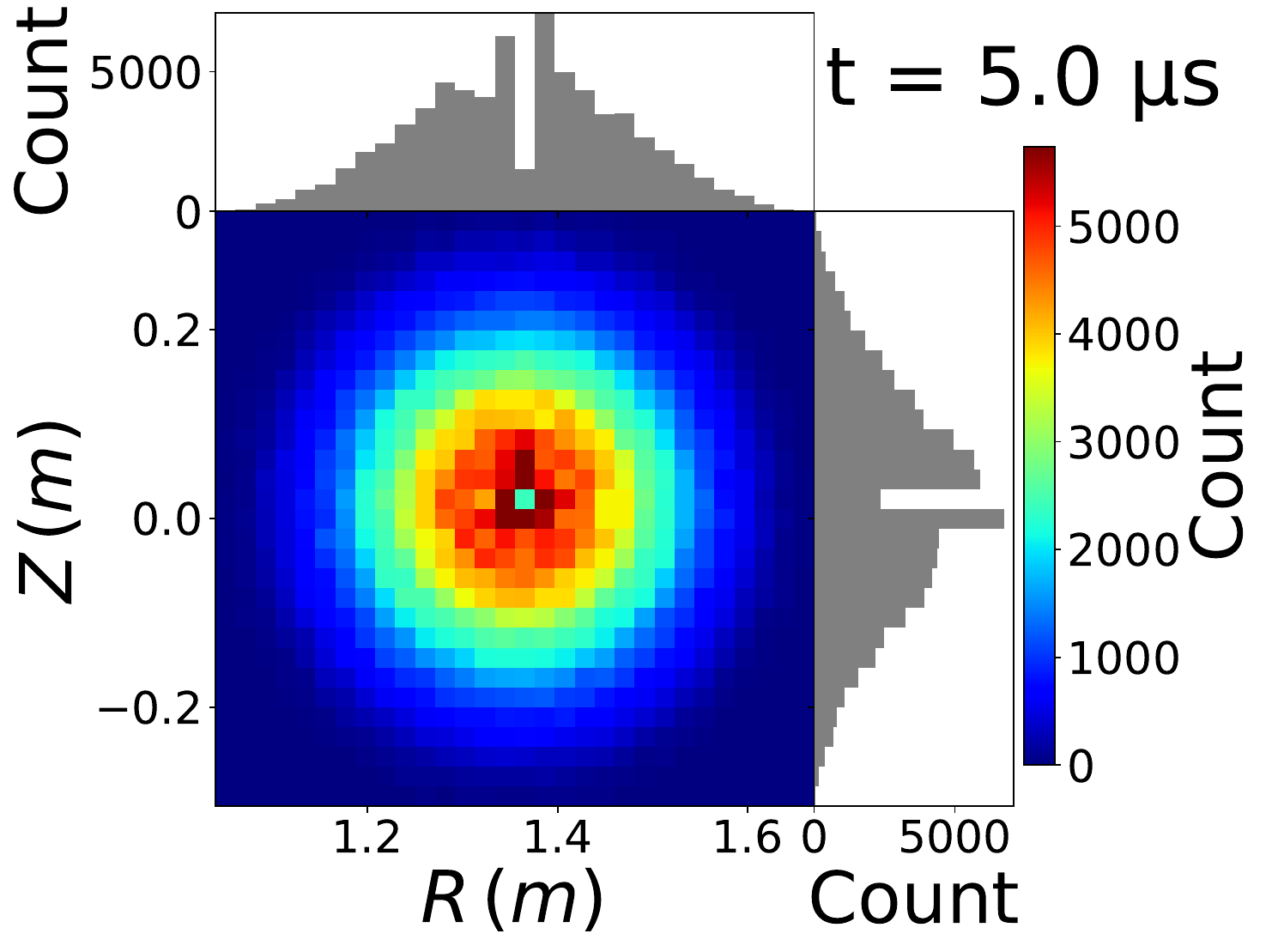}
            \put(20,65){\small{(e)}}
        \end{overpic}
        \label{fig:2d_sample_e}
    }
    \hspace{0.005\textwidth}
    \subfloat{%
        \begin{overpic}[width=0.4\columnwidth]{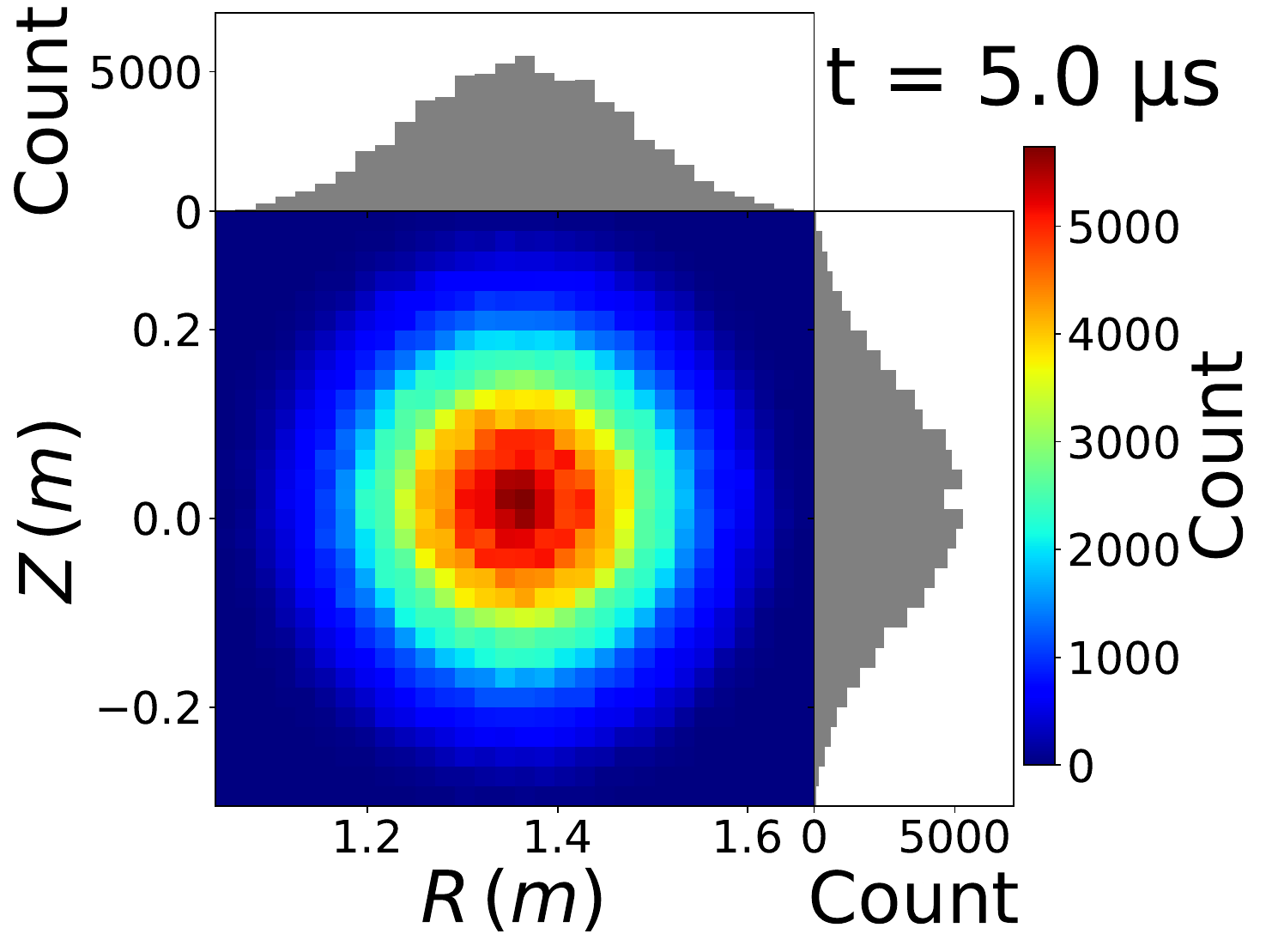}
            \put(20,65){\small{(f)}}
        \end{overpic}
        \label{fig:2d_sample_f}
    }
    \caption{Histograms of the particle count at $t=0$, $1\,\mu s$, and $5\,\mu s$ for two mesh representations. 
    Panels (a,c,e) are generated from the coarse mesh ($\text{poly-deg}=1$), while (b,d,f) correspond to the refined mesh ($\text{poly-deg}=1$). 
    In each panel, particle positions $(R,Z)$ are binned into a 2D histogram using 30 linearly spaced bins in each direction. 
    Superimposed on the 2D histogram are two 1D projections: the top projection histograms the $R$ values for particles within a window of width $2\,\delta Z=0.02\,\mathrm{m}$, 
    centered at the magnetic axis in the $Z$-direction, 
    and the right projection is constructed from the $Z$ values for particles within a window of width $2\,\delta R=0.02\,\mathrm{m}$ about the magnetic axis in the $R$-direction. 
    The window widths $2\,\delta R$ and $2\,\delta Z$ are approximately equal to the corresponding bin widths (i.e. the spatial range divided by 30). 
    Particle count under- and overshoots are absent at $t=0$, but by $t=5\,\mu s$ the coarse mesh clearly exhibit them near the magnetic axis.}
    \label{fig:2d_sample}
\end{figure}

\begin{figure*}[htbp]
    \begingroup                      
      \captionsetup[subfloat]{skip=2pt} % tighten caption-image vertical gap
    
      % Local macro: one-line centred legend for a row
      \newcommand{\rowlegend}[1]{%
        \noindent\begin{minipage}{\textwidth}\centering\textbf{#1}\end{minipage}%
        \par\vspace*{-0.8ex}}
    
      %------------------------------ Row 1
      \rowlegend{Coarse mesh, $\text{poly-deg}=1$, RE at $\sim \boldsymbol{6\,\mathrm{cm}}$ from the magnetic axis}

      \subfloat{\begin{minipage}{0.2299\textwidth}\centering
          \begin{overpic}[width=\linewidth]{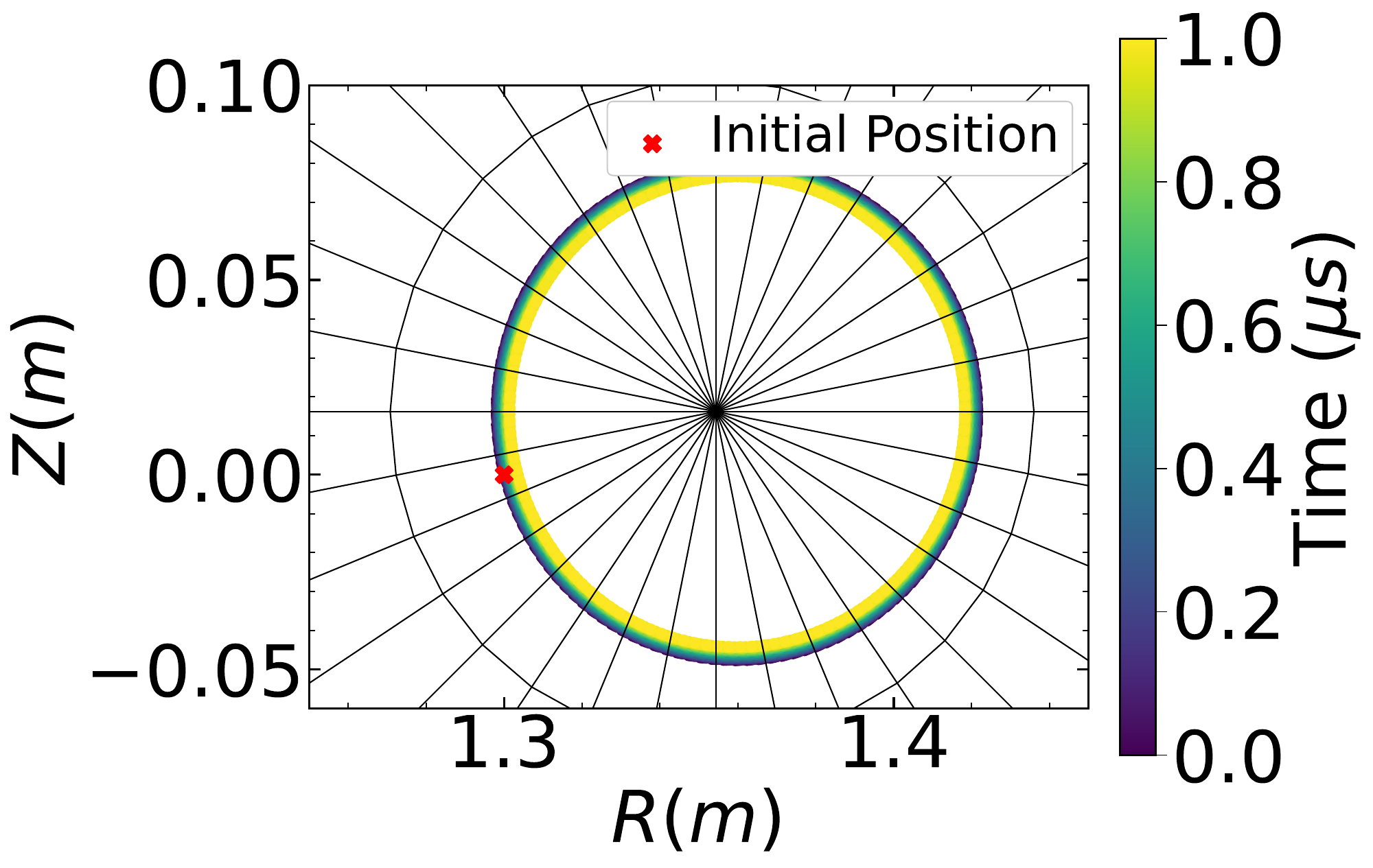}
            \put(82,69){\small (a)}
          \end{overpic}
          \label{fig:mx32_my32_pd1_far_orbit}
        \end{minipage}}%
      \hspace{0.005\textwidth}%
      \subfloat{\begin{minipage}{0.2299\textwidth}\centering
          \begin{overpic}[width=\linewidth]{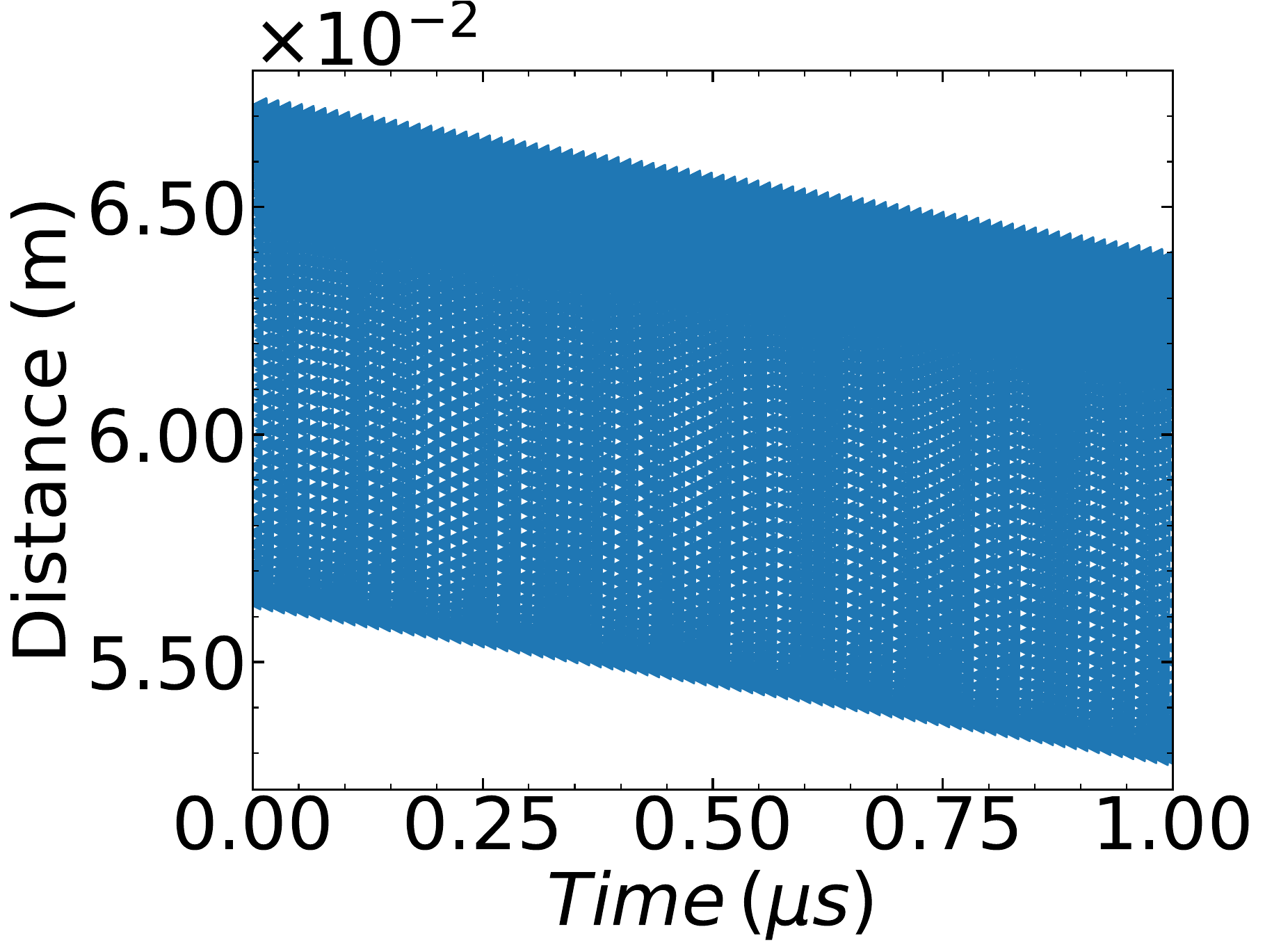}
            \put(84,71){\small (b)}
          \end{overpic}
          \label{fig:mx32_my32_pd1_far_distance}
        \end{minipage}}%
      \hspace{0.005\textwidth}%
      \subfloat{\begin{minipage}{0.2299\textwidth}\centering
          \begin{overpic}[width=\linewidth]{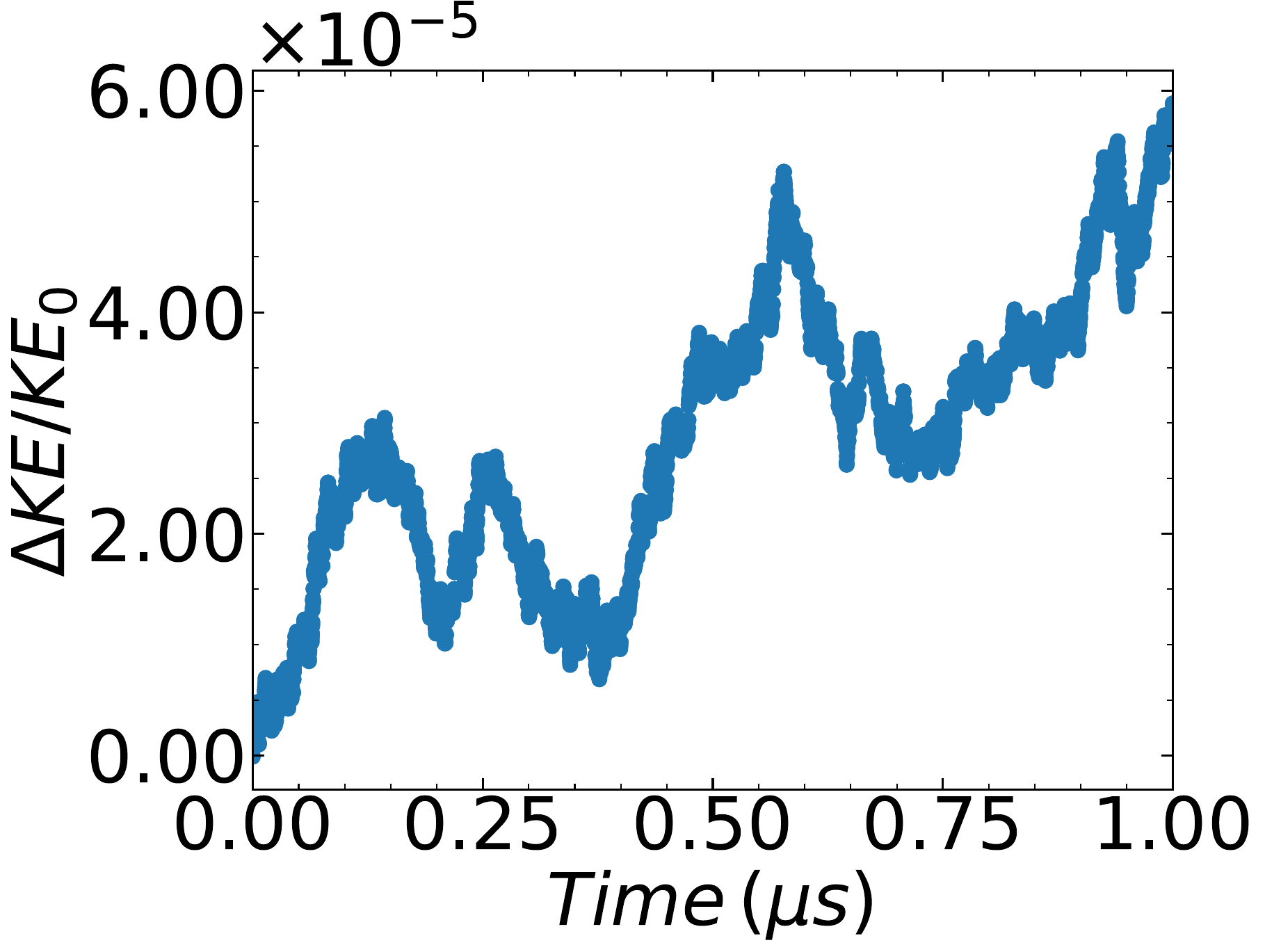}
            \put(84,71){\small (c)}
          \end{overpic}
          \label{fig:mx32_my32_pd1_far_energy}
        \end{minipage}}%
      \hspace{0.005\textwidth}%
      \subfloat{\begin{minipage}{0.2299\textwidth}\centering
          \begin{overpic}[width=\linewidth]{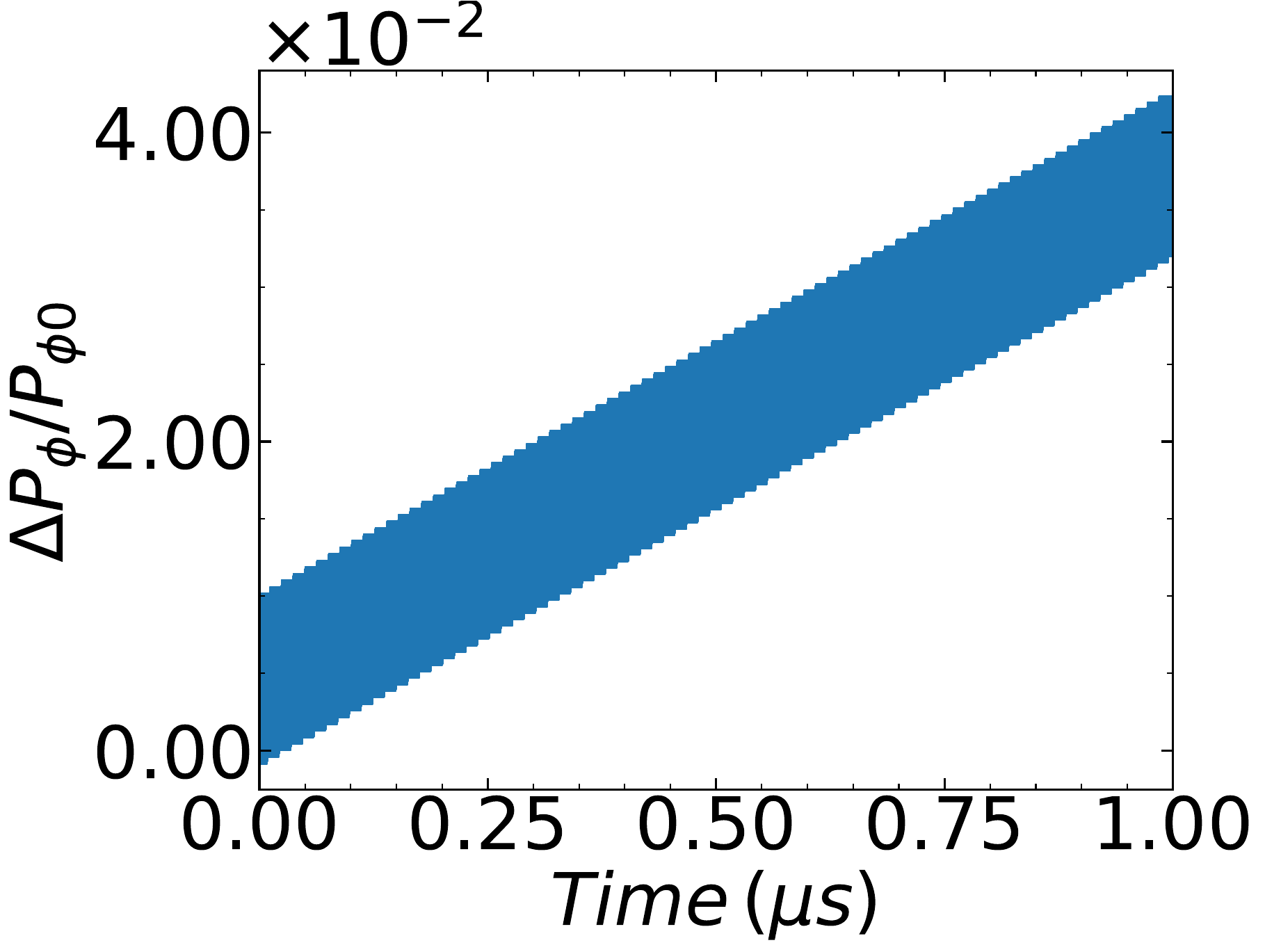}
            \put(84,71){\small (d)}
          \end{overpic}
          \label{fig:mx32_my32_pd1_far_momentum}
        \end{minipage}}%
      \\[-0.532ex]
    
      %------------------------------ Row 2
      \rowlegend{Refined mesh, $\text{poly-deg}=1$, RE at $\sim \boldsymbol{6\,\mathrm{cm}}$ from the magnetic axis}
    
      \subfloat{\begin{minipage}{0.2299\textwidth}\centering
          \begin{overpic}[width=\linewidth]{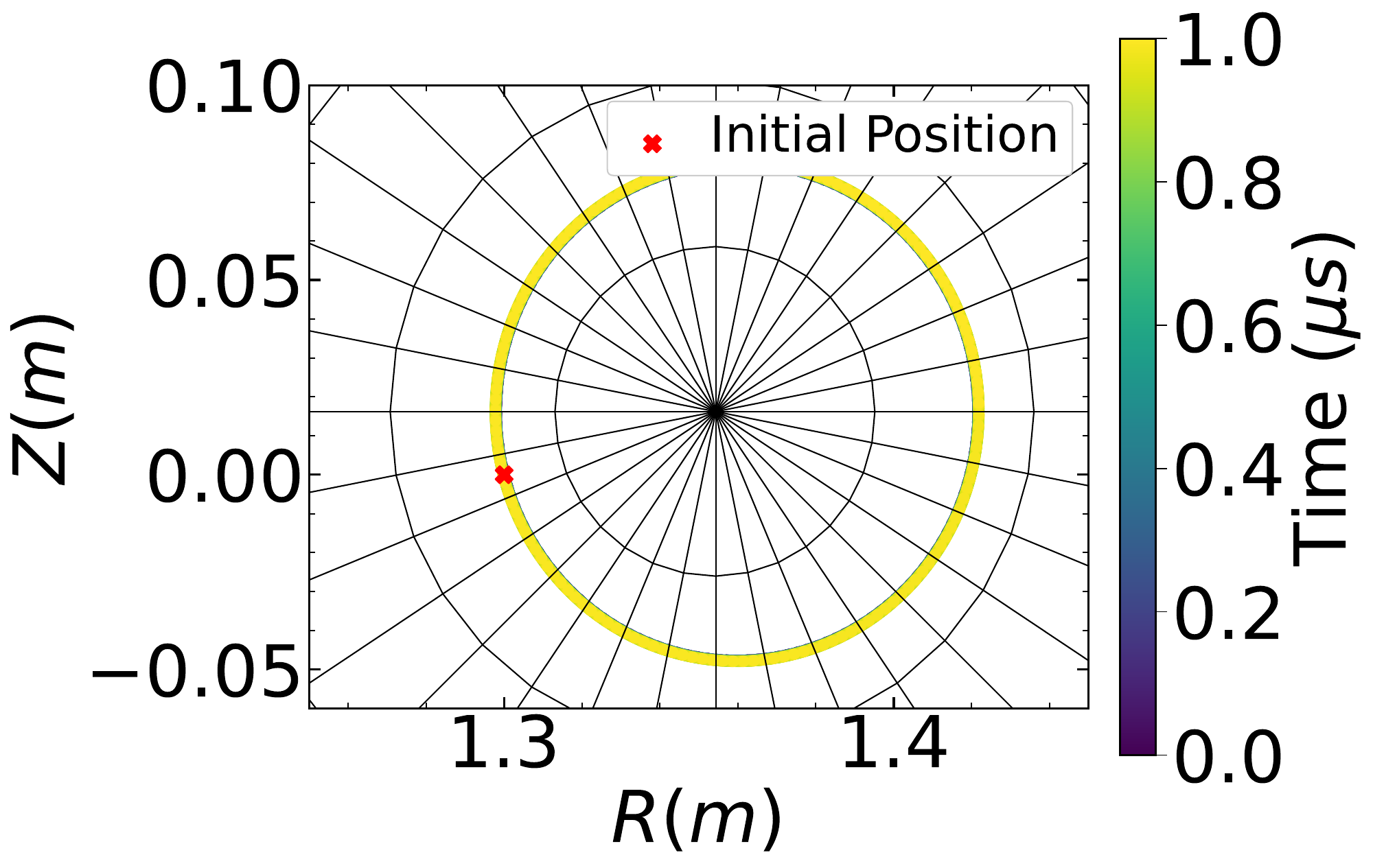}
            \put(72,60){\small (e)}
          \end{overpic}
          \label{fig:mx64_my32_pd1_far_orbit}
        \end{minipage}}%
      \hspace{0.005\textwidth}%
      \subfloat{\begin{minipage}{0.2299\textwidth}\centering
          \begin{overpic}[width=\linewidth]{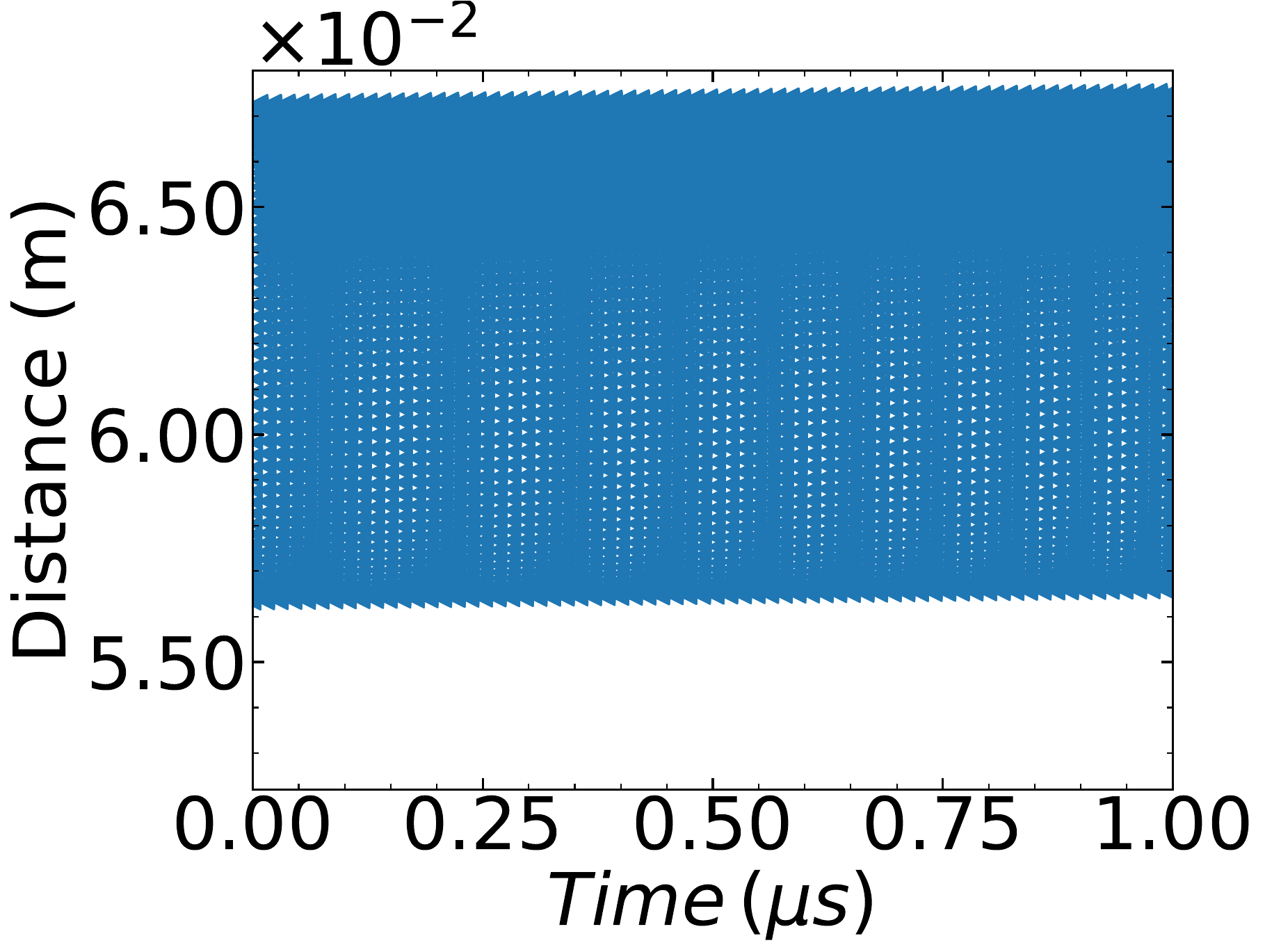}
            \put(84,71){\small (f)}
          \end{overpic}
          \label{fig:mx64_my32_pd1_far_distance}
        \end{minipage}}%
      \hspace{0.005\textwidth}%
      \subfloat{\begin{minipage}{0.2299\textwidth}\centering
          \begin{overpic}[width=\linewidth]{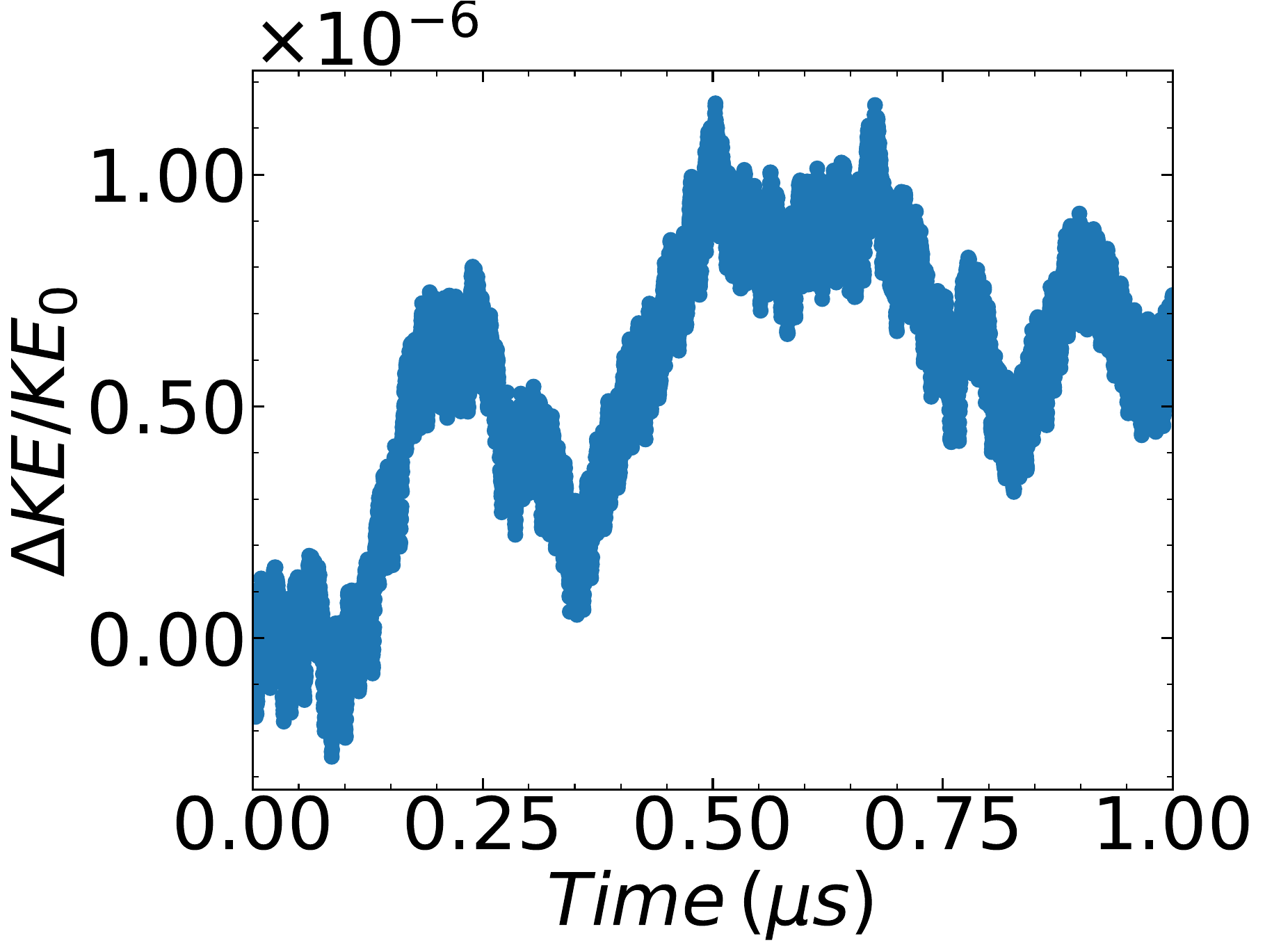}
            \put(84,71){\small (g)}
          \end{overpic}
          \label{fig:mx64_my32_pd1_far_energy}
        \end{minipage}}%
      \hspace{0.005\textwidth}%
      \subfloat{\begin{minipage}{0.2299\textwidth}\centering
          \begin{overpic}[width=\linewidth]{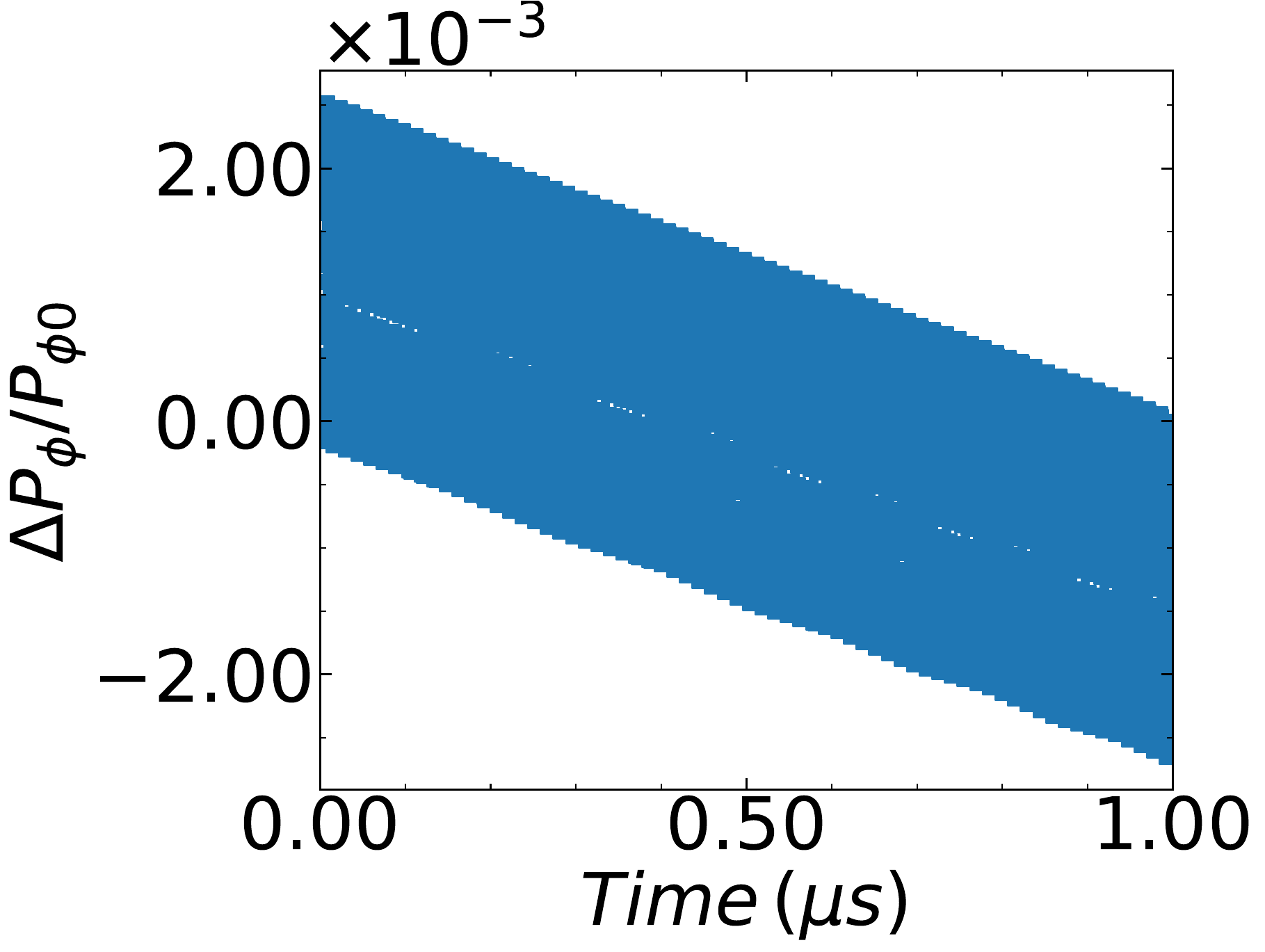}
            \put(84,71){\small (h)}
          \end{overpic}
          \label{fig:mx64_my32_pd1_far_momentum}
        \end{minipage}}%
      \\[-0.532ex]
    
      %------------------------------ Row 3
      \rowlegend{Coarse mesh, $\text{poly-deg}=6$, RE at $\sim \boldsymbol{6\,\mathrm{cm}}$ from the magnetic axis}
    
      \subfloat{\begin{minipage}{0.2299\textwidth}\centering
          \begin{overpic}[width=\linewidth]{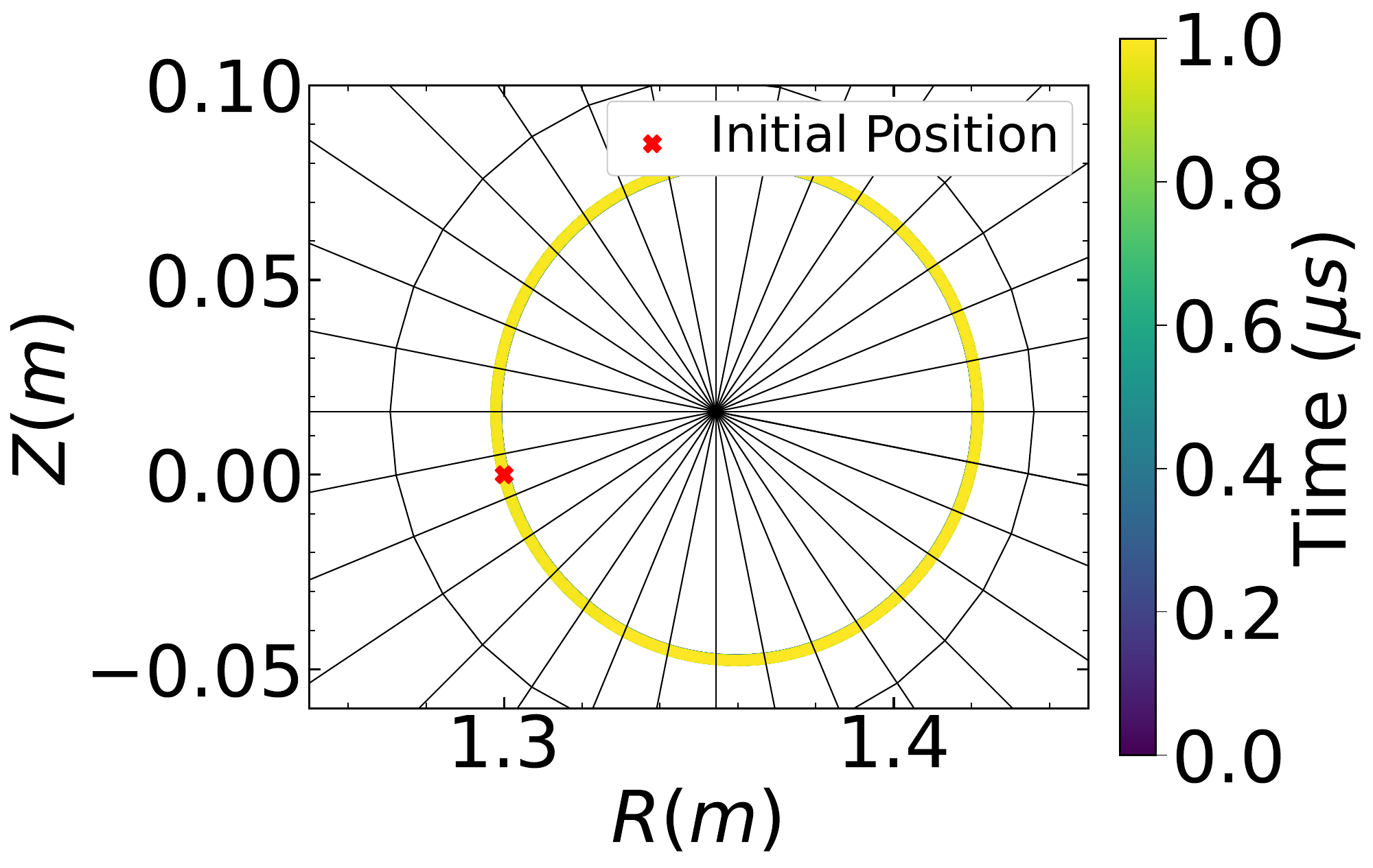}
            \put(72,60){\small (i)}
          \end{overpic}
          \label{fig:mx32_my32_pd6_far_orbit}
        \end{minipage}}%
      \hspace{0.005\textwidth}%
      \subfloat{\begin{minipage}{0.2299\textwidth}\centering
          \begin{overpic}[width=\linewidth]{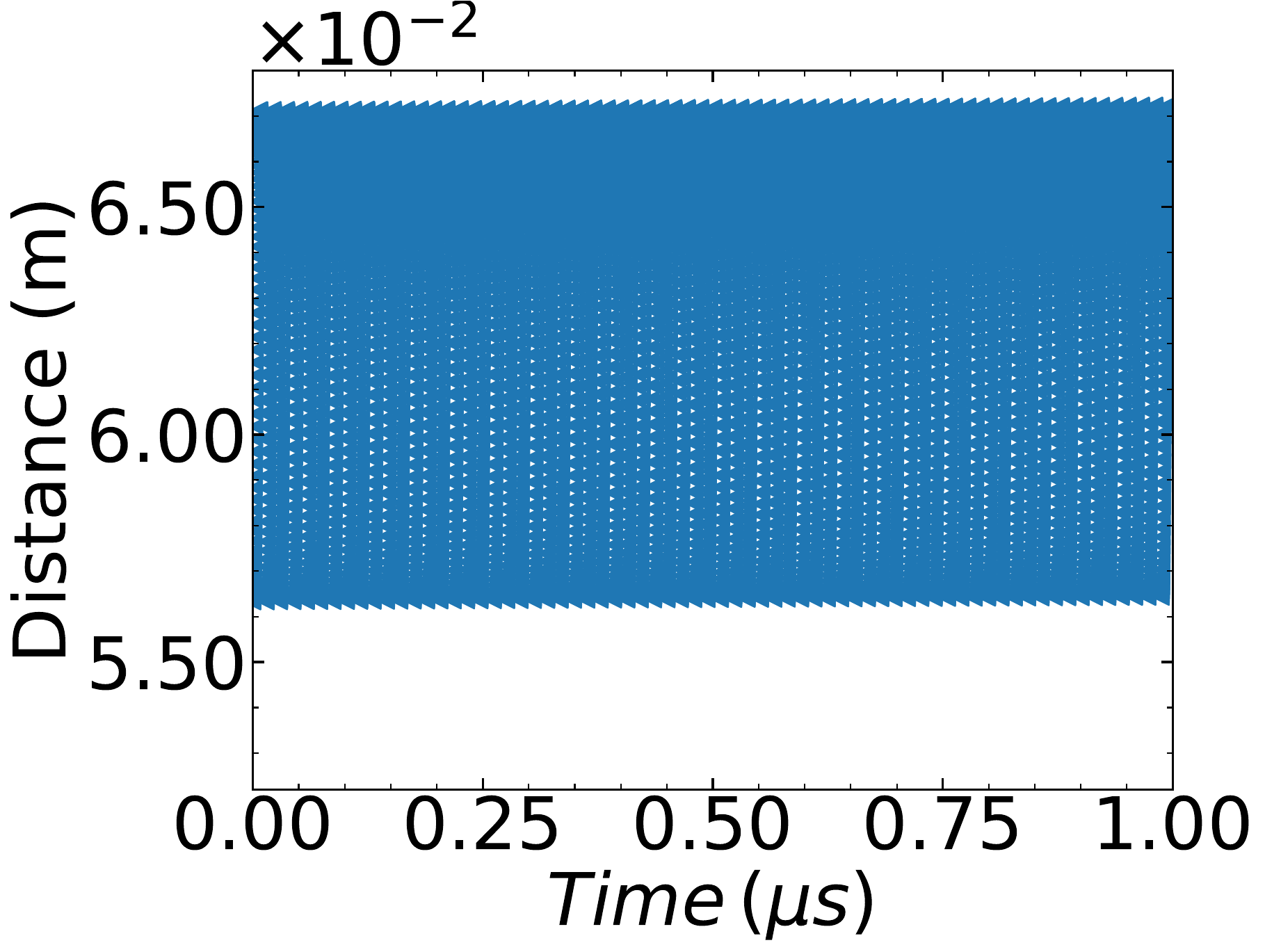}
            \put(84,71){\small (j)}
          \end{overpic}
          \label{fig:mx32_my32_pd6_far_distance}
        \end{minipage}}%
      \hspace{0.005\textwidth}%
      \subfloat{\begin{minipage}{0.2299\textwidth}\centering
          \begin{overpic}[width=\linewidth]{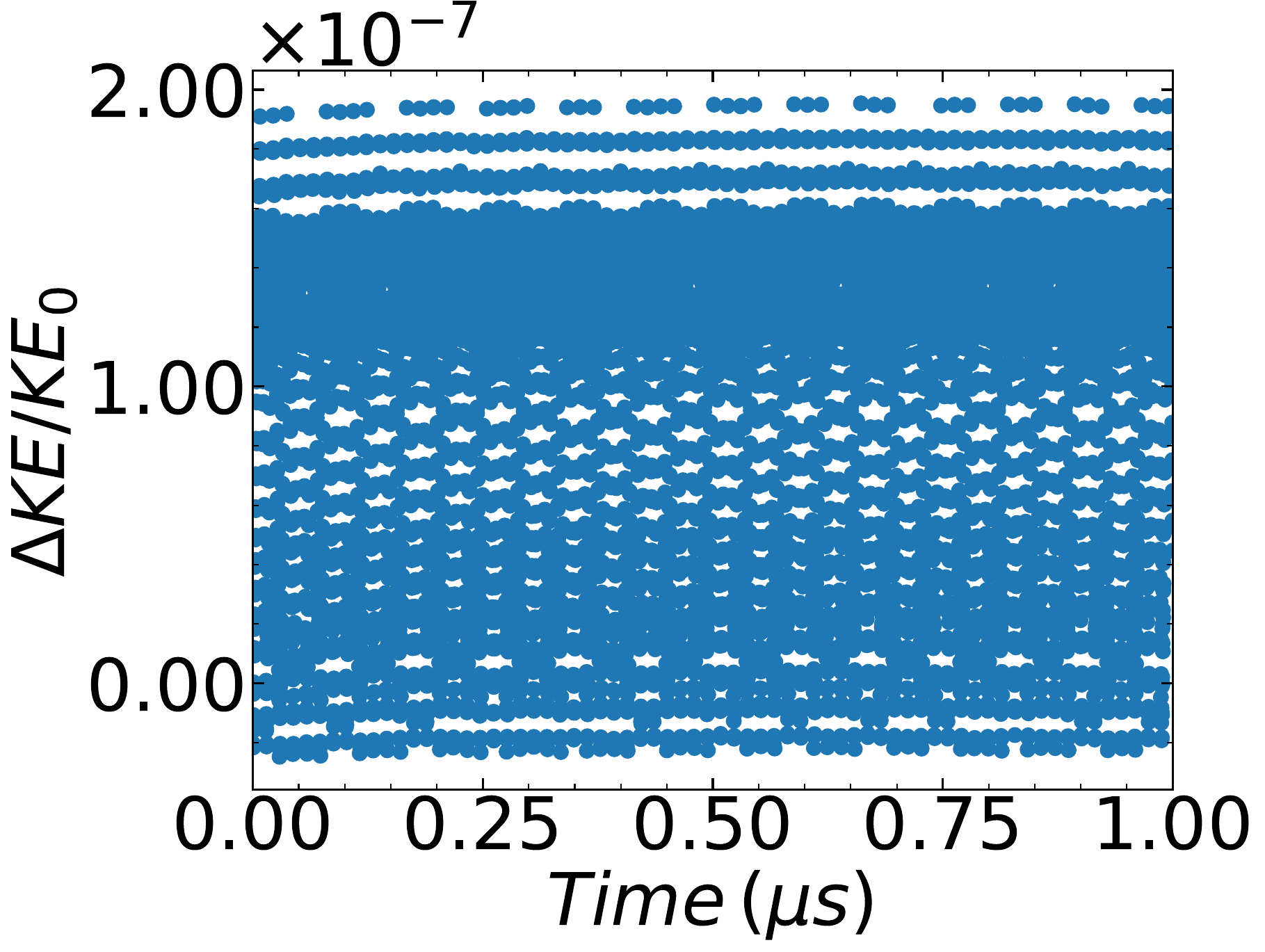}
            \put(84,71){\small (k)}
          \end{overpic}
          \label{fig:mx32_my32_pd6_far_energy}
        \end{minipage}}%
      \hspace{0.005\textwidth}%
      \subfloat{\begin{minipage}{0.2299\textwidth}\centering
          \begin{overpic}[width=\linewidth]{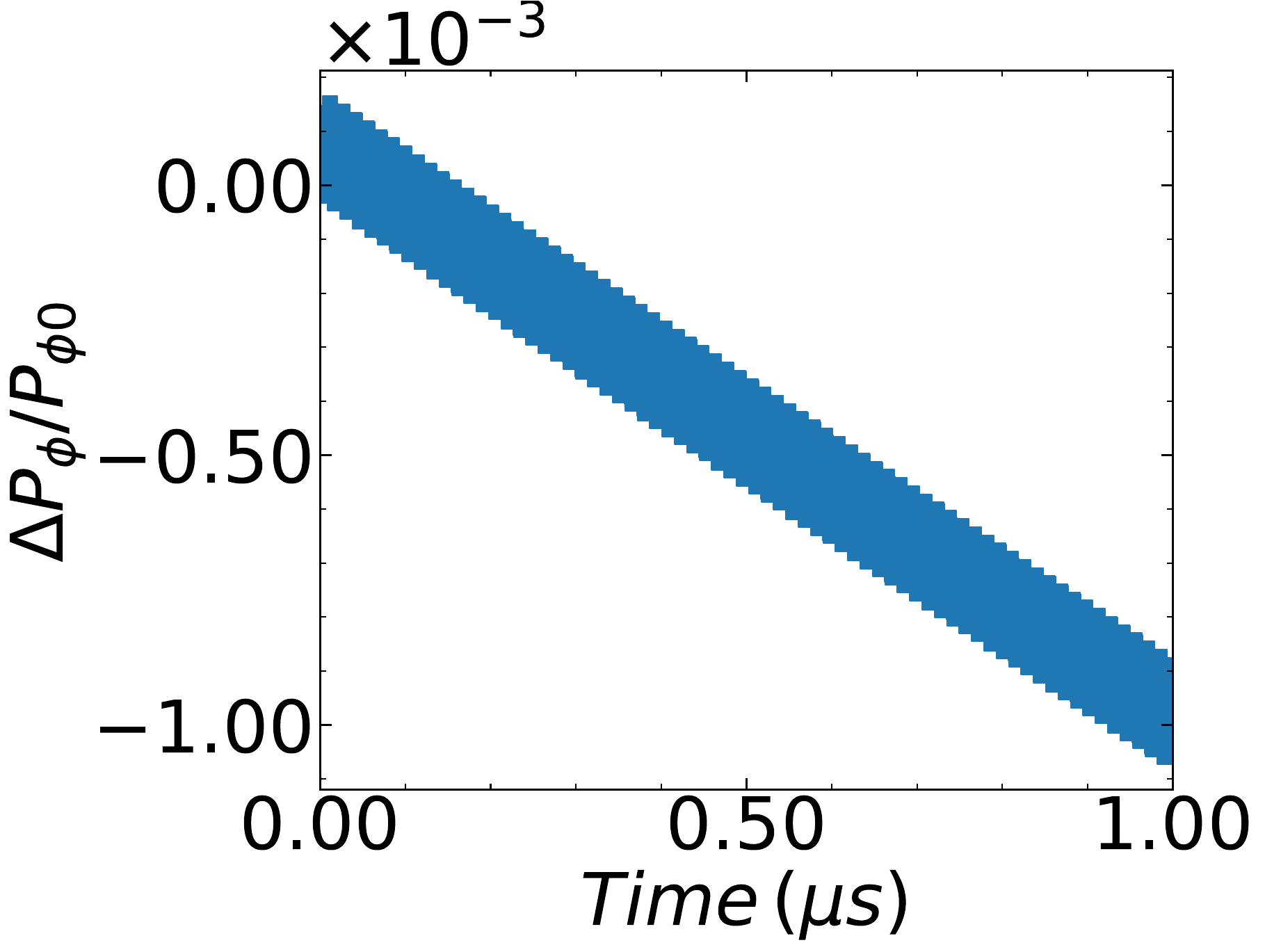}
            \put(84,71){\small (l)}
          \end{overpic}
          \label{fig:mx32_my32_pd6_far_momentum}
        \end{minipage}}%
      \\[-0.532ex]
    
      %------------------------------ Row 4
      \rowlegend{Refined mesh, $\text{poly-deg}=1$, RE at $\sim \boldsymbol{2\,\mathrm{cm}}$ from the magnetic axis}
    
      \subfloat{\begin{minipage}{0.2299\textwidth}\centering
          \begin{overpic}[width=\linewidth]{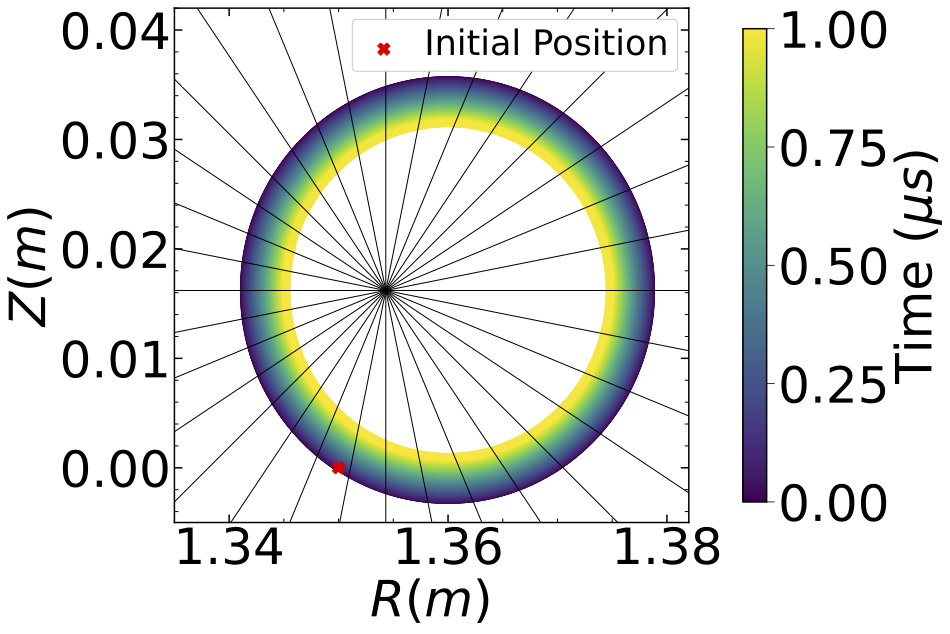}
            \put(70,60){\small (m)}
          \end{overpic}
          \label{fig:mx64_my32_near_orbit}
        \end{minipage}}%
      \hspace{0.005\textwidth}%
      \subfloat{\begin{minipage}{0.2299\textwidth}\centering
          \begin{overpic}[width=\linewidth]{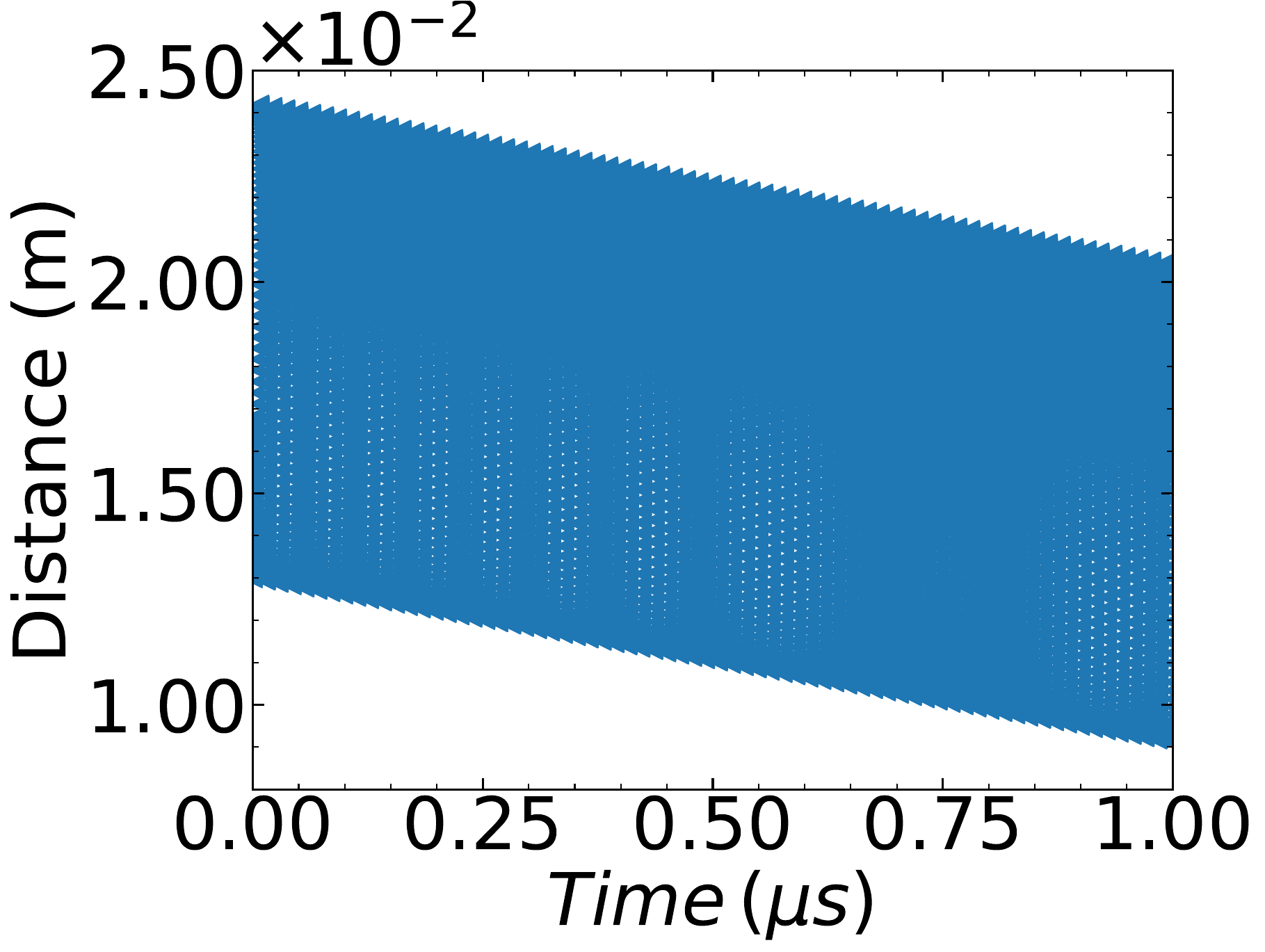}
            \put(84,71){\small (n)}
          \end{overpic}
          \label{fig:mx64_my32_near_distance}
        \end{minipage}}%
      \hspace{0.005\textwidth}%
      \subfloat{\begin{minipage}{0.2299\textwidth}\centering
          \begin{overpic}[width=\linewidth]{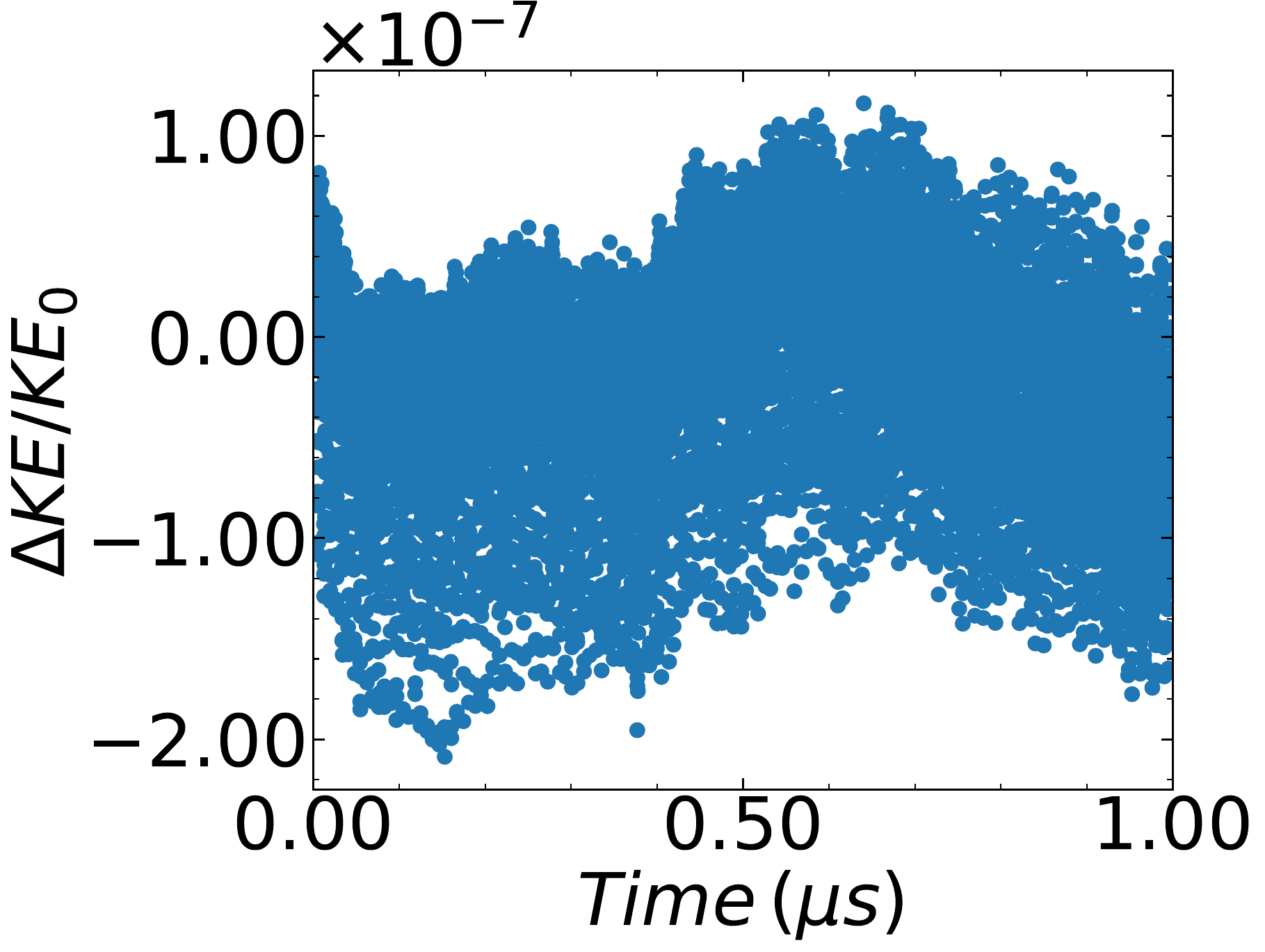}
            \put(84,71){\small (o)}
          \end{overpic}
          \label{fig:mx64_my32_near_energy}
        \end{minipage}}%
      \hspace{0.005\textwidth}%
      \subfloat{\begin{minipage}{0.2299\textwidth}\centering
          \begin{overpic}[width=\linewidth]{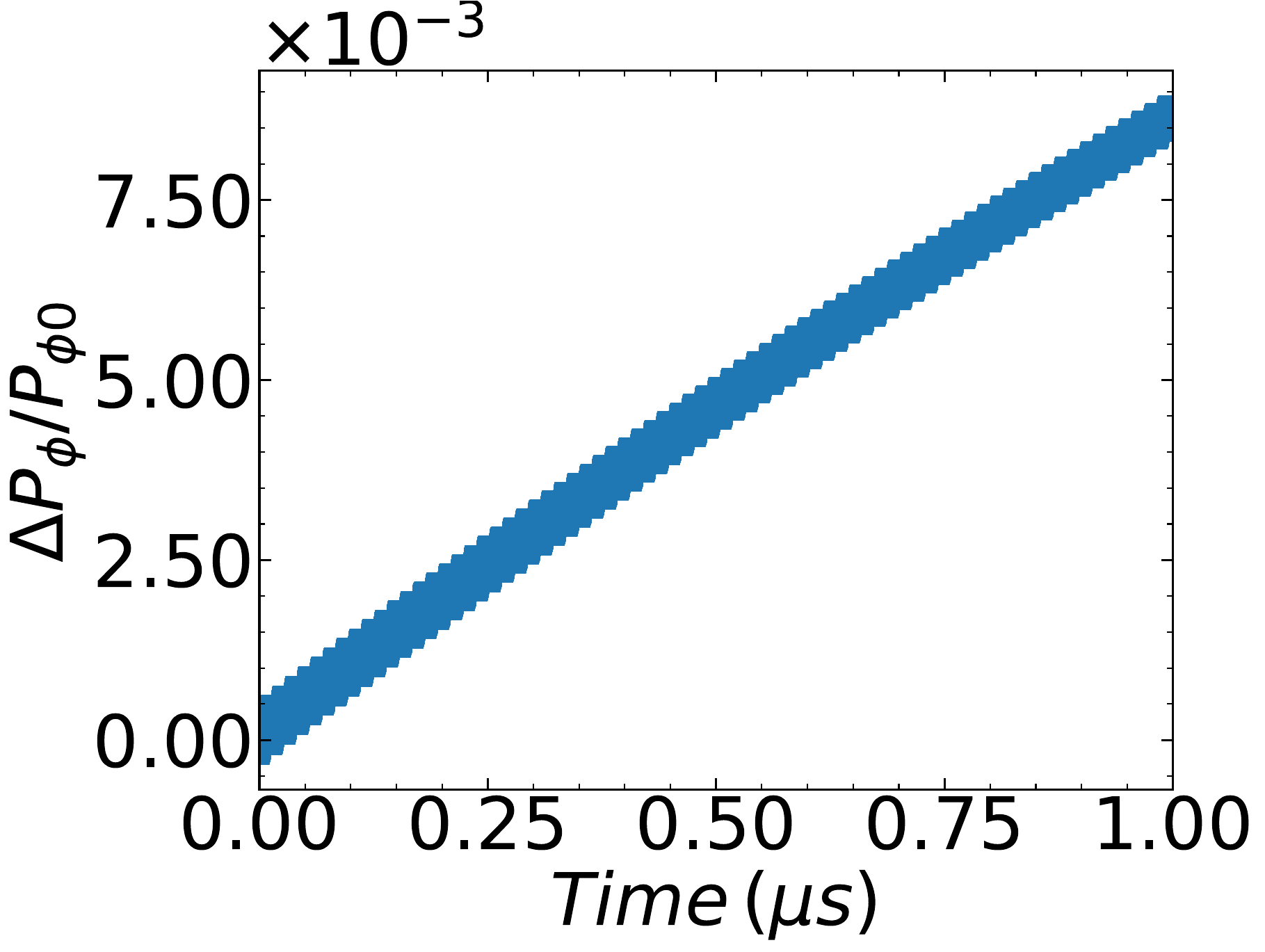}
            \put(84,71){\small (p)}
          \end{overpic}
          \label{fig:mx64_my32_near_momentum}
        \end{minipage}}%
      \\[-0.532ex]
    
      %------------------------------ Row 5
      \rowlegend{Coarse mesh, $\text{poly-deg}=6$, RE at $\sim \boldsymbol{2\,\mathrm{cm}}$ from the magnetic axis}
    
      \subfloat{\begin{minipage}{0.2299\textwidth}\centering
          \begin{overpic}[width=\linewidth]{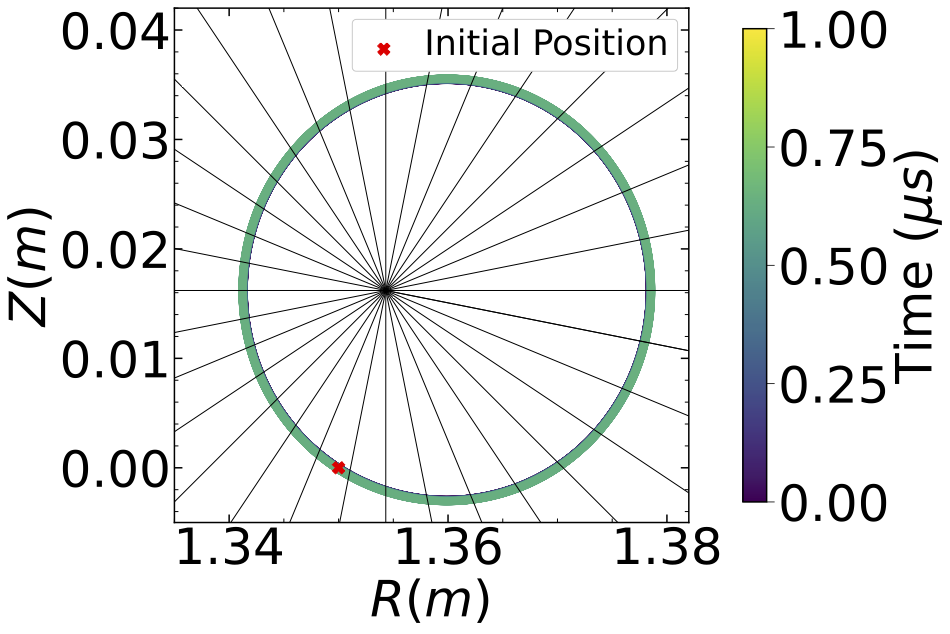}
            \put(70,60){\small (q)}
          \end{overpic}
          \label{fig:pd6_near_orbit}
        \end{minipage}}%
      \hspace{0.005\textwidth}%
      \subfloat{\begin{minipage}{0.2299\textwidth}\centering
          \begin{overpic}[width=\linewidth]{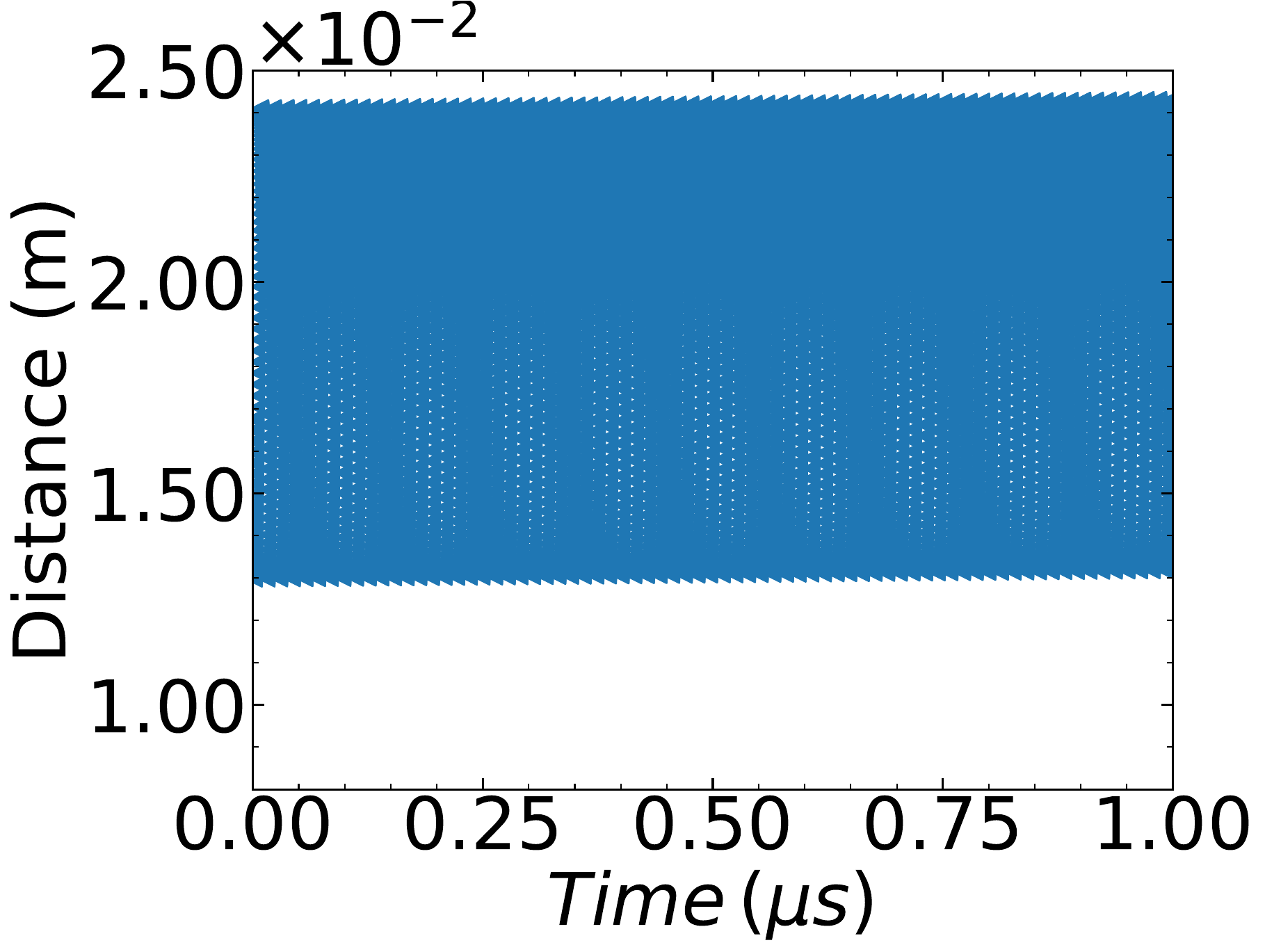}
            \put(84,71){\small (r)}
          \end{overpic}
          \label{fig:pd6_near_distance}
        \end{minipage}}%
      \hspace{0.005\textwidth}%
      \subfloat{\begin{minipage}{0.2299\textwidth}\centering
          \begin{overpic}[width=\linewidth]{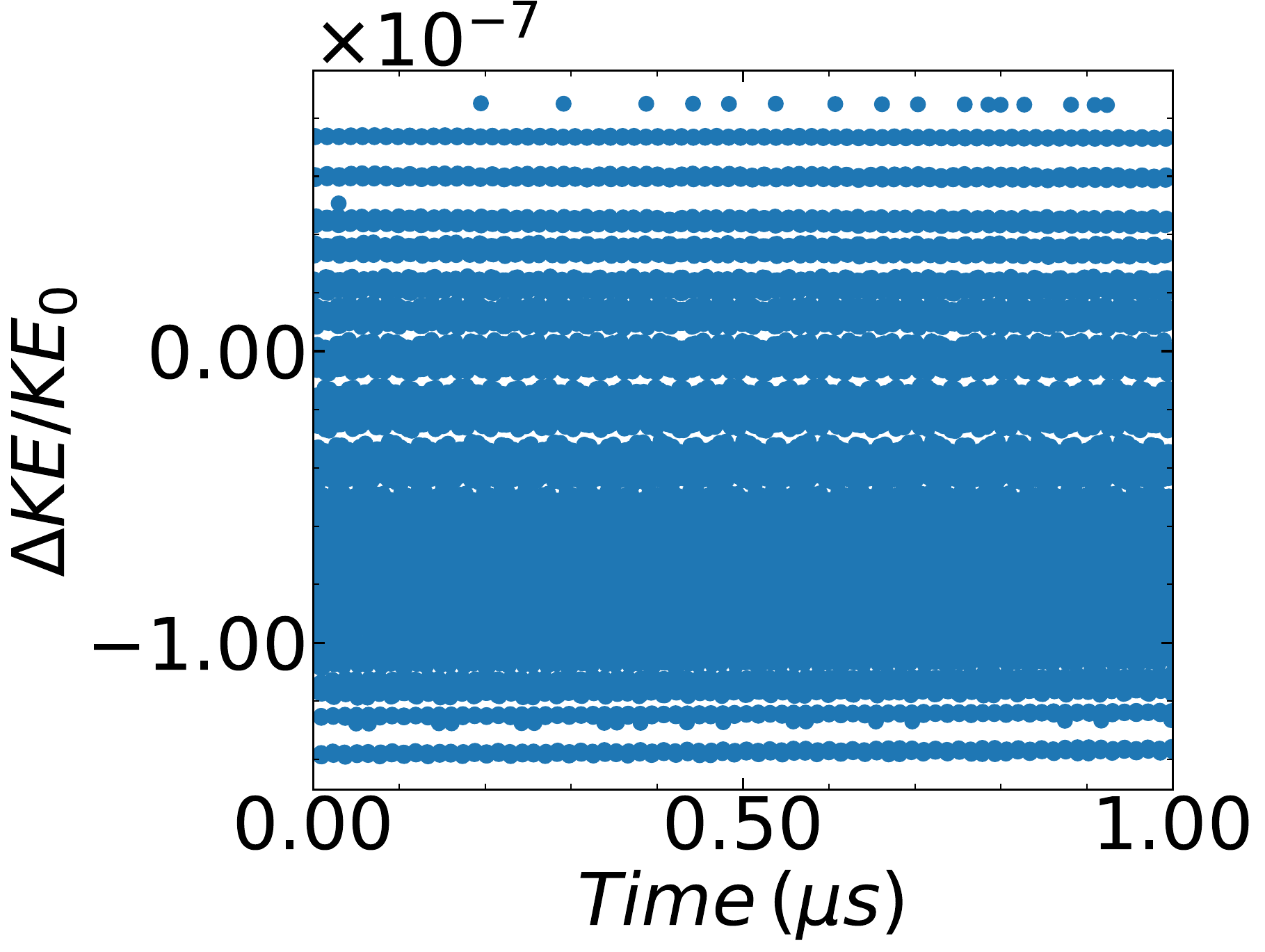}
            \put(84,71){\small (s)}
          \end{overpic}
          \label{fig:pd6_near_energy}
        \end{minipage}}%
      \hspace{0.005\textwidth}%
      \subfloat{\begin{minipage}{0.2299\textwidth}\centering
          \begin{overpic}[width=\linewidth]{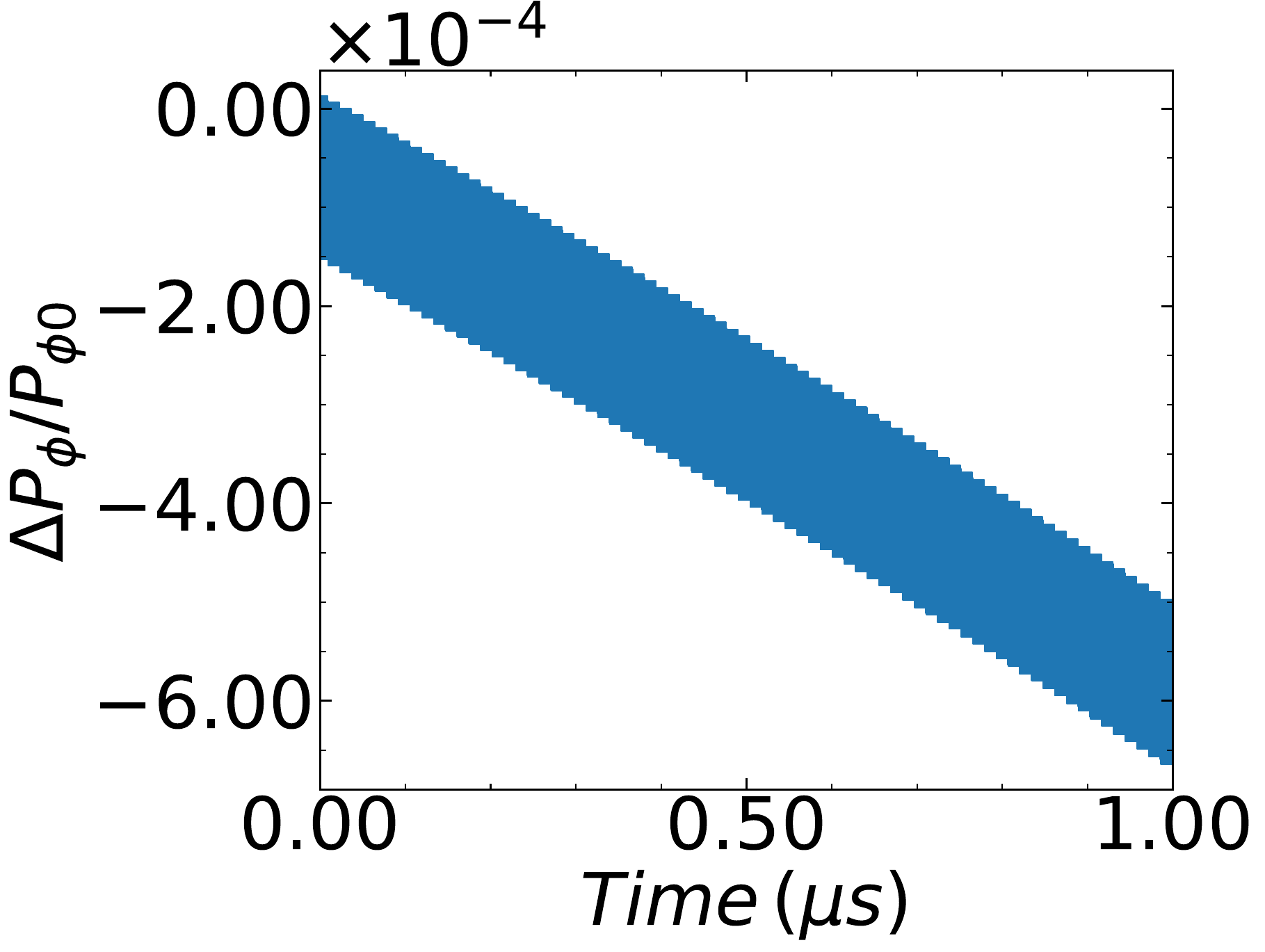}
            \put(84,71){\small (t)}
          \end{overpic}
          \label{fig:pd6_near_momentum}
        \end{minipage}}%
        \caption{RE orbits and time-series for varying mesh resolutions, polynomial degrees, and initial proximity to the magnetic axis. The first three rows correspond to REs ``far'' from the magnetic axis ($\sim 6\,\mathrm{cm}$), while the last two rows represent REs ``near'' the axis ($\sim2\,\mathrm{cm}$). Across the rows, different mesh resolutions ($n=32, m=32$ vs. $n=64, m=32$) and polynomial degrees ($\text{poly-deg}=1,6$) are used. From left to right, columns illustrate (i) the RE orbit projected onto the $R$-$Z$ plane, with color indicating time evolution. (ii) distance to the magnetic axis, (iii) normalized kinetic energy variation, and (iv) normalized canonical toroidal angular momentum variation.}
        \label{fig:combined_comparison}
    \endgroup               
    \end{figure*}

The two $10^6$ particle count cases previously considered in Sec.\,\ref{sec:error_analysis_parallel_current} are used in this subsection to analyze the temporal evolution of the parallel current profile with no loop-voltage ($\boldsymbol{E}=0$). Simulations reveal that low-accuracy field representations (coarse meshes with low polynomial degrees) significantly distort the temporal evolution of the parallel current density profile. Evidence for this distortion is presented in three steps. Figure\,\ref{fig:evol_dep_mesh_comp} contrasts deposited $J_{\parallel}$ obtained with coarse and refined meshes, exposing the spurious peaks that arise when the field representation is inadequate. Figures\,\ref{fig:1d_hist} and \ref{fig:2d_sample} connect those macroscopic errors to mesh-dependent modifications of the particle ensemble. Finally, Fig.\,\ref{fig:combined_comparison} isolates the root cause: spurious radial drifts driven by low accuracy in the magnetic field representation on coarse meshes. The analysis proceeds in this logical progression.

Figure\,\ref{fig:evol_dep_mesh_comp} presents four cross-sectional cuts of deposited $J_{\parallel}$ at $0$, $1$, and $5\,\mathrm{\mu s}$. 
Panels (a) and (c) use a coarse mesh; panels (b) and (d) use a refined mesh. Panels (a,b) show radial slices through the magnetic axis at $Z = Z_{mag}$ while (c,d) are slices at $Z = 0.0 \,\mathrm{m}$.
For the coarse mesh (a,c), current peaks develop near the magnetic axis, reaching $-15 \,\mathrm{MA/m^2}$ and $+5 \,\mathrm{MA/m^2}$ at $t = 5 \,\mathrm{\mu s}$. 
The refined mesh (b,d) largely suppresses these unphysical peaks at early times, although residual artifacts start to emerge near the axis by $5\,\mathrm{\mu s}$ in Fig.\,\ref{fig:evol_dep_mesh_comp}(b).

Figures\,\ref{fig:1d_hist} and~\ref{fig:2d_sample} illustrate how the coarse- and refined-mesh influence the evolution of the particle-phase-space distributions. Figure\,\ref{fig:1d_hist} shows 1D histograms (kinetic energy, pitch angle, parallel momentum, and magnetic moment) from the coarse-mesh simulation. Figure\,\ref{fig:1d_hist_a} reveals that, although all particles were initialized with 10\,MeV, a small spread is visible at the initial time ($t=0$). This kinetic energy spread at $t=0$ emerges due to finite precision in the storage format: particle data are stored in $(p_{\parallel}, \mu)$ coordinates rather than directly in $(KE, \eta)$ space and transforming back to $(KE, \eta)$ coordinates introduces minor rounding errors. Figures\,\ref{fig:1d_hist_b} and \ref{fig:1d_hist_c} confirm that the nominally Dirac-delta function pitch angle (10$^\circ$) and parallel momentum ($5.525\times10^{-21}$\,kg\,m/s) also broaden slightly in time. The magnetic moment remains unchanged, as expected (Fig.\,\ref{fig:1d_hist_d}).

Figure\,\ref{fig:2d_sample} tracks particle distributions in real space $(R,Z)$ for coarse- and refined-mesh cases at $t=0$, $1\,\mu s$, and $5\,\mu s$. Each subfigure overlays the 2D histogram with 1D projections along the $R$ and $Z$ directions. Panels~(a,c,e) are for the coarse mesh, while (b,d,f) are for the refined one. Figs.\,\ref{fig:2d_sample_a} and \ref{fig:2d_sample_b} show similar initial particle distributions for both meshes, but by $t=5\,\mu s$ the coarse mesh produces visible under- and overshoots near the magnetic axis (Fig.\,\ref{fig:2d_sample_e}), creating steep radial gradients in particle density and parallel current density. The under- and overshoots artifacts of Fig.\,\ref{fig:2d_sample_e} arise due to the coarse, low-order finite element representation of the magnetic field, which distorts the guiding-center orbits. Once such steep gradients emerge, bilinear elements on a coarse mesh cannot accurately capture them, producing the large parallel current density under- and overshoots discussed earlier in Figs.\,\ref{fig:evol_dep_mesh_comp_a} and \ref{fig:evol_dep_mesh_comp_c}. The refined mesh mitigates these artifacts over the $t=5\,\mu s$  time window (Figs.\,\ref{fig:evol_dep_mesh_comp_b}, \ref{fig:evol_dep_mesh_comp_d} and \ref{fig:2d_sample_f}).

To examine how differing accuracy in field representations impacts particle dynamics, we analyze single particle orbits through time-series of kinetic energy and canonical momentum conjugate to the toroidal angle, 
$p_\phi = p_{\parallel}  B_{\phi}  R / B + q_e\psi_p $. In these orbit integrations, the kinetic time step is $\Delta t_{\mathrm{KORC}} = 1.0 \times 10^{-11}\,\mathrm{s}$ which is about one quarter of the inverse gyro-frequency 
$\gamma\,m/(|q_e|\,B_0)=4.2 \times 10^{-11} \,\mathrm{s}$, set by a $10\,\mathrm{MeV}$ RE with with $10^{\circ}$ pitch angle at the magnetic axis. Lowering the time step to $\Delta t_{\mathrm{KORC}} = 1.0 \times 10^{-12} \,\mathrm{s}$ does not alter the trends in Fig.\,\ref{fig:combined_comparison}.  
We omit results for a time step of $\Delta t_{\mathrm{KORC}} = 1.0 \times 10^{-12} \,\mathrm{s}$ for clarity.
For an axisymmetric system without collisions and in the absence of externally applied electric fields, both the canonical momentum and kinetic energy should remain constant. 
We define $\Delta \mathrm{KE}$ and $\Delta p_\phi$ as the differences between their respective values at time $t$ and their initial values: 
$\Delta \mathrm{KE}(t) = \mathrm{KE}(t) - \mathrm{KE}_0$ and $\Delta p_\phi(t) = p_\phi(t) - p_{\phi,0}$, and we monitor the normalized variations 
$\Delta \mathrm{KE}(t)/\mathrm{KE}_0$ and $\Delta p_\phi(t) / p_{\phi,0}$.

Figure\,\ref{fig:combined_comparison} compiles five rows of subplots, each row consisting of four panels:  
(i) the RE orbit in the $R$-$Z$ plane, (ii) the distance from the magnetic axis vs.\ time, (iii) the normalized kinetic energy variation, 
and (iv) the normalized canonical momentum variation.  
From top to bottom, the initial RE radial position is varied between $\sim 6\,\mathrm{cm}$ (``far'') and $\sim 2\,\mathrm{cm}$ (``near'') from the axis, combined with changes in mesh resolution (coarse vs.\ refined) and polynomial degree ($\text{poly-deg}=1$ or $\text{poly-deg}=6$).  
For the particle at $\sim 6\,\mathrm{cm}$ from the axis, low-resolution cases (coarse mesh or $\text{poly-deg}=1$) exhibit kinetic energy variations $\Delta \mathrm{KE}/\mathrm{KE}_0$ at the $10^{-5}$ level  
and momentum variation $\Delta p_\phi / p_{\phi,0}$ approaching $10^{-2}$ within $1\,\mu\mathrm{s}$.  
Refined meshes or higher-order polynomials suppress these variations by one to two orders of magnitude, 
pushing $\Delta \mathrm{KE}/\mathrm{KE}_0$ below $10^{-6}$ and $\Delta p_\phi / p_{\phi,0}$ below $10^{-3}$.  
The refined mesh with $\text{poly-deg}=1$ and the coarse mesh with $\text{poly-deg}=6$ produce comparable orbits in the $R$-$Z$ plane (compare Figs.\,\ref{fig:mx64_my32_pd1_far_orbit} vs.\,\ref{fig:mx32_my32_pd6_far_orbit}), 
indicating that either increased spatial resolution or higher polynomial degree leads to similar orbits trajectories far from the magnetic axis ($\sim 6\,\mathrm{cm}$). Near the axis ($\sim 2\,\mathrm{cm}$), orbits with a higher polynomial degree ($\text{poly-deg}=6$) on a coarse mesh (Fig.\,\ref{fig:pd6_near_orbit}) differ from those with a lower polynomial degree ($\text{poly-deg}=1$) on a coarse mesh (Fig.\,\ref{fig:mx64_my32_near_orbit}).
The higher polynomial degree outperforms the refined mesh with $\text{poly-deg}=1$ in preserving both energy and momentum. This indicates that near the magnetic axis polynomial degree is more critical than mesh refinement. Lower mesh resolution or polynomial degree can cause inaccurate inward or outward radial drifts (compare panels in second column of Fig.\,\ref{fig:combined_comparison}), resulting in the under- and overshoots of particle count near the axis of Fig.\,\ref{fig:2d_sample_e}, and ultimately leading to the large under- and overshoots of the deposited RE parallel current density of Figs.\,\ref{fig:evol_dep_mesh_comp_a} and \ref{fig:evol_dep_mesh_comp_c}.

One way to improve field accuracy near the axis, which would help preventing the under- and overshoots in the parallel current shown in Fig.\,\ref{fig:evol_dep_mesh_comp} (at least within a certain time window), 
is to re-solve the Grad-Shafranov equation in NIMROD. However, we presently bypass re-solving and directly convert EFIT data into a NIMROD initialization file. 
A higher polynomial degree may also suppress near-axis artifacts even with a relatively coarse mesh. 
Although NIMROD supports higher-order polynomials, we restrict ourselves to bilinear fields to match DepoPy's Magnetic-Axis-Basis approach, which is presently verified only for bilinear elements (Sec.\,\ref{sec:DepoPy}). We plan to adopt higher-order polynomials for deposition in future work.

All simulations exhibit a secular increase in $\Delta p_\phi/p_{\phi,0}$ on the order of $10^{-3}$--$10^{-2}$ by $t\sim1\,\mu\mathrm{s}$, 
even with refined meshes and higher-order basis functions. It is plausible that the guiding-center Cash-Karp integrator's reliance on continuous field derivatives contributes to this issue, 
since NIMROD's piecewise $C^0$-continuous finite elements do not provide the required smoothness. Full-orbit simulations in KORC bypass this potential issue since they do not depend on field derivatives; 
and its Boris integrator contributes to excellent conservation properties in the millisecond time scale \cite{carbajal17}. 

%%%%%%%%%%%%%%%%%%%%%%%%%%%%%%%%%%%%%%%%%%%%%%%%%%%%%%%%%%%%%%%%%%%%%%%%%%%%%%%
%%%%%%%%%%%%%%%%%%%%%%%%%%%%%%%%%%%%%%%%%%%%%%%%%%%%%%%%%%%%%%%%%%%%%%%%%%%%%%%

\subsection{Orbit-averaging method}\label{sec:orbit-averaging}

For the simulations in this paper, the MHD fields remain fixed over a time interval $\Delta t_{\mathrm{dump}}$, after which the mass matrix system in Eq.\,\eqref{eq:mass_matrix_system} is solved to update the RE parallel current. Upon implementation of a self-consistent KORC-NIMROD coupling, $\Delta t_{\mathrm{dump}}$ will be replaced by the actual NIMROD time step ($\Delta t_{\mathrm{NIMROD}}$) at which the MHD fields will be updated.

\begin{figure}[ht]
    \centering
    \subfloat{%
        \begin{minipage}{0.7\columnwidth}
            \centering
            \textbf{Cumulative Deposition Step=1} \\[0.5em]
            \begin{overpic}[width=\linewidth]{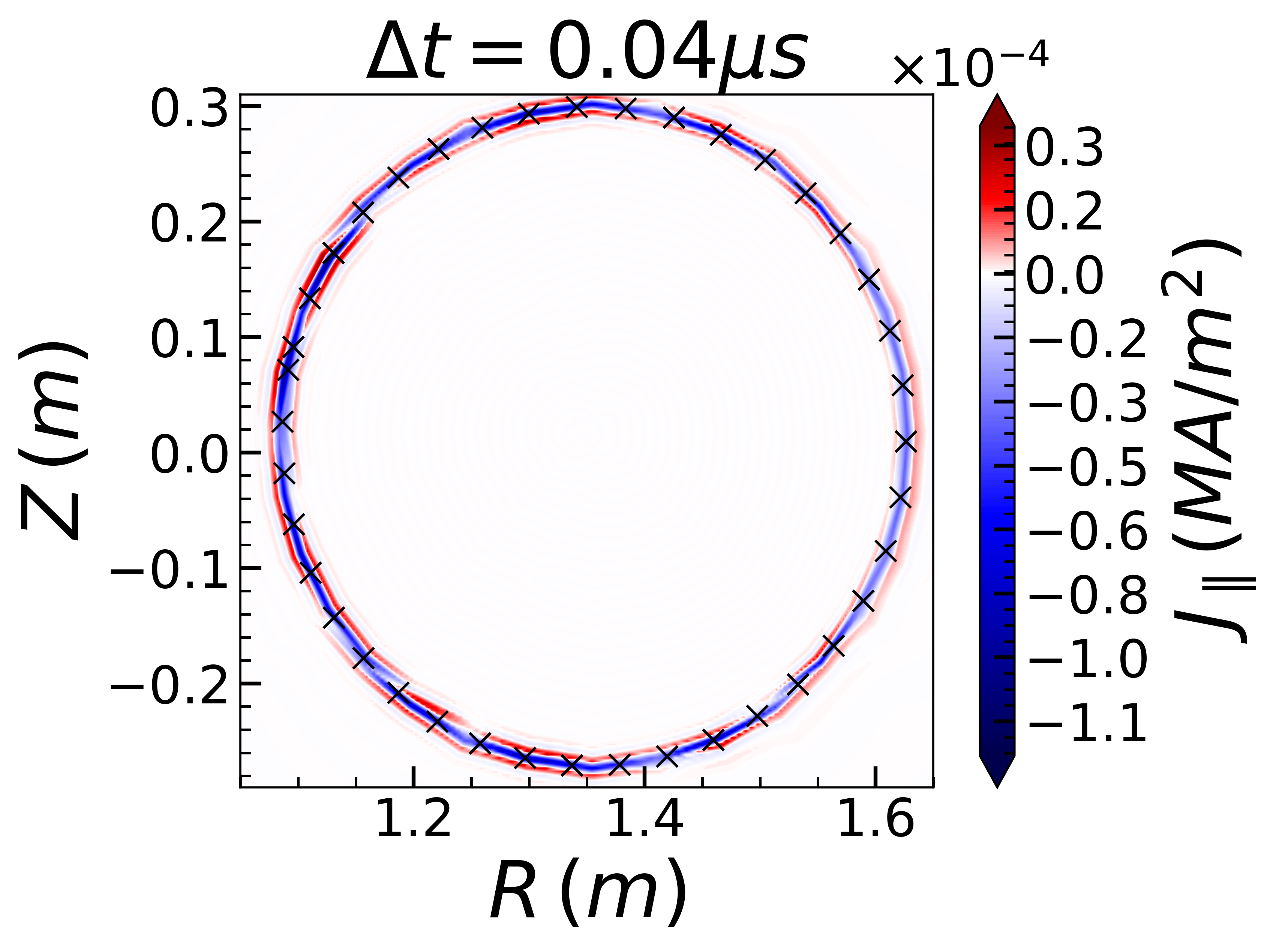}
                \put(22,62){\small{(a)}}
            \end{overpic}
            \label{fig:cumulative_deposition_a}
        \end{minipage}
    }
    \hspace{0.005\textwidth}
    \subfloat{%
        \begin{minipage}{0.7\columnwidth}
            \centering
            \textbf{Cumulative Deposition Step=10} \\[0.5em]
            \begin{overpic}[width=\linewidth]{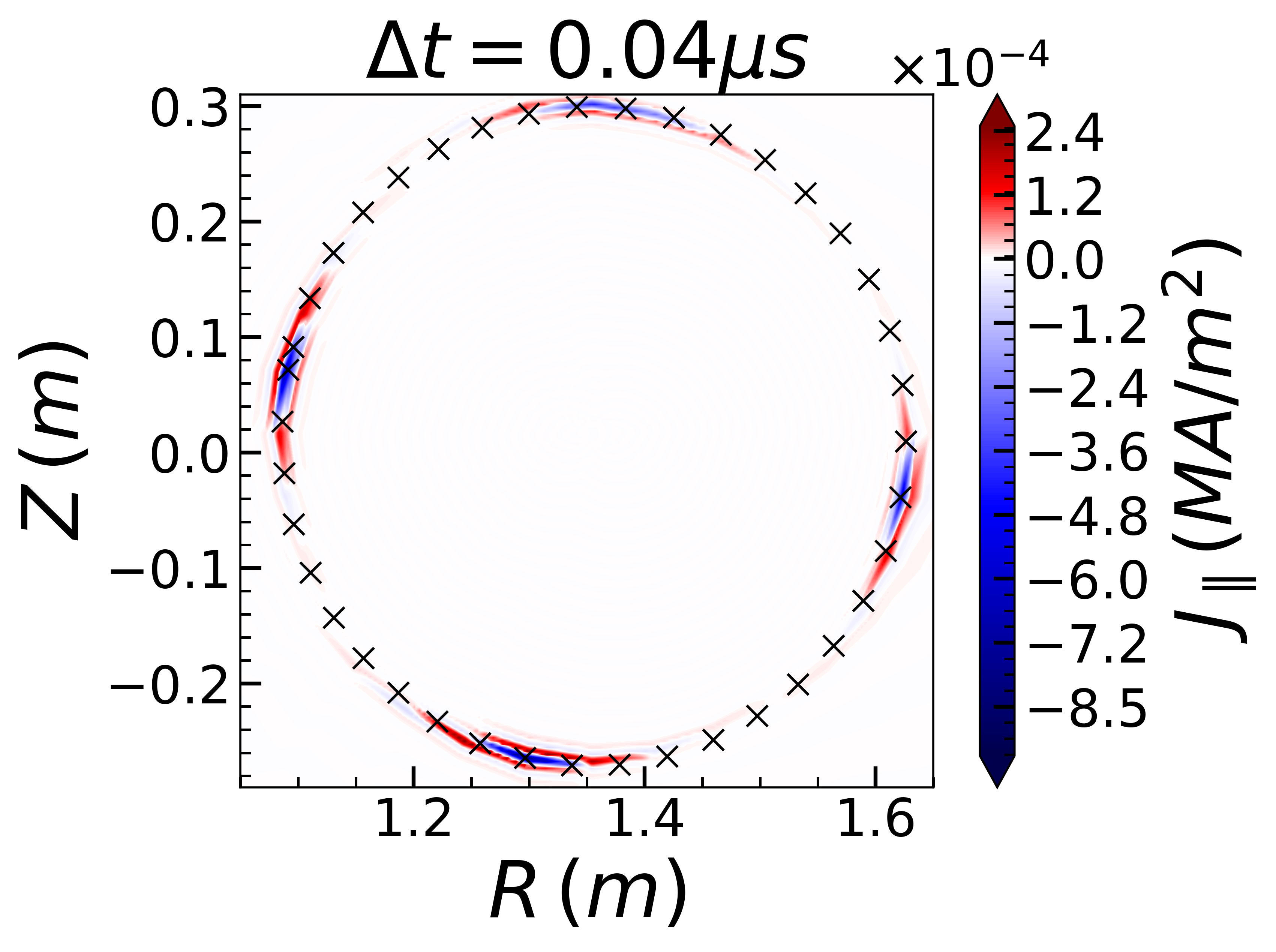}
                \put(22,62){\small{(b)}}
            \end{overpic}
            \label{fig:cumulative_deposition_b}
        \end{minipage}
    }
    \caption{Orbit-averaged $J_{\parallel}$ after one MHD-dump of duration 
    $\Delta t_{\mathrm{MHD}} = 4 \times 10^{-8} \mathrm{s}$ (40 kinetic intervals at 
    $\Delta t_{\mathrm{KORC}} = 1 \times 10^{-9} \mathrm{s}$). 
    In panel~(a), $c_\textsubscript{step}=1$ accumulates current every kinetic step (40 total). 
    In panel~(b), $c_\textsubscript{step}=10$ accumulates current every 10 kinetic steps (4 total). 
    Kinetic steps are marked with an $\times$.}
    \label{fig:cumulative_deposition}
\end{figure}

We introduce the \textit{cumulative deposition step} counter, $c_\textsubscript{step}$, to determine how often, within the total of $M$ kinetic steps in one MHD dump interval, a fractional deposition of the current onto the mesh takes place. Specifically, we assume $\Delta t_{\mathrm{dump}}$ is an integer multiple of $\Delta t_{\mathrm{KORC}}$, so that
\begin{equation}
M = \frac{\Delta t_{\mathrm{dump}}}{\Delta t_{\mathrm{KORC}}} \in \mathbb{Z}^{+}.
\end{equation}
We then require $M \bmod c_\textsubscript{step} = 0$ and define the dimensionless parameter
\begin{equation}
\alpha = \frac{c_\textsubscript{step}}{M}.
\end{equation}
The RE current is deposited onto the mesh $\alpha^{-1}$ times per MHD dump, each time, the weights $w_l$ in Eq.\,\eqref{eq:weight_parallel_current} are scaled by $\alpha$:
\begin{equation}
w_l = \alpha \, g_\textsubscript{RE}\Bigl(\tfrac{q_e\,p_{\parallel}}{m\,\gamma} 
- \mu\,(\hat{\mathbf{b}}\cdot\nabla\times\hat{\mathbf{b}})\Bigr)_l,
\label{eq:weight_parallel_current_alpha}
\end{equation}
where $g_\textsubscript{RE}$ is a constant factor (Sec.\,\ref{sec:error_analysis_parallel_current}).
After all fractional depositions over $\Delta t_{\mathrm{dump}}$ accumulate, the mass-matrix system (Eq.\,\eqref{eq:mass_matrix_system}) is solved once. This \emph{orbit-averaging method} forms a weighted average of the deposited current over each particle's trajectory, taking into account both peaks in instantaneous current and regions where particles spend more time. For particle number density deposition, the weights are $w_l=\alpha$, and integrating the density recovers the total number of particles, which has been verified numerically.

In Fig.\,\ref{fig:cumulative_deposition}, we show orbit-averaged current density profiles over one MHD-dump interval for two different deposition frequencies for an RE with $KE = 10\,\mathrm{MeV}$ and $\eta = 10^\circ $. The dump interval $\Delta t_{\mathrm{MHD}} = 4\times 10^{-8}\,\mathrm{s}$ is chosen so that a RE orbit completes roughly one poloidal rotation. For visualization, we use a larger kinetic step size ($\Delta t_{\mathrm{KORC}} = 1\times 10^{-9}\,\mathrm{s}$) than in other parts of this study (where $\Delta t_{\mathrm{KORC}} = 1\times 10^{-11}\,\mathrm{s}$). With 40 kinetic steps per MHD dump (marked as $\times$), the deposited current accumulates every kinetic step in Fig.\,\ref{fig:cumulative_deposition_a} ($c_\textsubscript{step} = 1$) and every 10 kinetic steps in Fig.\,\ref{fig:cumulative_deposition_b} ($c_\textsubscript{step} = 10$). Sparse deposition frequency (larger $c_\textsubscript{step}$) introduces additional granularity in the resulting profiles, while frequent deposition smooths the orbit-averaged current.

\begin{figure}[ht]
    \centering
    \subfloat{%
        \begin{minipage}{0.7\columnwidth}
            \centering
            \textbf{Cumulative Deposition = 1} \\[0.5em]
            \begin{overpic}[width=\linewidth]{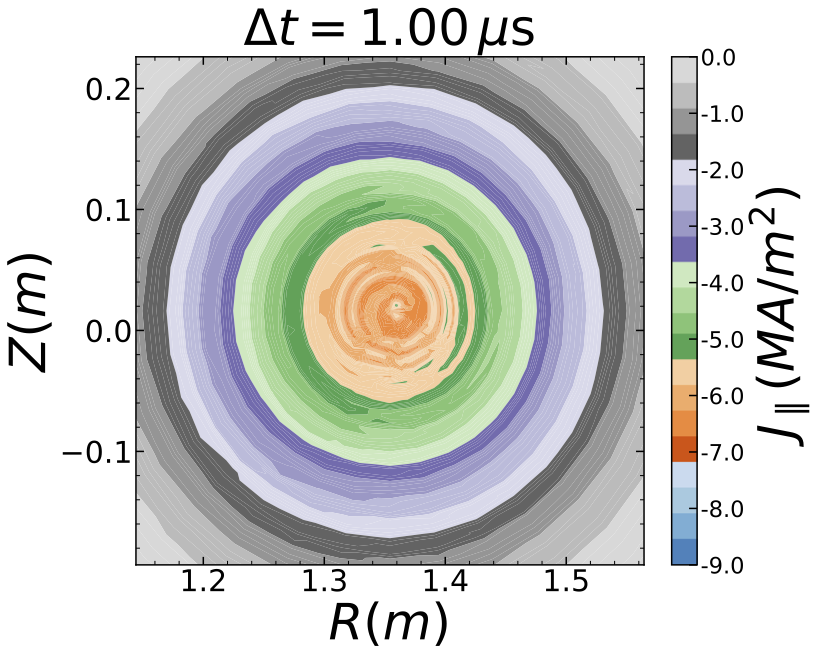}
                \put(18,66){\small{(a)}}
            \end{overpic}
        \end{minipage}
    }
    \hspace{0.005\textwidth}
    \subfloat{%
        \begin{minipage}{0.7\columnwidth}
            \centering
            \textbf{Cumulative Deposition = 1000} \\[0.5em]
            \begin{overpic}[width=\linewidth]{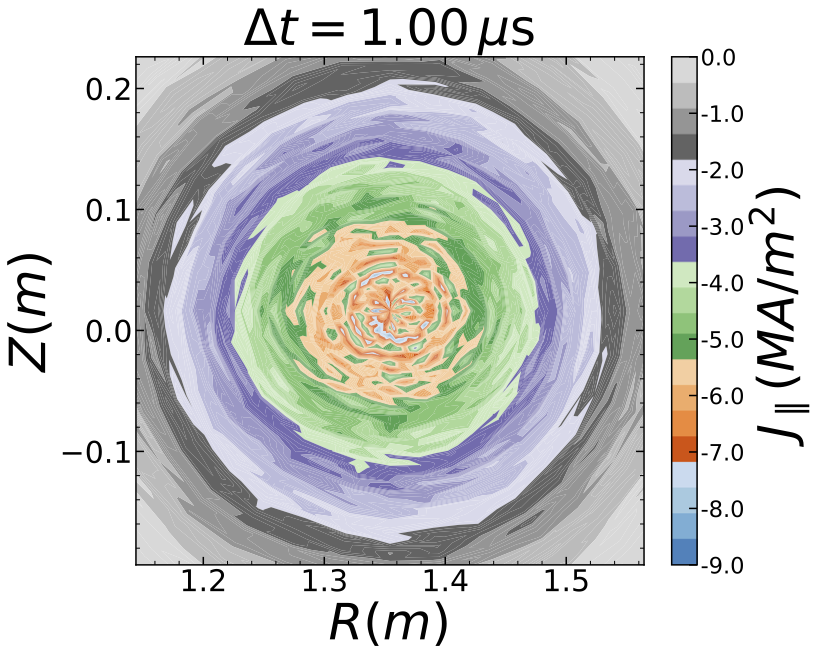}
                \put(18,66){\small{(b)}}
            \end{overpic}
        \end{minipage}
    }
    \caption{Comparison of the deposited $J_{\parallel}$ at the end of a 1-$\mu\mathrm{s}$ simulation, 
    which is divided into 100 MHD-dump intervals. Here, $\Delta t_{\mathrm{dump}}=1.0 \times 10^{-8}\,\mathrm{s}$ and $\Delta t_{\mathrm{KORC}}=1.0 \times 10^{-11}\,\mathrm{s}$. 
    These profiles are orbit-averaged only in the final MHD-dump step.
    Panel~(a) shows $c_\textsubscript{step} =1$ (highest frequency), and panel~(b) shows $c_\textsubscript{step}=1000$ (lowest frequency).
    Large $c_\textsubscript{step}$ introduces granular noise in the current profile.
    }
    \label{fig:cumulative_deposition_comparison}
\end{figure}

Although the RE in Fig.\,\ref{fig:cumulative_deposition} is initialized with a small pitch angle ($\eta = 10^\circ$) so that its parallel current is negative (counter-current flow), consistent with experiments (Fig.\,\ref{fig:j_par_184602}), Fig.\,\ref{fig:cumulative_deposition} exhibits unphysical positive current regions. These positive current regions  arise from the steep gradients imposed by the Dirac-delta function representation of a single particle. In contrast, Fig.\,\ref{fig:cumulative_deposition_comparison} employs $10^6$ particles, which smooths out these artificial positive contributions. As in Fig.\,\ref{fig:cumulative_deposition}, sparse deposition ($c_\textsubscript{step}  = 1000$) in Fig.\,\ref{fig:cumulative_deposition_comparison} amplifies pointwise noise relative to higher-frequency updates ($c_\textsubscript{step}  = 1$), but no longer exhibits the co-current artifacts, because summation over the $10^6$-particle ensemble suppresses them. Even with a large ensemble, spurious sign reversals may persist in regions with insufficient mesh refinement because the finite element representation, at a fixed polynomial degree, cannot resolve arbitrarily steep gradients. We are developing sign-preserving regularization methods to resolve these issues; details will appear in future work.

\begin{figure}[ht]
    \centering
    \includegraphics[scale=0.38]{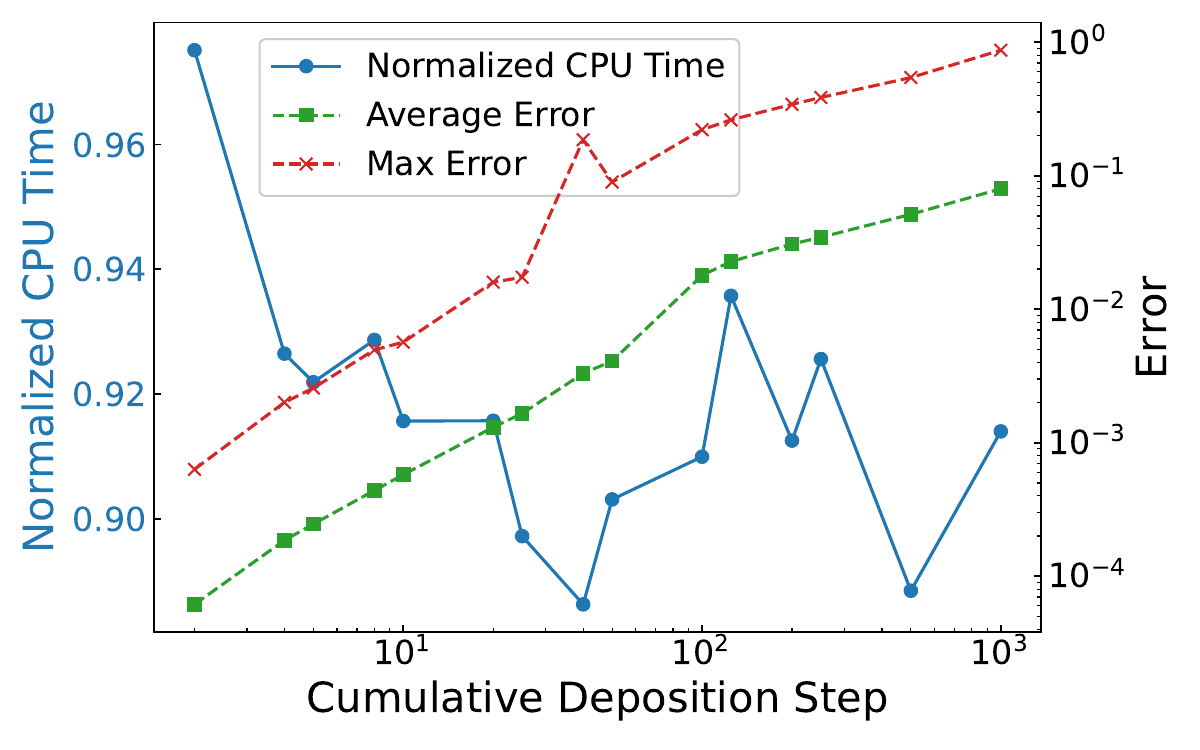}
    \caption{Variation of normalized CPU time (left axis) and global errors (right axis) versus the cumulative deposition step.  
    The error is measured relative to the high-frequency fractional-deposition baseline $c_\textsubscript{step}=1$.  
    Results are the median of 10 runs, each using one MHD step of $1.0 \times 10^{-8}\,\mathrm{s}$ and a KORC step of $1.0 \times 10^{-11}\,\mathrm{s}$. 
    Fewer depositions (larger $c_\textsubscript{step}$) yield negligible computational savings but significantly increase the error.
    }
    \label{fig:timings_errors}
\end{figure}

Figure\,\ref{fig:timings_errors} illustrates the trade-off between numerical cost and accuracy in a previously considered scenario with $10^6$ REs on the refined mesh (Fig.\,\ref{fig:compare_mesh}(b)).
We fix $\Delta t_{\mathrm{dump}}= 1.0 \times 10^{-8}\,\mathrm{s}$, $\Delta t_{\mathrm{KORC}}= 1.0 \times 10^{-11}\,\mathrm{s}$, and vary $c_\textsubscript{step}$ from 1 to 1000. 
The $c_\textsubscript{step}=1$ run serves as the baseline. The average and maximum errors, computed via Eqs.\,\eqref{eq:mean_error}--\eqref{eq:max_error}, 
grow approximately as 
\begin{equation}
\mathrm{Error} \propto c_\textsubscript{step}^{1.2}.
\end{equation}
Although reducing the deposition frequency lowers the number of fractional depositions, the total CPU time decreases by $\sim 10\%$ at most, 
while the global errors in $J_{\parallel}$ can grow by three orders of magnitude (e.g., $c_\textsubscript{step}=1000$ vs.\ $c_\textsubscript{step}=2$). 

The orbit-averaging method provides a tunable way to control numerical cost vs.~accuracy. Since large intervals between fractional depositions significantly
degrade accuracy with negligible speedup more frequent cumulative deposition is preferable. Orbit-averaging will be employed in future self-consistent KORC-NIMROD simulations to minimize particle count while maintaining accuracy.

%%%%%%%%%%%%%%%%%%%%%%%%%%%%%%%%%%%%%%%%%%%%%%%%%%%%%%%%%%%%%%%%%%%%%%%%%%%%%%%
%%%%%%%%%%%%%%%%%%%%%%%%%%%%%%%%%%%%%%%%%%%%%%%%%%%%%%%%%%%%%%%%%%%%%%%%%%%%%%%

\subsection{Major radius shift}

\begin{figure}[ht]
    \centering
    \includegraphics[scale=0.28]{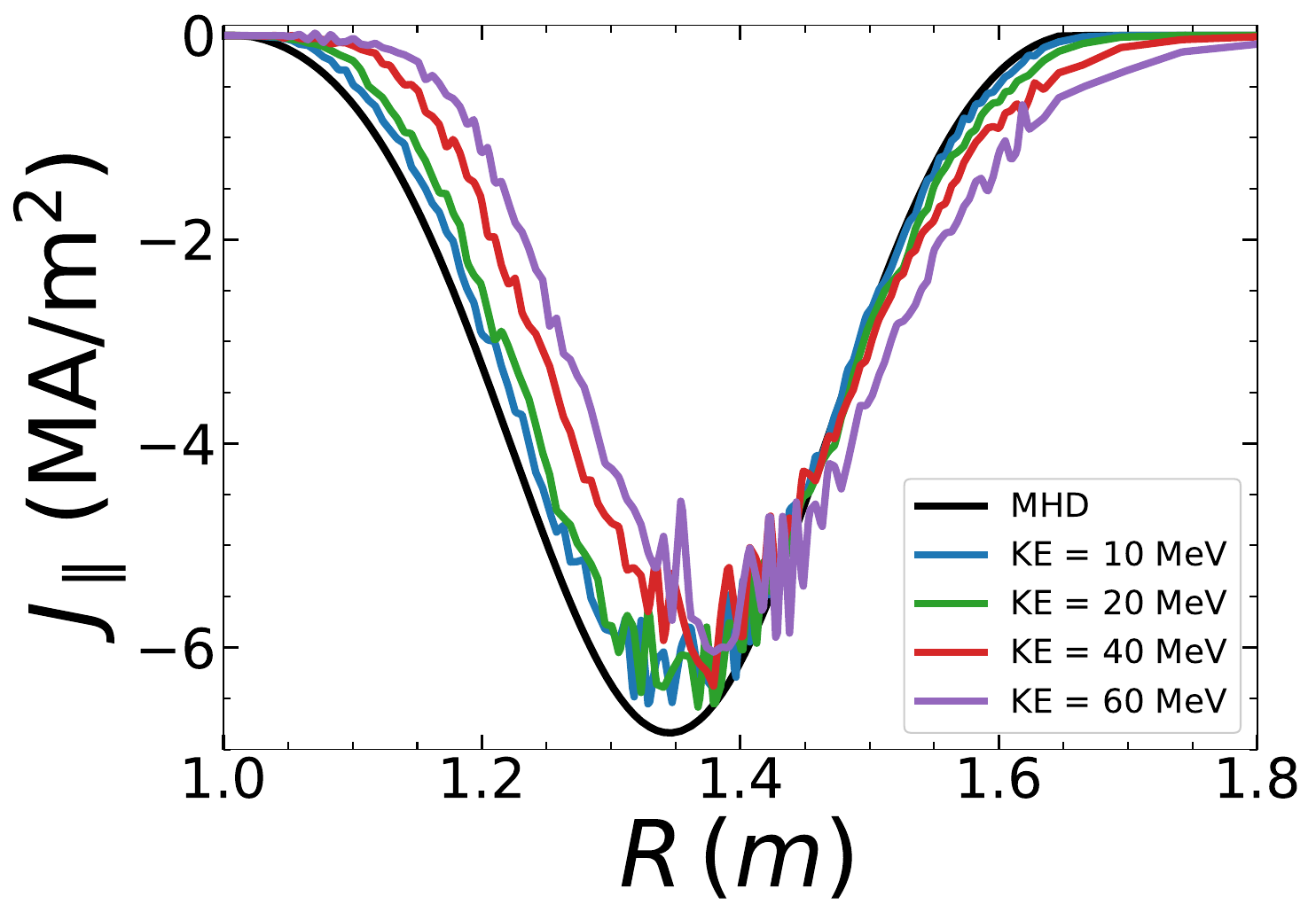}
    \caption{Outward shifts for a mono-pitch RE beam as a function of its initial kinetic energy, 
    holding an initial pitch of $\eta=10^\circ$. RE beams have been evolved for $1\,\mu\mathrm{s}$. 
    Deposition uses a high-frequency orbit-averaging method with $c_\textsubscript{step}=1$.}
    \label{fig:shift_deltaR}
\end{figure}

In the previous 10\,MeV simulations, the net major-radius shift was negligible.  However, when we increase RE energy beyond 10\,MeV, an outward radial drift emerges on the microsecond timescale.  As shown in Fig.\,\ref{fig:shift_deltaR}, 20--60\,MeV beams develop progressively larger centroid displacements.  We fit a 2D Gaussian to the RE sample at $1\,\mu$s to measure this shift $\Delta$, summarized in Table\,\ref{tab:energy_comparison_shift}.

We adopt a simplified constant-$q$ model to compute a proxy for the major-radius shift, $\Delta$. 
When variations in the toroidal velocity are ignored, as discussed in Refs. \inlinecite{carbajal17,diego18}, 
the guiding-center orbits in the $R-Z$ plane can be approximated as circles offset by $\Delta = q_0\,V_{\phi}/\Omega_{e}$, where $q_0$ is the safety factor at the magnetic axis,
$V_{\phi}$ is the toroidal velocity, and $\Omega_{e}$ is the gyrofrequency. 
In the DIII-D shot 184602 under analysis, the safety factor is monotonically increasing from an on-axis value of $q_0=0.5$ (see Fig.\,\ref{fig:q_184602}), 
and the guiding-center velocity generally varies. Hence, $\Delta$ only serves as an order-of-magnitude indicator.

Table\,\ref{tab:energy_comparison_shift} compares the fitted shifts to the analytical proxy. 
The proxy values match the fitted shifts in order of magnitude, specially at higher energies. 
The 10\,MeV case shows a small negative fit result, likely due to numerical noise and consistent with an effectively negligible shift. 
Overall, these findings confirm that energetic beams experience progressively larger displacements in the major radius as their kinetic energy increases.

\begin{table}[htbp]
    \centering
    \resizebox{0.85\columnwidth}{!}{%
    \begin{tabular}{|c|c|c|}
    \hline
    \textbf{Energy (MeV)} & \textbf{Gaussian-Fit $\Delta$ (m)} & \textbf{Proxy $\Delta$ (m)} \\
    \hline
    10 & $-4.8 \times 10^{-3}$ & $6.2 \times 10^{-3}$ \\
    20 & $2.9 \times 10^{-3}$  & $1.2 \times 10^{-2}$ \\
    40 & $1.9 \times 10^{-2}$  & $2.4 \times 10^{-2}$ \\
    60 & $3.9 \times 10^{-2}$  & $3.6 \times 10^{-2}$ \\
    \hline
    \end{tabular}
    }
    \caption{Shift in the major radius for the RE beam at different initial energies ($\eta=10^\circ$). The analytical proxy $\Delta$ uses a constant-$q$ model with $q_{0}=0.5$.}
    \label{tab:energy_comparison_shift}
\end{table}

%%%%%%%%%%%%%%%%%%%%%%%%%%%%%%%%%%%%%%%%%%%%%%%%%%%%%%%%%%%%%%%%%%%%%%%%%%%%%%%
%%%%%%%%%%%%%%%%%%%%%%%%%%%%%%%%%%%%%%%%%%%%%%%%%%%%%%%%%%%%%%%%%%%%%%%%%%%%%%%
%%%%%%%%%%%%%%%%%%%%%%%%%%%%%%%%%%%%%%%%%%%%%%%%%%%%%%%%%%%%%%%%%%%%%%%%%%%%%%%
%%%%%%%%%%%%%%%%%%%%%%%%%%%%%%%%%%%%%%%%%%%%%%%%%%%%%%%%%%%%%%%%%%%%%%%%%%%%%%%
\section{Conclusions}\label{sec:conclusions}

In this study, we advanced toward fully self-consistent, time-dependent kinetic-MHD modeling by focusing on RE dynamics during the RE plateau. This setting allowed rigorous verification of numerical methods without the additional complexity of evolving fields. 
Methodological enhancements included accurate initialization of RE distributions from experimental parallel current profiles, 
robust particle-to-mesh mapping via a barycentric search algorithm, and an orbit-averaging deposition strategy that reduces numerical noise efficiently. Collectively, these developments establish the essential numerical foundation for future time-dependent kinetic-MHD coupling between KORC and NIMROD, ultimately supporting the modeling of RE mitigation strategies.

In this work, a sampling procedure initializes energetic RE beams from EFIT-based $J_{\parallel}(R,Z)$ in the post-thermal-quench DIII-D shot 184602. A barycentric search provides initial guesses for the Newton-Raphson logical-to-physical coordinates solver, ensuring robust particle-to-mesh mapping.

Cross-verification with DepoPy (Sec.\,\ref{sec:DepoPy}) validates NIMROD's axisymmetric deposition, with $\sim 10^{-4}$ global errors for $\sim 10^6$ particles. At the initial time (Sec.\,\ref{sec:error_analysis_parallel_current}), refined meshes require more particles for comparable accuracy, but both coarse and refined meshes keep parallel current average errors under a few percent with $\sim10^7$ particles.

Over the $1\,\mu\mathrm{s}$ time scale, low mesh resolution and low polynomial degree ($\text{poly-deg}=1$) induce significant under- and overshoots in the deposited parallel current density near the magnetic axis (Fig.\,\ref{fig:evol_dep_mesh_comp}). These artifacts result from inaccurate single-particle orbits caused by low-order magnetic field representations, which generate spurious inward or outward radial drifts (Fig.\,\ref{fig:combined_comparison}, second column). Such spurious drifts radially misplace particles, creating artificial undershoots and overshoots of particle counts near the axis in the 2D histograms of Fig.\,\ref{fig:2d_sample_e}. This particle misplacement consequently distorts the spatial profile of the deposited RE parallel current (Figs.\,\ref{fig:evol_dep_mesh_comp_a}, \ref{fig:evol_dep_mesh_comp_c}). Mesh refinement partially suppresses these artifacts. Near-axis orbits computed with high polynomial degree ($\text{poly-deg}=6$) preserve energy and canonical momentum more accurately than those from refined meshes at low order polynomial degree. Although increasing the polynomial degree for deposition in NIMROD could further mitigate near-axis under- and overshoots, it awaits verification. Future work will pursue high-order deposition schemes.

An orbit-averaging procedure (Sec.\,\ref{sec:orbit-averaging}) accumulates partial current deposits across multiple sub-steps within each MHD dump, lowering statistical noise at minimal extra cost (Figs.\,\ref{fig:cumulative_deposition_comparison}--\ref{fig:timings_errors}). Specifically, each interval 
$\Delta t_{\mathrm{dump}}$ is subdivided into $M$ kinetic steps, yet only one inversion of the 
mass-matrix system (Eq.\,\eqref{eq:mass_matrix_system}) is required at the end of the interval. 
Although sparser deposits yield only marginal speedup, they greatly increase noise. By using a high deposition frequency, one could significantly reduce 
the number of particles required, while still achieving error levels comparable to larger ensembles 
with less frequent deposits. The orbit-averaging method offers a robust way to balance accuracy and cost, and it will be essential 
for future KORC-NIMROD coupling. Finally, we have shown that for a $10\,\mathrm{MeV}$ beam, the major-radius shift is effectively negligible, whereas at $60\,\mathrm{MeV}$ it reaches $\sim4\,\mathrm{cm}$ after $1\,\mu\mathrm{s}$ (Fig.\,\ref{fig:shift_deltaR}, 
Table\,\ref{tab:energy_comparison_shift}), consistent with a constant-$q$ orbit proxy. 

Future work will incorporate refinements motivated by Ref.~\inlinecite{bandaru2023}, including the role of RE curvature drift and deviations from force-free equilibria at high energies, as well as by Ref.~\inlinecite{marini2024}, whose diagnostic-constrained RE current profiles indicate significant modifications as compared with RE currents determined from EFIT-based reconstructions. In parallel, we will verify NIMROD's higher-degree polynomial depositions using the Magnetic-Axis-Basis (MAB) approach, ensuring continuity at the magnetic axis, implement sign-preserving deposition schemes, and continued advancement toward fully self-consistent KORC-NIMROD coupling.

%%%%%%%%%%%%%%%%%%%%%%%%%%%%%%%%%%%%%%%%%%%%%%%%%%%%%%%%%%%%%%%%%%%%%%%%%%%%%%%
%%%%%%%%%%%%%%%%%%%%%%%%%%%%%%%%%%%%%%%%%%%%%%%%%%%%%%%%%%%%%%%%%%%%%%%%%%%%%%%
%%%%%%%%%%%%%%%%%%%%%%%%%%%%%%%%%%%%%%%%%%%%%%%%%%%%%%%%%%%%%%%%%%%%%%%%%%%%%%%
%%%%%%%%%%%%%%%%%%%%%%%%%%%%%%%%%%%%%%%%%%%%%%%%%%%%%%%%%%%%%%%%%%%%%%%%%%%%%%%

\section{Acknowledgements}

We thank Dr. E. C. Howell for discussions on constructing NIMROD equilibrium files from EFIT files, and initial assistance with code compilation. Dr. V. Izzo for sharing an initial particle data file and discussing the original implementation of particle advancement in NIMROD. Dr. C. C. Kim for early discussions on the original $\delta f$ PIC implementation in NIMROD. 

This research used resources of the Experimental Computing Laboratory (ExCL) at the Oak Ridge National Laboratory, which is supported by the Office of Science of the U.S. Department of Energy under Contract No. DE-AC05-00OR22725.
This research used resources of the National Energy Research Scientific Computing Center (NERSC), a US Department of Energy Office of Science User Facility located at Lawrence Berkeley National Laboratory, operated under Contract No. DE-AC02-05CH11231.  
This work uses the DIII-D National Fusion Facility, a DOE Office of Science user facility, under Award DE-FC02-04ER54698.  

\section{Disclaimer}

This report was prepared as an account of work sponsored by an agency of the United States Government. Neither the United States Government nor any agency thereof, nor any of their employees, makes any warranty, express or implied, or assumes any legal liability or responsibility for the accuracy, completeness, or usefulness of any information, apparatus, product, or process disclosed, or represents that its use would not infringe privately owned rights. Reference herein to any specific commercial product, process, or service by trade name, trademark, manufacturer, or otherwise, does not necessarily constitute or imply its endorsement, recommendation, or favoring by the United States Government or any agency thereof. The views and opinions of authors expressed herein do not necessarily state or reflect those of the United States Government or any agency thereof.
%%%%%%%%%%%%%%%%%%%%%%%%%%%%%%%%%%%%%%%%%%%%%%%%%%%%%%%%%%%%%%%%%%%%%%%%%%%%%%%
%%%%%%%%%%%%%%%%%%%%%%%%%%%%%%%%%%%%%%%%%%%%%%%%%%%%%%%%%%%%%%%%%%%%%%%%%%%%%%%
%%%%%%%%%%%%%%%%%%%%%%%%%%%%%%%%%%%%%%%%%%%%%%%%%%%%%%%%%%%%%%%%%%%%%%%%%%%%%%%
%%%%%%%%%%%%%%%%%%%%%%%%%%%%%%%%%%%%%%%%%%%%%%%%%%%%%%%%%%%%%%%%%%%%%%%%%%%%%%%

\appendix

\section{RE parallel current in the RGC approximation}\label{sec:CurrentApp}

This Appendix derives the single-particle parallel RE current, $I_{\parallel}^{(1e)}$ (Eq.\,\ref{eq:single_j}), 
using the relativistic guiding-center (RGC) model from Sec.\,\ref{subsec:RelativisticGuidingCenter}. 
This current is crucial for both particle initialization via Metropolis-Hastings sampling (Sec.\,\ref{sec:MH}) 
and setting deposition weights on the NIMROD mesh (Eqs.\,\ref{eq:weight_parallel_current} and \ref{eq:weight_parallel_current_alpha}).

Each RE is represented by a \emph{single-particle PDF},
$f^{(1e)}(\mathbf{X},p_{\parallel},\mu,\chi)$, where $\chi$ is the gyro-phase, 
and the remaining quantities are defined in Sec.\,\ref{subsec:RelativisticGuidingCenter}. 
The function $f^{(1e)}$ integrates to unity over phase space:
\begin{equation}
\int d^3V d^3X \, f^{(1e)}(\mathbf{X},p_{\parallel},\mu,\chi) = 1.
\end{equation}
Here, $ d^3X $ is the volume element in configuration space, 
and $d^3V $ is the volume element in velocity space. 
Since $ f^{(1e)} $ is expressed in the velocity-space coordinates $(p_{\parallel},\mu,\chi)$, the velocity-space volume element is transformed accordingly:
\begin{equation}\label{eq:volumenElement}
d^3V=\frac{B}{\gamma^5\,m^2} dp_{\parallel} d\mu d\chi.
\end{equation}

For a RE guiding-center located at $(R_j, \phi_j, Z_j)$ with parallel momentum $p_{\parallel,j} $ and magnetic moment $ \mu_j $, 
we introduce the \emph{single-particle guiding-center PDF}, expressed in terms of Dirac-delta functions, as:
\begin{align}\label{eq:rePDF}
    f_{RGC}^{(1e)}(\mathbf{X},p_{\parallel},\mu)
    &= \frac{m^2 \gamma^5}{2\pi B}
    \,\frac{\delta(R - R_j)\,\delta(Z - Z_j)\delta(\phi - \phi_j)}{R} 
    \nonumber \\
    &\quad \times \delta\bigl(p_{\parallel}-p_{\parallel j}\bigr)
    \,\delta\bigl(\mu - \mu_j\bigr),
\end{align}
where the factor $ 2\pi $ in the denominator of Eq.\,\eqref{eq:rePDF} accounts for the uniform distribution of the gyro-phase $ \chi $ over the interval $ [0,2\pi)$, ensuring proper normalization of the probability density function.

Following Eq.\,(14) of Ref.~\inlinecite{todo98}, the RE current density is:
\begin{equation}\label{eq:single_j_vector}
\mathbf{J}^{(1e)} = \int \left( \mathbf{V}_\parallel^* + \mathbf{V}_B \right) f_{RGC}^{(1e)} d^3V - \nabla \times \int \mu \hat{\mathbf{b}} f_{RGC}^{(1e)} d^3V,
\end{equation}
where
\begin{equation}
\mathbf{V}_\parallel^* = \frac{p_{\parallel}\mathbf{B}^*}{\gamma m \hat{\mathbf{b}} \cdot \mathbf{B}^*},
\end{equation}
and
\begin{equation}
\mathbf{V}_{B} = \frac{\mu}{\gamma \hat{\mathbf{b}} \cdot \mathbf{B}^*} \hat{\mathbf{b}}\times\nabla B.
\end{equation}
The quantity $\mathbf{B}^*$ is defined in Eq.\,\eqref{eq:Bstar}. Ref.~\inlinecite{todo98} and the subsequent work in 
Ref.~\inlinecite{todo04} introduce a $\delta f$ hybrid kinetic-MHD model for energetic particles using the RGC model. 
This formulation was later extended to full-$f$ hybrid kinetic-MHD simulations of energetic ions in the MEGA code \cite{bierwage13}. 
In Sec.\,\ref{subsec:RelativisticGuidingCenter}, we adopt the same RGC model, adapting it for REs.

We recall from Sec.\,\ref{sec:MH} that the RE parallel current alone is sufficient to consistently initialize the RE beam from experimental reconstructions. This simplification arises because pressure gradients are negligible during the RE plateau phase, implying that the Grad-Shafranov equilibrium is determined by the parallel current. By projecting Eq.\,\eqref{eq:single_j_vector} onto $\hat{\mathbf{b}}$, the cross-product term $\hat{\mathbf{b}}\times\nabla B$ vanishes. Integrating over velocity space using Eqs.\,\eqref{eq:volumenElement} and \eqref{eq:rePDF} yields:
\begin{equation}
    J_{\parallel}^{(1e)}(\mathbf{X}) = I_{\parallel}^{(1e)}(R,Z)\,\delta(R - R_j)\,\delta(Z - Z_j)\,\delta(\phi - \phi_j),
\end{equation} 
where the scalar quantity $I_{\parallel}^{(1e)}(R,Z)$ represents the magnitude of the single-particle current along the magnetic field (with units of amperes), given explicitly by:
\begin{equation}\label{eq:single_j2}
    I_{\parallel}^{(1e)}(R,Z)=\frac{1}{2\pi R}\left(\frac{q_e p_{\parallel}}{m\,\gamma}-\mu\,(\hat{\mathbf{b}}\cdot\nabla\times\hat{\mathbf{b}})\right).
\end{equation}
This expression corresponds exactly to Eq.\,\eqref{eq:single_j}, serving as the definition of the single-particle contribution to the parallel current.

%%%%%%%%%%%%%%%%%%%%%%%%%%%%%%%%%%%%%%%%%%%%%%%%%%%%%%%%%%%%%%%%%%%%%%%%%%%%%%%
%%%%%%%%%%%%%%%%%%%%%%%%%%%%%%%%%%%%%%%%%%%%%%%%%%%%%%%%%%%%%%%%%%%%%%%%%%%%%%%
%%%%%%%%%%%%%%%%%%%%%%%%%%%%%%%%%%%%%%%%%%%%%%%%%%%%%%%%%%%%%%%%%%%%%%%%%%%%%%%
%%%%%%%%%%%%%%%%%%%%%%%%%%%%%%%%%%%%%%%%%%%%%%%%%%%%%%%%%%%%%%%%%%%%%%%%%%%%%%%

\section{Barycentric search algorithm}\label{sec:BarycentricApp}

This appendix presents pseudocode for the barycentric particle-to-element mapping described in Sec.\,\ref{sec:barycentric}. 
The poloidal index is computed via binary search over poloidal sectors, and the radial index through barycentric coordinates
within nested triangles. 

Here, the poloidal sector boundaries of Sec.\,\ref{sec:poloidal_index} are collected in $\Theta = \{\theta_{0},\theta_{1},\ldots,\theta_{m-1}\}$.
Without loss of generality we set $\theta_{0}=0$ and measure all angles counter-clockwise from $\theta_{0}$, so that $\Theta$ is strictly ascending; 
$\theta^{*}$ (the poloidal angle of a particle at $\mathbf{P}$) is measured in the same manner.
Figure\,\ref{fig:poloidal_radial} illustrates these quantities.
Algorithm\,\ref{alg:bisectsort} returns the index $j^{*}\in\{0,\ldots,m-1\}$ satisfying either
$\theta_{j^{*}}\le\theta^{*}<\theta_{j^{*}+1}$ for $(0\leq j^{*}<m-1)$, or $\theta_{m-1}\le\theta^{*}<\theta_{0}$ in the case $j^{*}=m-1$; i.e. the insertion point of $\theta^{*}$ in the ordered list $\Theta$.
The procedure is a classic binary search with $\mathcal{O}(\log m)$ complexity.

\begin{algorithm}[h]
    \caption{Poloidal index determination}
    \KwIn{A sorted array of poloidal angles $\Theta$ and the poloidal angle $\theta^*$ of particle at $\mathbf{P}$}
    \KwOut{Index where $\theta^*$ should be inserted}

    Initialize $low \gets 0,\; high \gets \mathrm{m}$\

    \While{$low$ < $high-1$}{
        $mid$ $\gets$ $\left\lfloor \frac{low + high}{2}  \right\rfloor$\

        \If{$\theta^*$ < $\theta_{mid}$}{
            $high$ $\gets$ $mid$\
        }
        \Else{
            $low$ $\gets$ $mid$\
        }
    }
    \Return{ $low$}\
    \label{alg:bisectsort}
\end{algorithm}

Radial localization follows Algorithms\,\ref{alg:barycentric}-\ref{alg:barycentricsort}.  
Algorithm\,\ref{alg:barycentric} evaluates the barycentric coordinates of $\mathbf{P}$ relative to the triangle $(\mathbf{A},\mathbf{B},\mathbf{C})$.  
Algorithm\,\ref{alg:barycentricsort} then performs a binary search through the $n$ nested triangles extending from the magnetic axis to the mesh boundary. 
At each iteration it checks, via the barycentric coordinates, whether $\mathbf{P}$ lies inside the current triangle. 
Throughout this search $\mathbf{A}=(R_{mag},Z_{mag})$ remains the magnetic-axis vertex, while $\mathbf{B}$ and $\mathbf{C}$ are updated as the algorithm narrows the radial index (see Fig.\,\ref{fig:barycentricsort}).

\begin{algorithm}[h]
    \caption{Barycentric coordinates determination}
    \KwIn{Three triangle vertices $\mathbf{A}=(a_x,a_y), \mathbf{B}=(b_x,b_y), \mathbf{C}=(c_x,c_y)$ and a point $\mathbf{P}=(p_x,p_y)$}
    \KwOut{Barycentric coordinates $(\alpha, \beta, \gamma)$ of $\mathbf{P}$}

    \tcp{Calculate triangle areas using determinant-based formula}
    $\mathcal{A}_{\mathbf{ABC}} \gets \frac{1}{2} \left|\begin{matrix}
    b_x - a_x & c_x - a_x \\ 
    b_y - a_y & c_y - a_y
    \end{matrix}\right|$\

    $\mathcal{A}_{\mathbf{PBC}} \gets \frac{1}{2} \left|\begin{matrix}
    p_x - b_x & c_x - b_x \\
    p_y - b_y & c_y - b_y
    \end{matrix}\right|$\

    $\mathcal{A}_{\mathbf{PCA}} \gets \frac{1}{2} \left|\begin{matrix}
    p_x - c_x & a_x - c_x \\
    p_y - c_y & a_y - c_y
    \end{matrix}\right|$\

    \tcp{Compute Barycentric coordinates}
    $\alpha \gets \frac{\mathcal{A}_\mathbf{PBC}}{\mathcal{A}_\mathbf{ABC}}$\;
    $\beta \gets \frac{\mathcal{A}_\mathbf{PCA}}{\mathcal{A}_\mathbf{ABC}}$\;
    $\gamma \gets 1 - \alpha - \beta$\tcp*{Since $\alpha + \beta + \gamma = 1$}

    \Return $(\alpha, \beta, \gamma)$\
    \label{alg:barycentric}
\end{algorithm}

\begin{algorithm}[h]
    \caption{Radial index determination}
    \KwIn{Triangle vertices $\mathbf{A}$, $\mathbf{B}$, $\mathbf{C}$ and a particle at $\mathbf{P}$}
    \KwOut{Index where $\mathbf{P}$ should be inserted based on barycentric coordinates}

    Initialize $low \gets 0,\; high \gets \mathrm{n}$\

    \While{$low < high - 1$}{
        $mid \gets \left\lfloor \frac{low + high}{2}  \right\rfloor$\;
        $\mathbf{B} \gets \bigl(\mathcal{R}(\xi_{mid},\, \upsilon_{j^*}),\; \mathcal{Z}(\xi_{mid},\, \upsilon_{ j^*})\bigr)$\;
        $\mathbf{C} \gets \bigl(\mathcal{R}(\xi_{mid},\,\upsilon_{j^*+1}),\; \mathcal{Z}(\xi_{mid},\,\upsilon_{ j^*+1})\bigr)$\;
        $(a,b,c)\gets\mathrm{barycentric~coordinates}(\mathbf{A},\mathbf{B},\mathbf{C},\mathbf{P})$\;
        
        \If{$a \geq 0$ \textbf{and} $b \geq 0$ \textbf{and} $c \geq 0$}{
            $high \gets mid$\
        }
        \Else{
            $low \gets mid$\
        }
    }
    \Return $low$\
    \label{alg:barycentricsort}
\end{algorithm}

\bibliographystyle{unsrtnat}
\bibliography{references}    % external file reference.bib

\end{document}